\documentclass[acmsmall]{acmart}
\usepackage{enumitem}
\usepackage{listings}
\usepackage{physics}
\usepackage{wasysym}
\usepackage{marvosym}
\usepackage{pythontex} 
\usepackage{multicol}
\usepackage{multirow}
\usepackage{hyperref}
\usepackage{cancel}
\usepackage{color}
\usepackage{bm}
\usepackage[para]{footmisc}

\AtBeginDocument{%
  \providecommand\BibTeX{{%
    \normalfont B\kern-0.5em{\scshape i\kern-0.25em b}\kern-0.8em\TeX}}}

\setcopyright{acmcopyright}
\copyrightyear{2020}
\acmYear{2020}

\acmJournal{JACM}
\acmVolume{XX}
\acmNumber{XX}
\acmArticle{XX}

\begin{document}

\title{Testing Feedforward Neural Networks Training Programs}

\author{Houssem Ben Braiek}
\affiliation{%
  \institution{SWAT Lab., Polytechnique Montreal}
%   \streetaddress{8600 Datapoint Drive}
  \city{Montreal}
%   \state{Quebec}
  \country{Canada}
%   \postcode{78229}
  }
\email{houssem.ben-braiek@polymtl.ca}
\author{Foutse Khomh}
\email{foutse.khomh@polymtl.ca}
\affiliation{%
  \institution{SWAT Lab., Polytechnique Montr\'{e}al}
  \streetaddress{2500 Chemin de Polytechnique} 
  \city{Montr\'{e}al}
  %\state{Qu\'{e}bec}
  %\postcode{H3T 1J4}
  \country{Canada}
}
\definecolor{cadmiumgreen}{rgb}{0.0, 0.42, 0.24}
\newcommand{\Houssem}[1]{\textcolor{cadmiumgreen}{{\it [Houssem says: #1]}}}
\newcommand{\Foutse}[1]{\textcolor{red}{{\it [Foutse says: #1]}}}
\newcommand{\name}{{\it TheDeepChecker}}
\newcommand{\ie}{{\it i.e.,}}
\renewcommand{\shortauthors}{Ben Braiek and Khomh}

\begin{abstract}
Nowadays, we are witnessing an increasing effort to improve the performance and trustworthiness of Deep Neural Networks (DNNs), with the aim to enable their adoption in safety critical systems such as self-driving cars or aircraft collision-avoidance systems. Multiple testing techniques are proposed to generate test cases that can expose inconsistencies in the behavior of DNN models. 
These techniques assume implicitly that the training program is bug-free and appropriately configured. However, satisfying this assumption for a novel problem requires significant engineering work to prepare the data, design the DNN, implement the training program, and tune the hyperparameters in order to produce the model for which current automated test data generators search for corner-case behaviors. All these model training steps can be error-prone. Therefore, it is crucial to detect and correct errors throughout all the engineering steps of DNN-based software systems and not only on the resulting DNN model. 

In this paper, we gather a catalog of training issues and based on their symptoms and their effects on the behavior of the training program, we propose practical verification routines to detect the aforementioned issues, automatically, by continuously validating that some important properties of the learning dynamics hold during the training. Then, we design, \name{}, an end-to-end property-based debugging approach for DNN training programs and implement it as a TensorFlow-based library. As an empirical evaluation, we conduct a case study to assess the effectiveness of \name{} on synthetic and real-world buggy DL programs and compare its performance to that of the Amazon SageMaker Debugger (\textit{SMD}). 
Results show that \name{}'s on-execution validation of DNN-based program's properties through three sequential phases (pre-, on-, and post-fitting), succeeds in revealing several coding bugs and system misconfigurations errors, early on and at a low cost. Moreover, our property-based approach outperforms the \textit{SMD}'s offline rules verification on training logs in terms of detection accuracy for unstable learning issues and coverage of additional DL bugs.
\end{abstract}



\maketitle

\section{Introduction}
Nowadays, we are witnessing an increase in efforts to improve the reliability of Deep Neural Networks (DNNs), and consequently the trustworthiness of deep learning based systems such as self-driving cars or aircraft collision-avoidance systems~\cite{DLTestReview}. Multiple approaches are being developed to search for test cases that can expose inconsistencies in the behavior of DNN models; we refer to them as model testing approaches. These approaches implicitly assume that the trained DNN is already performing well on the original test dataset compared to known results, i.e., the training program is bug-free and appropriately configured. However, recent research works~\cite{Comp_DL_Bugs}~\cite{DLfaults}~\cite{DL_bugs_1} report that bugs exist in DL programs, which invalidates this assumption made by model testing approaches. Their main finding is that faults that lead to program crashes represent only a fraction of real bugs found in DNN training programs. The vast majority of bugs are due to hidden logic errors and configuration inconsistencies, leading to silent failures that are difficult to detect (since they do not prevent the program from running and producing a model). In fact, a DNN is trained using the back-propagation algorithm that relies on a loss function to estimate the distance between actual predictions and the ground truth, and then, the estimated error is back propagated through the DNN's learnable parameters to adjust their values in the opposite direction of the loss gradient. In practice, components of the training algorithm are provided as ready-to-use configurable routines by DL libraries, however, reusing these routines to implement a training program for a designed DNN is not straightforward and it can be error-prone. From a fundamental point of view, the backpropagation algorithm can be considered as a leaky abstraction since the details of its implementation are not trivial. To illustrate this point let's consider the basic rule of weights initialization which states that values should be small random numbers. Setting weights' values is not simple and straightforward because the use of dummy random initialization could prevent the DNN from training. In practice, depending on developers' design choices, there is a set of custom weight initializations that have been formally proven to be optimal choices, and hence should be adopted by the developers. 

DL developers also leverage generic-purpose libraries for scientific computations and DL routines that support a rich set of APIs, incorporating many assumptions in order to make them ready-to-use for common problems. These DL routines can be configured and adapted for new problems. However, their misuses, i.e., the usage of APIs without fully understanding their inner functioning and--or checking that their assumptions are fulfilled, are likely to result in erroneous behavior. The implementation of the training steps can also contain errors leading to erroneous behavior. It is therefore very important to test the source code of the training program, to ensure that it accurately reflects the algorithm's mathematical formulation.\\
The current trend in testing the resulting DNN model is not sufficient to ensure that the training process was bug-free and consequently that the DNN model is reliable. Because all models are imperfect abstractions of the reality and subject to residual error, it is not possible to rely on the observable relations between the inputs and outputs of the model to hunt for errors in the training program; if the trained model produces a wrong prediction, it does not necessarily mean that the training program is buggy. Moreover, the level of performance of a model (in terms of precision, recall, or accuracy) cannot serve as a reliable predictor of the reliability of its training process, since it is unclear what should be an appropriate level of performance for a model, to discard any coincidental high probabilistic measure while taking into consideration various assumptions regarding model capacity and data complexity. It is unclear how much precision, recall, or accuracy may be indicative of errors in the training process. D'amour et al.~\cite{d2020underspecification} have investigated the credibility of various practical ML systems, using examples from computer vision, medical imaging, natural language processing, etc. They conclude the limitation of model testing on the held-out data, called iid performance, and show that plenty of ML models share strong iid performance on test datasets, however, they often exhibit unexpectedly distant behaviors when they are deployed in real-world domains. Hence, the iid performance test is insufficient to separate between effective and non-effective models that are trained using, respectively, clean and buggy training programs. Therefore, we need more indicative measures than probabilistic correctness measures. It is crucial to develop efficient debugging techniques that developers can use to detect errors in the training programs of DNNs. Besides the loss- and accuracy evolution curves, researchers~\cite{viz_review} have recently proposed advanced visual analytics systems to help DL developers debug and refine the design of their DNNs. Oftentimes, this requires monitoring individual computational units of the model during the training, such as activation maps or error gradients produced at each layer, and after training, an instance-based analysis is performed to identify misclassified instances from a handful of chosen data instances. 
However, such visualization based diagnosis techniques require significant human intervention and good expertise on deep learning concepts. Given the internal complexity of a DNN, it is always challenging to select a handful of drawing representations (i.e., suitable for screen display) that should be watched and analyzed, interactively. Additionally, these visualizations can hardly be used automatically to track for regressions after a program update. We believe that full-fledged software debuggers are needed to support DL developers. \\
In this paper, we propose \name{}, the first end-to-end automated debugging approach for DNN training programs. To develop \name{}, we gathered a catalog of fundamental algorithmic and development issues in relation with the DNN training programs. Then, we inferred the properties that are violated by these identified training program issues, in order to develop the verification routines that can be used to detect their occurrences during the training process. Next, we developed a property-based testing method that activates the derived verification routines, in order to drive an end-to-end automated debugging process for DNN training programs. To assess the effectiveness of \name{}, we implemented it as a TensorFlow-based testing framework that enables the automatic detection of the identified issues in DNN training programs developed with Tensorflow (TF) library. We rely on the taxonomy of real DL faults elaborated in~\cite{DLfaults} and further searching on Stackoverflow, to identify the structure of faults occuring in DNN training programs. Then, we inject these faults in clean DNN training programs to create a set of synthetic buggy programs, each one containing a particular fault, aiming at challenging the \name{} in identifying the faults or steering the users to them by detecting precise fault-indicative symptoms. Moreover, we assess the performance of \name{} on a mixed selection of $20$ real-world TF buggy programs~\cite{DL_bugs_1} splitted equally between snippets of code shared on StackOverflow (SO) and bug-fixing commits from DL projects hosted on GitHub (GH). As a DL debugging baseline, we manage to run all the studied buggy DL programs on Amazon SageMaker Cloud ML service while activating its internal Debugger's built-in rules. Results show that \name{} can successfully support DL practitioners in detecting earlier a wide range of coding bugs and system misconfigurations through reported violations of essential DL program's properties. Additionally, the comparison with SageMaker Debugger (\textit{SMD}) highlights the ascendancy of \name{}'s on-execution validation of DL properties over \textit{SMD}'s rules verification on training logs in terms of detection accuracy and DL bugs' coverage.\\
This paper makes the following contributions.
\begin{itemize}
\item We investigate former DL faults' empirical studies and SO posts about buggy TF training programs to identify common errors made by DL practitioners and provide a catalog outlining DL program's development and configuration pitfalls;
\item We present a property-based DL software debugging approach, which implements verification routines associated to the discovered DL pitfalls across three sequential phases: pre-, on-, and post-fitting of the DNN.
\item We implement our debugging approach based on TF concepts and features to demonstrate the effectiveness of our verification routines in detecting real bugs that occur in TF-based training programs. 
\item By dissecting the root causes and symptoms of collected DL bugs, we simulate these bugs in DNN architectures to create a benchmark of synthetic buggy programs. In addition, a real-world database of buggy DL programs is compiled from snippets of code shared on SO posts and bug-fixing commits in GH projects.
\item We evaluate the effectiveness of \name{} in debugging synthetic and real-world TF programs and compare its detection bugs' accuracy and coverage with Amazon Sagemaker Debugger (\textit{SMD}). Besides, we conduct a usability study on \name{} in collaboration with two experienced DL engineers from the industry, focusing on the usefulness and precision of \name{}'s debugging reports to steer users towards finding and fixing the actual faults in real DL buggy programs.
\end{itemize}
\textbf{The remainder of this paper is organized as follows.} Section~\ref{background} provides background information about DL, property-based software testing, and TensorFlow. Section~\ref{catalog} presents common training program pitfalls discovered by DL researchers and experienced by DL developers. Section~\ref{approach} introduces our proposed property-based debugging approach for DNN training programs alongside their derived verification mechanisms and its TF-based implementation. Section~\ref{evaluation} reports about the empirical evaluation of our proposed debugging approach. Sections~\ref{threats},~\ref{limits_and_future},and~\ref{relatedWork} discuss, respectively, the threats to validity, the limitations and future works, and the related literature. Finally, Section~\ref{conclusion} concludes the paper.
\section{Background}
\label{background}
This section provides background information about DL, property-based software testing, and TensorFlow.
\subsection{Deep Neural Networks}
A deep neural network (DNN) is an artificial neural network with a stack of multiple computational layers, hence the adjective ``deep''. DNNs are often much harder to train than shallow neural networks~\cite{glorot2010}. However, they are endued with a hierarchical features learning that lets them capture increasingly complex patterns directly from the data when they are appropriately designed and trained.
DNNs include many variants of architectures that have found success in several domains. We present in the following the Deep Feed-Forward Neural Network and its popular variant Convolutional Neural Network.
\subsubsection{Deep Feed-Forward Neural Networks}
\label{FNN}
\begin{figure}[ht]
\centering
%\captionsetup{justification=centering}
\includegraphics[scale=0.65]{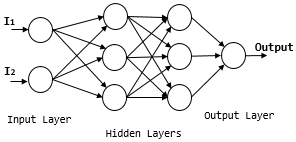}
\caption{Schema of feedforward neural network}
\label{fig:fnn}
%\vspace{-15pt}
\end{figure}

Feedforward neural network (FNN) architecture is the quintessential and the most used neural network~\cite{DL_ebook_2016}. The objective of FNNs is to learn the mapping of a fixed-size input (for example, a signal vector) to a fixed-size output (for example, a probability for each label). Apart from the input and output layers, FNN contains a cascade of multiple intermediate layers, which are called hidden layers because the ground truth data does not provide the desired values for these layers. Last, the name feedforward arises from the fact that information flows through the processing layers in a feed-forward manner, i.e., from input layer, through hidden layers and finally to the output layer. Figure \ref{fig:fnn} illustrates a simple FNN.

A FNN encapsulates a mapping function $f$ that maps the input $x$ to its corresponding output $y$ as presented in Equation \ref{Eq1}. If $y$ is a category label, the FNN solves a classification problem and if $y$ is a continuous value, the FNN performs a regression task. 
\begin{equation}\label{Eq1}
    y = f(x; W, b)
\end{equation}
Where $W$: weights and $b$: biases represent the learnable parameters.\\
Given the training data $D=\{(x_i,t_i); \forall i \in [1,n]\}$, the FNN training algorithm aims to find the best approximation function $f^*$ by learning the optimal values of its inner parameters $W^*$ and $b^*$. To do that, a loss function, $loss(f, x, t)$ is leveraged to measure how well the prediction $y = f(x)$ matches the ground truth output $t$. Considering the example of multinomial classification problem, we have $y$, a vector of probabilities where $y(j)$ represents the probability that $x$ belongs to the label $j$ and $t$, a one-hot encoding label where $t(j) = 1$ if $j$ is the true label; otherwise $t(j)=0$. The most common loss function is the cross entropy $loss(f, x, t) = - \sum_{j}{t(j)\log(y(j))}$.\\
To estimate the loss over all the training data $loss(f, D)$, we use an expectation, as formulated in Equation \ref{Eq2}, that can be either the average or the sum of data instances' losses.
\begin{equation}\label{Eq2}
    loss(f, D) = E_i[loss(f, x_i, y_i)] = E_D[loss(f,D)]
\end{equation}
Therefore, the optimal parameters $W^*$ and $b^*$ result in the best function approximation that spawns the minimum possible estimated loss in the training data $D$. 
\begin{equation}
    W^*, b^* = \underset{W,b}{\mathrm{argmin}}{E_D[loss(f^*, D)]}
\end{equation}
The loss minimization problem can be solved, iteratively, using gradient descent algorithm, where the following equations represent an iteration's updates :
\begin{equation}
    W^{(t+1)} = W^{(t)} - \eta \pdv{E_D[loss(f^{(t)}, D)]}{W} ;\quad
    b^{(t+1)} = b^{(t)} - \eta \pdv{E_D[loss(f^{(t)}, D)]}{b}
\end{equation}
As introduced, the FNN assembles multiple computational layers and each layer $l$ perform a computation; so it encapsulates a kind of sub-function $f_l(x; W_l, b_l)$ with inner parameters $W_l$ and $b_l$. Thus, the approximate mapping function $f$ of an FNN with $L$ layers is a composite function formulated as below :
\begin{equation}
    f(x; W, b) = f_L(f_{L-1}(...(f_1(x;W_1, b_1)...);W_{L-1}, b_{L-1});W_L, b_L)
\end{equation}
Where $W=\{W_l, \forall l\in[1,L]\}$ and $b=\{b_l, \forall l\in[1,L]\}$.\\
Given the huge number of parameters to approximate, the computation of gradients would be very time-consuming. 
Deep learning relies on a fast algorithm, named backpropagation, that applies the derivative chain rule principle to compute, sequentially, all these derivative backing from the output to the first hidden layer, while taking full advantage of the derivatives estimated w.r.t previous layers.
Backpropagation is based on two alternatives main phases, respectively, forward pass and backward pass, which are detailed below.\\
\textbf{Forward Pass. }Each hidden layer $l$ contains a set of computation units, called neurons, that perform a linear transformation $z_l$ of their inputs from previous layers and pass the result through an activation function $\Phi_l$. The latter is a non-linear transformation $a_l$ that allows adding non-linearity in the approximated mapping function in order to be insensitive to irrelevant variations of the input. The layer's computation can be written as:
\begin{equation}
    z_l = W_la_{l-1} + b_l, \forall l\in[1,L]
\end{equation}
\begin{equation}
a_l = \Phi_l(z_l), \forall l\in[1,L]
\end{equation}
We note that $a_0$ which is the input layer activation is equal to the input data $x$. The hidden layers share generally the same activation function. We denote them  $\Phi_1=\Phi_2=...=\Phi_L=\Phi$. We denote the activation function of the output layer by $\Psi=\Phi_L$.\\
\textbf{Backward Pass. }First, we introduce an intermediate quantity, $\delta$, where $\delta_l$ is the vector of error associated to the layer $l$.  $\delta_l$ is computed as the gradient of the loss with respect to the weighted input $z_l$. In the following, we present the equations used by the backpropagation algorithm to compute the error for every layer. We refer the reader to the work of Goodfellow et al.~\cite{DL_ebook_2016} for the proof and in-depth details.
\begin{equation}
\label{Eq3}
    \delta_L = \pdv{loss}{z_L} = \nabla_a(loss)\odot\Psi'(z_L) ;\quad
    \delta_l = \pdv{loss}{z_l} = W^{(l+1)T}\delta_{l+1}\odot\Phi'(z_l)
\end{equation}
In Equation \ref{Eq3}, $\odot$ is the Hadamard product, which is an elementwise product of two vectors in a way that for each component $j$, $v\odot u$ means $(u\odot t)_j=s_jt_j$.\\
Second, we formulate the gradient of loss w.r.t the DNN parameters using the computed error term $\delta$.
\begin{equation}
\label{Eq4}
    \pdv{loss}{W_l} = \delta_la^{(l-1)T} ;\quad
    \pdv{loss}{b_l} = \delta_l
\end{equation}
Last, we perform both of weights and biases iteration updates in the opposite direction of their gradients w.r.t the loss.
\begin{equation}
\label{Eq5}
    W^{(i+1)} = W^{(i)} - \eta \delta_la^{(l-1)T} ;\quad
    b^{(i+1)} = b^{(i)} - \eta \delta_l
\end{equation}
In practice, the backpropagation algorithm does not loop over the training examples and perform the forward and backward passes on each example separately. Indeed, it relies on mini-batch stochastic gradient descent that computes both passes on a mini-batch of examples simultaneously. This is done by formulating the equations presented above as fully matrix-based formulas given the input matrix $X = [x_1,x_2,...,x_m]$ of a mini-batch containing $m$ examples. Thus, the parameters updates are based on the average of loss gradients estimated over all the examples of the batch, which leads to more stable gradient updates.
\subsubsection{Deep Convolutional Neural Networks}
\begin{figure}[ht]
\centering
%\captionsetup{justification=centering}
\includegraphics[scale=0.65]{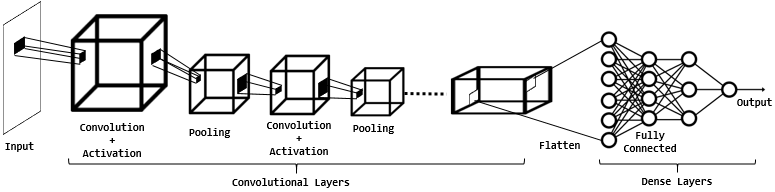}
\caption{Schema of convolutional neural network}
\label{fig:cnn}
%\vspace{-15pt}
\end{figure}

Convolutional Neural Network (CNN) represents a particular type of feedforward network that is designed to process data in the form of multiple arrays, such as 2D images and audio spectrograms, or 3D videos~\cite{DL_ebook_2016}. A CNN contains the following specialized layers that transform the 3D input volume to a 3D output volume of neuron activations: Convolutional Layer, Activation Layer, and Pooling Layer.\\
\textbf{Convolutional Layer. }The main building block of this type of transformation layer is the convolution. A convolution provides a 2D feature map, where each unit is connected to local regions in the input data or previous layer's feature map through a multi-dimensional parameter called a filter or kernel. These filters play the role of feature detectors. The feature map is produced by sliding the filter over the input data, then computing products between the filter entries and the local input region at each spatial position, to infer the corresponding feature map response. Different filters are performed in a layer and resulting feature maps are stacked in 3D volumes of output neurons. The separate filters aim to detect distinctive local motifs in the same local regions. However, all units in one feature map share the same filter, because motifs are generally invariant to location.\\ 
\textbf{Activation Layer. }As in any FNN layer, we use an activation function to add non-linearity to the computed value. The activation layer applies the activation function on the extracted feature map as an element wise operation (i.e., per feature map output). The resulting activation map indicates the state of each neuron, i.e., active or not.\\ 
For each hidden neuron $(i,j)$ in a feature map:
\begin{equation}
    a_{ij} = \Phi(z_{ij}); 
\end{equation}
\textbf{Pooling Layer. }The pooling layer ensures spatial pooling operation to reduce the dimensionality of each feature map and to retain the most relevant information by creating an invariance to small shifts and distortions. Depending on the chosen spatial pooling operations, it can be average or max pooling that computes, respectively, the average or max of all elements in a pre-defined neighboring spatial window size. Therefore, these neighboring pooling makes the resulting activation maps smaller and robust to irrelevant variance, which helps shortening the training time and controlling the overfitting.\\
Figure \ref{fig:cnn} shows a typical architecture of CNN with two main stages: (1) multiple stack of convolution, activation, and pooling that ensure the detection of relevant features from the input data; (2) the final activation map is flatten to be a vector of features and is fed to a fully-connected neural network that performs the prediction on top of these extracted features to estimate the labels' scores or the predict value.
\subsection{Regularization of Deep Neural Network Training}
Given the high capacity of DNNs, developing regularization techniques that prevent the model from overfitting the training data, has been one of the major research efforts in the deep learning field. In general, regularization tolerates a relatively small increase of training loss in favor of reducing the error on the test data with the hope of having a model that generalizes well on the future coming data.   
\subsubsection{Standard Regularization Techniques}
The standard regularization techniques consist in adding restrictions on the values of the trained parameters, i.e., by adding a penalty term $\Omega(W)$ in the loss function $loss(f^*,D)$ (see the regularized loss from Equation \ref{reg_loss}) that can be seen as a soft constraint on the magnitude of parameters to restrict and smooth their corresponding distributions. 
\begin{equation}
    \tilde{loss}(W,b,D) = loss(W,b,D) + \lambda\Omega(W)
    \label{reg_loss}
\end{equation}
Where $\lambda$ fixes the relative contribution of the norm penalty omega; so setting $\lambda=0$ means no regularization and larger values of $\lambda$ correspond to more regularization. The most popular penalty consists in penalizing the weights of the linear computations; keeping their values closer to the origin by using L2-norm ($\Omega(W)=\frac{1}{2}\norm{W}_2^2$) and--or enhancing the sparsity of the weights by using L1-norm ($\Omega(W)=\norm{W}_1$). With the gradient-based learners, we compute the gradient of the weight penalty term $\Omega(w_i)$ w.r.t weight $w_i$ as described in the following formula.
\begin{equation}
\frac{\partial \Omega}{\partial \mathbf{w}_{i}}=\left\{\begin{array}{ll}
2 \times \mathbf{w}_{i}^{(t)}, & \text { if } \Omega(\mathbf{w})=\|\mathbf{w}\|_{2}^{2} \\
1 \times \operatorname{sign}\left(\mathbf{w}_{i}^{(t)}\right), & \text { if } \Omega(\mathbf{w})=\|\mathbf{w}\|_{1}
\end{array}\right.
\end{equation}
From the above formula, we can see that L2-norm penalty continuously reduces the magnitude of the weights proportionally to it while L1-norm reduces the magnitude by a constant. Thus, L2-norm and L1-norm pushes, respectively, the magnitude of weights towards increasingly lower and zero values. Moreover, the defined hyperparameter, $\lambda$, controls the strength of the applied regularization. It is therefore important to adjust it appropriately, taking into account the design of the model and the complexity of the target problem but concerning changes w.r.t size of batches, it has been shown that scaling the $\lambda$ by $1/m$, where $m$ is the size of the batch, can make it comparable across different size of batches~\cite{CourseDNN}, which avoids tuning manually $\lambda$. Intuitively, the training algorithm tries to approximate an unknown distribution by minimizing the empirical error on a sample of data; so a large sample is likely to be more representative of this unknown distribution, and as a result, less regularization might be needed in order to capture the maximum information about the target data distribution.
\subsubsection{Advanced Regularization Techniques}
The advanced regularization techniques for DNNs include normalization and stochasticity to the DNN inner computations~\cite{garbin2020}. In fact, manipulating randomly the DNN architecture, over training passes, minimizes the risks that the learned parameters are highly customized to the underlying training data. Furthermore, normalizing the input of each layer, not only the input data, improve furthermore the smoothness of the loss landscape towards more stable gradient-based optimization, and fortunately, higher chances to avoid saddle points and ineffective local minima.\\
\textbf{Dropout.} One of the most frequently used regularization techniques is dropout~\cite{dropout}, which randomly deactivates a subset of neurons from the dropped dense or convolutional layer at every training iteration. Indeed, the degree of randomness of dropout should be adjusted with respect to the width of the layer using a hyperparameter, $pkeep$, which represents the neurons' retention probability. At inference time, dropping neurons is stopped and compensated by multiplying all the weights in the layer by $pkeep$ in order to keep the distribution of the layer outputs (i.e., results of the layer's affine transformation) during inference time close to the distribution during training time.
Mathematically, for each hidden layer $l$, we define a binary mask $m^l$, where each element can be $1$ with predefined probability $pkeep$. Thus, the dropped out version of the hidden layer output $z^l$ masks out units using element-wise production, $z_d^l = m^l\odot z^l$. 
Intuitively, dropout reduces the risk of overfitting by making the DNN robust against the deactivation of some neurons, which forces the DNN to rely on population behavior instead of the activity of other feature detectors unit (i.e., preventing the co-adaptation of feature detectors). Indeed, model ensembling, a well-known technique in statistical learning, consists in combining the output of multiple models, each trained differently in some
respect, to generate one final answer. The resulting improvements on the performance metrics explain its domination over recent machine learning competitions~\cite{DL_ebook_2016}, however, it requires a much larger training time by definition (compared to training only one model). More fundamentally, dropout~\cite{dropout_exp} simulates model ensembling without creating multiple neural networks by combining the predictions of multiple sub-DNNs resulting from dropping, randomly, different subset of neurons between every two consecutive training passes.\\
\textbf{Batch-normalisation.} Another interesting regularization strategy designed for DNN is Batch-normalisation (Batchnorm). It relies on continuously normalizing intermediate activations across batches during a mini-batch loss minimization~\cite{batchnorm}. Indeed, the proceeded normalization is based on the standardization of each intermediate feature using the pre-computed mean and standard deviation on the current batch, respectively, $\mu_B = \frac{1}{m}\sum_{i=1}^mx_i$ and $\sigma_B^2 = \frac{1}{m}\sum_{i=1}^m(x_i-\mu_B)^2$ given the batch data $B$ of size $m$. The normalized activations $\hat{x_i} = \frac{x_i-\mu_B}{\sqrt{\sigma_B^2 + \epsilon}}$ would have zero mean and unit standard deviation. However, the resulting reduction of magnitude may induce information loss caused by the distortions in the learned intermediary features, and consequently, the degradation of model's learning capacity. Therefore, batch-normalization performs a linear transformation to scale and shift the normalized activations, $a_i = \alpha\hat{x_i}+\beta$ with aim of preserving the expressiveness of the DNN through learning additionally both of the parameters $\alpha$ and $\beta$. In addition, batch-normalisation also computes two other statistics, $E[x]$ and $Var[x]$, which represent, respectively, the moving average and the moving variance of the flowing data during all the training. Thus, at the test time, we use population mean($E[x]$) and population variance($Var[x]$) to standardize the layer outputs instead of using batch mean($\mu_B$) and batch variance($\sigma_B^2$). 

Santurkar et al.~\cite{batchnorm_opt} investigated the fundamental reasons behind the effectiveness of batch normalization in regularizing modern DNNs. They found that normalizing continuously all the inputs of hidden layers (i.e., all the activations) instead of normalizing only the inputs, makes the surface of loss smoother, which ensures faster convergence and safer training using relatively higher learning rates for which unnormalized DNNs (i.e., standard variants without batchnorm) diverge. This positive effect on the loss landscape is called better conditioning of loss minimization problem, which is described explicitly in the following sub-section.
\subsection{Conditioning of Loss Minimization Problem}
Conditioning refers to how swiftly a function changes with respect to small variations in its entries~\cite{DL_ebook_2016}. High sensitive functions lead to poorly-conditioned scientific computations, such as numerical optimization. Deep learning algorithms rely on numerical cost minimization. The performance of DNN is estimated as its prediction ability with respect to unseen data (which represents future data). However, training a DNN consists in minimizing the loss on the training data, which means reducing the empirical error on a sample data with the aim of optimizing indirectly the true error on the unknown target distribution. Traditionally, machine learning algorithms design the loss function and constraints carefully; ensuring that the minimization problem is convex, where any found local minimum is guaranteed to be a global minimum. When training a deep neural network, the loss function is not only non-convex but also tends to have a large number of ``kinks'', flat regions, and sharp minima~\cite{loss_landscape}. This makes the empirical risk minimization difficult. The empirical risk may become non-representative of the true risk that we aim to reduce. Nevertheless, deep learning has achieved impressive results by training DNN using first-order gradient-based optimization. This practical trainability success is highly dependent on the design of DNN, the choice of optimizer, parameters initialization, normalization, regularization, and a variety of hyperparameters. However, finding adequate configurations and parameters can be very challenging and the optimization of neural networks is still an open problem. In practice, training a DNN for a novel problem, context or data requires a series of trial-and-error using different configuration choices and hyperparameters tuning. These configuration choices have a strong effect on the conditioning of the minimization problem. In fact, the Hessian matrix of the loss encapsulates all the second-derivatives that represent the curvature of the loss surface. The condition number which is the ratio between the largest and the smallest eigenvalue is very important because a high condition number indicates a situation of ill-conditioning, where some parameters have huge curvature while others have smaller one. Such a situation results in a pathological loss curvature, and a first-order gradient descent would have difficulty progressing. However, novel DNN design components like skip-connections of ResNet, advanced regularization and reparameterization techniques~\cite{batch_norm_exp, ba2016layer} have shown an ability to improve the Lipschitzness of the loss function, which results in the loss exhibiting a significantly better beta-smoothness. These smoothing effects impact the performance of the training algorithm in a major way because it provides more confidence that the estimated gradient direction for each training step remains a fairly accurate estimate of the actual gradient direction after taking that step. This enables performing update steps without high risk of running into a sudden change of the loss landscape; including flat region (corresponding to vanishing gradient) or sharp local minimum (causing exploding gradients). In other words, finding good ways to configure and parametrize the DNN ensures the stability of the loss function and better predictiveness of its computed gradients.\\ 
Li et al.~\cite{loss_landscape} have shown that sufficiently deeper neural networks encounter a high risk of sudden transition in their loss landscapes from being nearly convex to being highly chaotic, which is correlated to the lack of trainability and dramatic decrease of the generalization error. Therefore, the conditioning of the minimization problem and the loss landscape geometry have a substantial effect on the quality of both training and generalization abilities. 
\subsection{Property-Based Software Testing}
Property-based testing (PBT) is a practical testing method~\cite{paraskevopoulou2015foundational} that provides a systematic way for reasoning about the properties of the appropriate program’s behaviors instead of the correct outcomes. For example, one may validate that a random data generator can produce probabilities within $[0, 1]$, while abstracting away from the actual probabilities. It follows the philosophy of invariant detection, which defines a set of invariant properties that allow aligning an incorrect execution against the expected execution~\cite{ernst2001dynamically}. Invariant detection belongs to the broader class of pseudo-oracle techniques~\cite{barr2014oracle} that bypass the lack of genuine oracle and can be used to distinguish a program's correct behavior from an incorrect behavior.
Thus, DNN training programs are considered programs that are mainly written to determine the answer (i.e., the learned parameters). As a result, such programs cannot have a conventional oracle~\cite{DLTestReview}. Invariant detection belongs to the broader class of pseudo-oracle techniques that bypass the lack of genuine oracle and can be used to distinguish a program's correct behavior from an incorrect behavior. Indeed, PBT defines the essential properties that the under-test program must respect in any possible execution scenario, rather than searching for the exhaustive set of all valid input-output pairs. These properties should represent high-level specifications, describing the program's correctness. First, PBT requires the collection of sufficient properties about the component under test (such as its function, program, or system). Then, the verification process starts by generating inputs for the component relying on specific heuristics to cover the equivalent classes of data inputs. Afterward, it validates for all the generated inputs that all preconditions, invariant properties on intermediary results, and postconditions associated to the component under test are totally true. When a property is failed, the counterexample is shrunk by searching the minimal combination of input elements that causes the property to fail. In fact, large inputs may cause the failure of multiple properties; so shrinking the input allows developers to extract the smallest part of inputs that is capable of reproducing a particular failure, which is essential for fixing the bug.
\subsection{TensorFlow Library: Concepts and Features}
TensorFlow (TF)~\cite{tf2016} is an open source DL library released by Google to help DL practitioners construct different architectures of DNNs. TF is based on a Directed Acyclic Graph (DAG) that contains nodes, which represent mathematical operations, and edges, which encapsulate tensors (i.e., multidimensional data arrays). This dataflow graph offers a high-level of abstraction to represent all the computations and states of the DNN model, including the mathematical operations, the parameters and their update rules, and the input preprocessing. Thus, the dataflow graph establishes the communication between defined sub-computations; making it easy to partition independent computations and execute them in parallel on multiple distributed devices. Given a TF dataflow graph, the TF XLA compiler is able to generate an optimized and faster code for any targeted hardware environment including conventional CPUs or high-performance graphics processing units (GPUs), as well as Google's custom designed ASICs known as Tensor Processing Units (TPUs). Regarding the running of graph computations, TF follows the principle of deferred execution or lazy evaluation that consists of two main phases in a TF program: (1) a construction phase that assembles a graph, including variables and operations; (2) an execution phase that uses a session to execute operations and evaluate results in the graph. In our work, we choose to implement our debugging approach on top of TF because of its high popularity in the ML community~\cite{braiek2018open}.
\section{DNN Training Program: Pitfalls}
\label{catalog}
Many issues can prevent the fitting process of a DNN from performing properly, i.e., finding the best-fitted model. In this section, we elaborate on some of the common pitfalls in designing and implementing DNNs, while pointing out and discussing concrete examples of their derived non-crashing bugs.
\subsection{DL Faults Investigation}
\label{real_faults_investigation}
\begin{figure}[ht]
\centering
\includegraphics[scale=0.4]{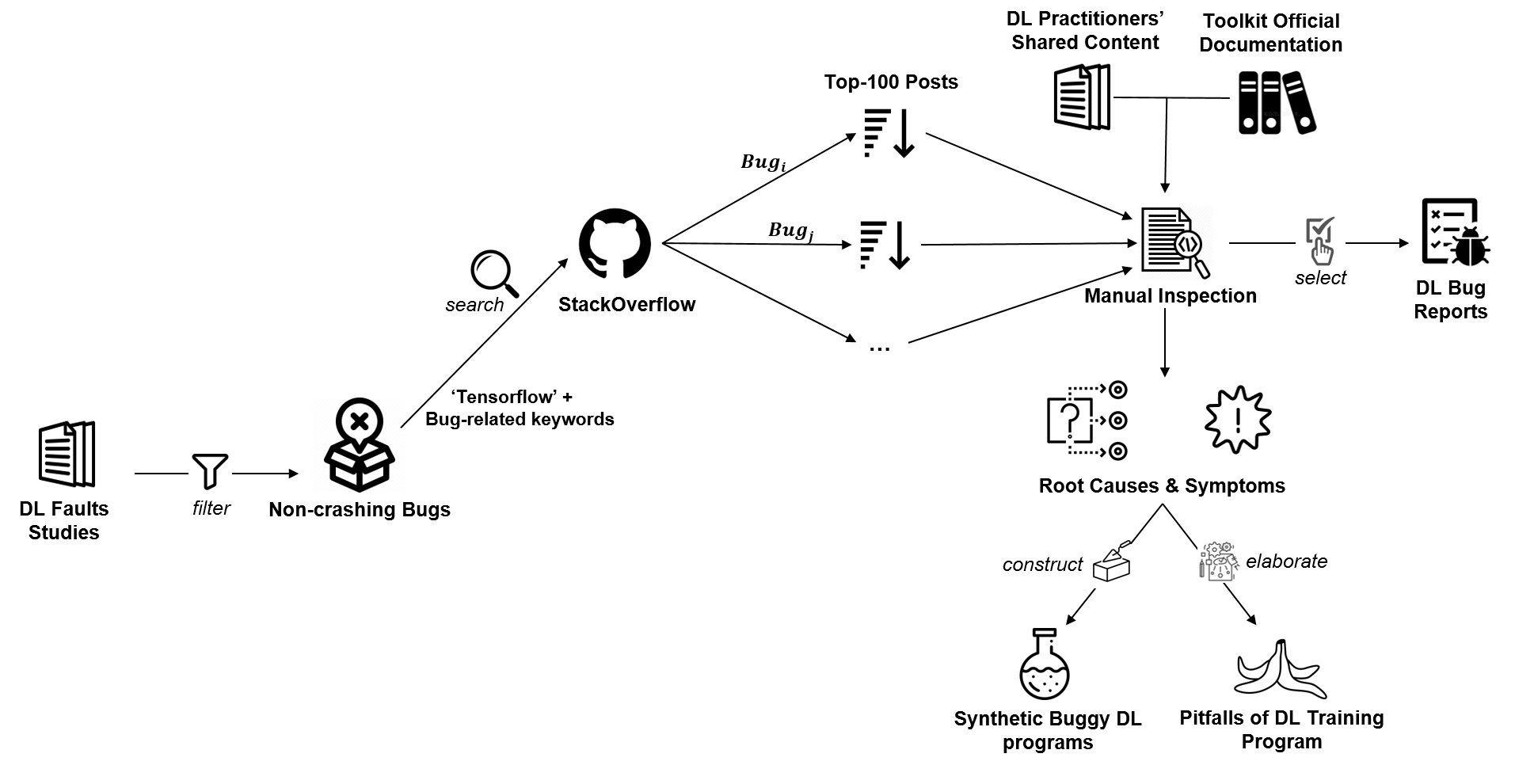}
\caption{Overview of DL Pitfalls Investigation Process}
\label{fig:Pitfalls_Process}
\end{figure}
Since deep learning has been increasingly leveraged in diverse real-world applications, researchers have been interested in studying the software development challenges for this next generation of software, including the faults’ taxonomy and bugs characteristics. Zhang et al.~\cite{DL_bugs_1} manually inspected the real-world SO and GH's Tensorflow programs and identified some of the DL bugs, their root causes and symptoms at high level. Then, Islam et al.~\cite{DL_bugs_2} extended the investigation by including DL programs written with other competitive libraries such as Pytorch and Caffe, and studied furthermore the categories of bugs and their relationships. More recently, Humbatova et al.~\cite{DLfaults} refined the former bug investigation \cite{DL_bugs_1, DL_bugs_2} into a taxonomy of real faults that occur in DL software systems. The taxonomy was deduced from 375 labeled buggy DL code examples built using three popular DL libraries: Tensorflow, Keras and Pytorch. Moreover, the construction of the taxonomy was built in collaboration with 20 DL developers and validated by a different set of 21 DL developers who confirmed the relevance and completeness of the identified categories. Indeed, a bunch of the reported bugs were caused by either coding mistakes, model design issues, or wrong configurations, that share a common high-level symptom of inefficient training. The latter manifests through convergence difficulties, preventing partially or even totally the training program from fitting the data. Hence, these non-crashing bugs are unique to the deep learning software systems that do not raise exceptions but adversely affect the training dynamics and results. Thus, we aim to identify and understand the development pitfalls and root causes behind the non-crashing DL bugs. First, we started by filtering them from the DL faults collected and reported in the former studies' datasets. Mainly, we discarded the two categories of Tensors\&Inputs~\cite{DLfaults} and GPU usage~\cite{DLfaults} that represent, respectively, crash-inducing bugs and GPU-related bugs. We focus on detecting the non-crashing bugs among the remaining three categories of Model~\cite{DLfaults}, Training~\cite{DLfaults} and API~\cite{DLfaults} that contain, respectively, different misconceptions of the model, multiple poor coding/configuration bugs in the training algorithm implementation, and misuses of the DL libraries' API. Then, we extend the selected subset of bugs with more Q\&A posts from Stackoverflow that are related to this family of bugs. We conduct a keyword-based search on StackOverflow (SO) with queries in the form of 'bug\_type-related keywords+Tensorflow' and we select, for each bug type/query, the top-100 SO posts (sorted by SO internal relevance criterion). Next, we inspect manually the SO post content including the shared code snippets and users' comments, with the aim of identifying more instances of the studied faults in Tensorflow. Therefore, we found $155$ bug reports in relation to occurrences of our targeted DL faults in Tensorflow DNN programs. Overall, only $8\%$ of the bugs in the three above-mentioned categories from~\cite{DLfaults} are included directly into our dataset, but as we searched using their keywords on SO, $38\%$ of them represent same bug types/root causes in our datasets. In fact, these buggy DL programs could not be included because they are either related to another DL framework (Pytorch, Keras, etc.) or missing necessary information for reproduction such as training examples or hyperparameters' values.

\begin{figure}[ht]
\centering
\includegraphics[scale=0.55]{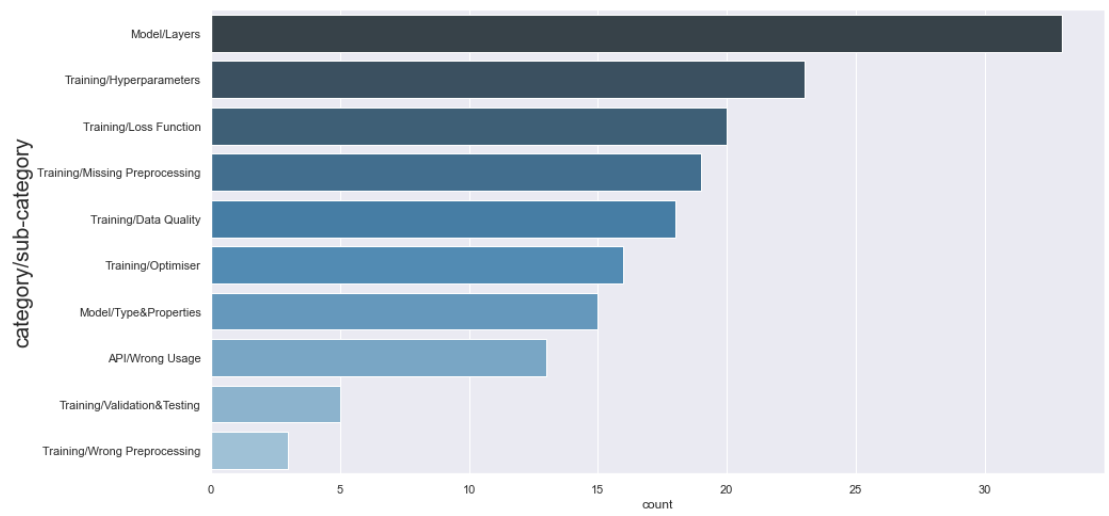}
\caption{Distribution of the collected Tensorflow Bugs over the Taxonomy of DL Faults~\cite{DLfaults}}
\label{fig:TF_bugs_distribution}
\end{figure}

Figure~\ref{fig:TF_bugs_distribution} shows the categories and subcategories of the bugs we collected in our datasets with respect to the taxonomy of DL faults~\cite{DLfaults}.
During the manual inspection of these bugs' instances, we abstract their main root causes and their typical symptoms (i.e., the negative effects observed on the training dynamics and the produced DL model), relying on the practitioners' shared content on troubleshooting and debugging the DL training algorithms including blog posts~\cite{slavReasons, ceceliaCheckList, dishankPitfalls, troubleshootingD4J, karpathyRecipe} and popular forum discussions~\cite{reddit,twitter}. Indeed, the decontextualization of DL bugs from the leveraged API version, the used data and the targeted application, allows recognizing firstly the design or implementation pitfalls that can be the origin of these training issues. Lately, the concrete occurrences of DL pitfalls would serve us in the creation of the synthetic buggy examples (section~\ref{synthetic_examples}), which have been used to evaluate the effectiveness of our debugging approach on detecting the targeted DL-specific bugs. Figure~\ref{fig:Pitfalls_Process} illustrates the schema of the above-mentioned steps to systematically enrich the datasets of non-crashing bug reports, as well as, identify their main root causes and symptoms. In the following, we present a comprehensive review of the DNN training pitfalls, organized in groups based on the main problematic component of the DNN training program.
\subsection{Input Data-related Issues} 
A DNN training program implements a data-sensitive algorithm whose inner logic is learned from the training data and generalized to future unseen data. Poor training data quality often translates into an unstable and inefficient training process. Below, we detail the training issues in relation to the input data and DNN components making use of them. 
\subsubsection{Unscaled Data}
The scale of DNN inputs and outputs~\cite{lecun1998gradient} is an important factor that affects the quality of the training. In fact, larger scale input features produce larger intermediate activations, and consequently, larger gradients regarding the weights connected to these over-scaled input features compared to others. Similarly, over-scaled predicted quantities would generate larger errors and gradients. Inversely, an abuse of data re-scaling penalizes the quantities with initially a small range of values. Both situations will induce a pathological loss curvature and an ill-conditioned loss minimization problem. As a result, the risk of gradient unstable phenomenon~\cite{uns_grad} increases. Modern deep neural networks deploy inner normalization techniques~\cite{batchnorm, ba2016layer} to overcome unstable distributions of the computed activations and gradients, however, their optimization routines should cope with high update oscillations during the early stage of training because of the unscaled data, and as a result, the DNN is firstly trained on how to scale and shift intermediate calculations into an appropriate range. This overhead complexity slows down the training procedure and might prevent the convergence towards the best-fitted model.
\subsubsection{Distribution-Shifting Augmentation}
Given their high learning capacity, DNNs require relatively large and sufficient training data to avoid simply overfitting the data. Since many application domains lack the access to big data and because gathering data is expensive, DL developers often resort to data augmentation techniques~\cite{perez2017effectiveness} to increase the quantity and diversity of their training data. Examples of data augmentation techniques for images include geometric transformations, color space augmentations, kernel filters, random erasing, and cropping. Nonetheless, the use of inappropriate augmentation rules, as shown in the SO posts \#57275278, \#48845354 and \#55786384, can induce a shift in the training data distribution that prevents the DNN from learning effectively. A DNN trained on noisy, shifted data is often hard to converge to a stable state and also incapable of predicting correctly on unseen data (i.e., validation or testing datasets).
\subsubsection{Corrupted Labels}
The data used for training supervised machine learning problems are composed of features $X$ (predictor inputs) and labels $y$ (supervised outputs to predict). The DNN's loss minimization problem is non-convex with several possible local minima; so standard gradient descent often falls into those minima because of the unchanged input data $X$ over all the training iterations. To overcome this problem, a mini-batch gradient descent with shuffling has been used to train DNNs, as introduced in~\ref{FNN}. Indeed, shuffling the data instances and performing the gradient estimation on only a subset of them, makes the batch inputs $X_b$ change with every iteration. This helps the optimizer to avoid sub-optimal local minima with relatively noisy and frequent updates to the DNN's parameters. In the implementation, we handle the features $X$ and labels $y$ in separate data structures because the labels should be used only for estimating the loss and performance metrics in supervised learning problems. A common bug in the data shuffler or mini-batch loader, as reported by SO users in \#47866803, \#46136553 and \#41864333, consists of inducing mistakenly a mismatch between features and labels. This represents a particular case of a more general DL issue of corrupted labeled data, which has been extensively studied in the machine learning community (e.g., ~\cite{van2017theory}).
\subsubsection{Unbalanced Data}
Very often in classification problems, there is an unequal number of instances for different labels. An unbalanced dataset biases the predictions towards the majority class or group of labels. Various mitigation techniques have been proposed to address this issue, e.g., over-sampling, under-sampling, or weighted loss function~\cite{more2016survey}. DL developers should be aware of this situation, whenever it exists, in order to use earlier a mitigation technique for class imbalance or improve the performance measure to capture fairly mispredictions for underrepresented classes. Otherwise, a biased model could be selected based on the overall accuracy among all classes, resulting in erroneous or unfair behavior when dealing with instances that belong to ones that are underrepresented.
\subsection{Connectivity and Custom Operation Issues}
To implement a DNN training program, DL developers use DL libraries that allow constructing the computational graph, where nodes and edges represent, respectively, operations and data paths. The operations represent the computational units that form the linear computations, activations, and gradient estimations. Data paths interconnect the operations and allow data to flow from one operation to the next, in order to successfully train and use the model. Through program code, DL developers use library's built-in and newly-implemented components for operations and connect them by either feeding one's outputs as inputs to another or by performing a math operation joining them. Configurable routines enable rapid development and expansion of reliable DNN programs, however they may lead to spaghetti code that becomes too large with scatter variables and glue code (build bindings between components), increasing the risk of coding errors.
\subsubsection{Network Disconnections} 
The most basic dependency is between the inputs and the outputs of the DNN. The DNN should predict the outputs based on the information distilled from the inputs. Thus, a DNN training program that does not consider the inputs when performing its internal computations is definitely erroneous. Moreover, disconnections can occur between the intermediate layers. Indeed, DL engineers can forget to connect some branches of the DNN or to pass the right inputs to the layers. When such omissions occur, one or more DNN layers are accidentally removed. A DNN with fewer layers than necessary can still be trained. A DNN can converge to an acceptable performance with only partial layers. If this occurs, however, the program will no longer comply with its specifications and its performance may be severely impacted. An illustrative example~\cite{unit_ML_test} of a connectivity bug occurs when cloning multiple times the code block for constructing a layer, the DL developer may forget to change the input and output for one of these constructed layers which makes it disconnected from the neural network.
\subsubsection{Incorrect Custom Operation}
The common abstractions used by computational units to encode numerical data are tensors, which are multidimensional arrays with supported algebraic operations. These tensors make it easy to manage high dimensional parameters and perform operations on them efficiently. However, the translation of math formulas from scientific pseudo-code to tensor-based operations can be error-prone. As an illustration, let's consider the cross-entropy loss which is a matrix-matrix operation that accepts the probabilities matrix and the matrix of one-hot encoding labels in order to estimate a particular distance. A buggy loss function may not correctly broadcast the operation if the reduction is done over the wrong axis (e.g., sum over rows instead of columns) and mix information between independent data instances of the batch. This introduces an incorrect dependency to the loss function. This issue can be difficult to detect since the DNN can still train and converge poorly and in the best case, can learn to ignore data coming from other batch elements. Besides, DL libraries include an automatic differentiation module that generates the analytical formula and computes the gradient automatically. However, DL developers can include non-differentiable or problematic operations in their custom function as shown in the SO posts \#41780344 and \#54346263, which negatively affect the gradients flowing over the newly-designed DNN. In similar way, DL developers can also hand-crafted the gradient calculation for their custom operations, but these gradients implemented from scratch should be tested carefully to avoid wrong computations as motivated by the SO posts \#46876063 and \#64172765.
\subsection{Parameters-related Issues}
DNN parameters represent the weights and biases of a DNN's layers. These parameters are randomly initialized, then, they are optimized during the training process. In the following, we discuss pitfalls in the initialization of parameters that can affect their learning dynamics.
\subsubsection{Poor Weight Initialization} 
An improper initialization of the weights for a DNN hampers the stability of the learning optimization problem, leading to unstable activation during the forward pass and unstable loss gradients of the backward flow. First, the constant weights induces a symmetry between hidden neurons of the same layer. Thus, the hidden units of the same layer share the same input and output weights, which makes them compute the same output and receive the same gradient. Hence, each layer's neurons perform the same update and remain identical; i.e., wasting capacity. Second, random sampling of initial weights breaks the symmetry between the neurons, however, the quality of training is strongly affected by the choice of initialization~\cite{DL_ebook_2016}. Indeed, the derivative equations~\ref{Eq3} and~\ref{Eq4} show that the estimated gradients include multiplication by weights, which makes their initial magnitude scale affects their growth or decay over iterations and might induce exploding or dead weights~\cite{dead_W_U}.
\subsubsection{Ineffective Bias Initialization} 
A bias is like the intercept added to a linear equation. Its main purpose is to allow degrees of freedom close to the origin, which improves the representation capacity of a neural network; so it can fit better to the given data. Generally, the initial biases are always set to zeros. Despite this, null bias for particularly-skewed data distributions (e.g., unbalanced datasets) slows down DNN training, which would do the bias calibration during its first few iterations. It means that non-zero bias could contribute significantly to fit the model if it is delicately set up to approximate the bias of data. For example, learning a classification problem with a rare label is a kind of bias already known; so the final layer's bias should be carefully initialized to accelerate the learning task.
\subsection{Activation-related issues}
Activation represents the intermediate computation that introduces non-linearity to filter the information computed by the previous layer. In the following, we discuss some problems related to activations.
\subsubsection{Activations out of Range}
Activation functions are nonlinear functions that determine the output of a neural network. The function is attached to each neuron and determines whether it should be activated (``fired'') or not, based on the relevance of neuron output for the model's prediction. Activation functions also help normalize the output of each neuron by transforming inputs into outputs that are within a predefined range of values. When DL developers implement an activation function from scratch, there is a risk of bugs that leads to a wrong or unbounded mathematical function yielding outputs within a range inconsistent with what is expected by the developer (e.g., sigmoid's outputs are between $[0,1]$ and  tanh's outputs are between $[-1,1]$).
\subsubsection{Inadequate Hidden Layer Activation}
Although the choice of hidden activation function is a design engineering problem, it is not an empirical and performance-driven selection because there are activations that are more suitable and even specialized for particular use cases rather than others. In fact, a non-linear activation is an essential design component; so a bottom-line inadequate choice would be keeping identity function for activation as evidenced by SO posts \#53138899 and \#46181692. Nevertheless, softmax is a special non-linear activation designed specially to transform the logits into probabilities; so mistakenly choosing it, as an activation for hidden layers, would likely hinder the parameters learning and often discourage a smooth flowing of gradient, as shown in the SO post \#52575271. Below, we detail training issues in relation with well-known hidden activations that could happen in certain design circumstances, where selecting an alternative activation function should be considered.
\paragraph{Saturation of Bounded Function}
Activation functions with a bounded sigmoidal curve, such as sigmoid or tanh, exhibit smooth linear behavior for inputs within the active range and become very close to either the lower or the upper asymptotes for relatively large positive and negative inputs. The phenomenon of neuron saturation occurs when a neuron returns only values close to the asymptotic limits of the activation functions. In this case, any adjustment of the weights will not affect the output of the activation function. As a result, the training process may stagnate with stable parameters, preventing the training algorithm from refining them. In fact, we can write the equation index-free to illustrate the gradient computation flow in general:
\begin{equation}
    \pdv{loss}{W} = a_{in} \times \delta_{out}
    \label{eq_loss_W_idx_free}
\end{equation}
where $a_{in}$ is the activation of the neuron input to the weight $W$ and $\delta_{out}$ is the error of the neuron output from the weight $W$.\\
When the activation function $\Phi$ is saturated, its outputs are in the flat region where $\Phi'\approx0$; so $\delta_{out}\approx0$ and $W$ freezes or learns slowly.\\
\paragraph{Dead ReLU Function}
ReLU stands for rectified linear unit, and is currently the most used activation function in deep learning models, especially CNNs. In short, ReLU is linear (identity) for all positive values, and zero for all negative values. Contrary to other bounded activation functions like sigmoid or tanh, ReLU does not suffer from the saturation problem because the slope does not saturate when $x$ gets large and the problem of vanishing gradient is less observed when using ReLU as activation function. Nevertheless, ReLU risks ``dead ReLU'' phenomenon~\cite{dead_W_U} because it nullifies equally all the negative values. A ReLU neuron is considered ``dead'' when it always outputs zero. Such neurons do not have any contribution in identifying patterns in the data nor in class discrimination. Hence, those neurons are useless and if there are many of them, one may end up with completely frozen hidden layers doing nothing. In fact, given the index-free Equation~\ref{eq_loss_W_idx_free}, we can see that when the activation is zero $a_{in}=0$, the loss gradient w.r.t weights becomes zero too ($\pdv{loss}{W}=0$); therefore $W$ freezes and no longer receives updates. This problem is often caused by a high learning rate or a large negative bias. However, recent ReLU variants such as Leaky ReLU and ELU are recommended as good alternatives when lower learning rates do not prevent this issue.
\subsubsection{Inadequate Output Layer Activation}
Concerning the output layer, the activation function should map the internal calculated results into valid predictions. In case of mismatch between the ranges of last activation layer's outputs and ground truth labels, the model could not learn a correct mapping function since it is not able to produce the full range of possible outcomes.\\
\paragraph{Classification Outputs}
The model is learning to predict probabilities, so sigmoid and softmax are the best candidates for, respectively, binary and multinomial classification. For instance, a missing softmax layer prior to the cross-entropy calculation can lead to performance degradation and numerical instability issues as evidenced by the SO post \#53254870. However, the use of softmax for a classifier model with $1$-dim output leads to the incapacity of outputting the full range of class labels, as evidenced by the SO posts \#59129802 and \#53971451 where sigmoid should be used to output the negative class $0$, or \#51993989 where tanh should be used to output both of labels: $0$ and $-1$. Other common pitfalls are stacking consecutive output activations, which add useless computation levels that may erase relevant learned information, obstruct the natural gradient flow, and adversely affect DNN performance. Indeed, the redundancy activations were often result from: \textit{(1)API misuse}, where recently-provided stable API loss functions with the logits activation, softmax or sigmoid, included (i.e., \textit{tf.nn.softmax\_cross\_entropy\_with\_logits} and \textit{tf.sigmoid\_cross\_entropy\_with\_logits}), mislead several DL practitioners that passed the result of last layer activation to these loss functions, which resulted in a double application of softmax or sigmoid on the outputs (e.g., we refer to SO posts \#36078411, \#46895949, and \#42521400); \textit{(2)misconception of abstractions}, the definition of a function that abstracts the creation layers with some parameters to facilitate stacking the neural network's layers, however, useless non-linear activation can be applied to the last layer before probabilities transformation by mistake, which restricts the range of outputs. For instance, a ReLU activation before applying softmax, as happened in the SO post \#44450841, would nullify negative values and make all their corresponding labels share the same probability after applying the softmax.
\paragraph{Regression Outputs}
The last activation should map the internal computations into a range of values that equals (or is the closest) to the actual interval of target outputs, in order to ease the optimization process. For instance, SO posts \#60801900 and \#64998875 show how the use of Relu, having an output range of $[0,+\infty]$, prevents the estimation of negative targets and the SO user in the post \#62313327 should switch from sigmoid (i.e., outputs values within $[0,1]$) to tanh (i.e., outputs values within  $[-1,1]$) to meet the real range of ground truth labels.
\subsubsection{Unstable Activation Distribution}
The activations encode the representation of features detected at that processing layer during the training process. Thus, the fired activations indicate that the DNN already detects low-level features, which could be relevant for following layers. That is why the stagnation of activations caused by saturation or dead phenomenon hinders the capacity of the DNN to learn useful patterns from the data. Similarly, over-activated layers that are active for all inputs and unstable activation layers that have high variability in their values can lead to numerical instability and--or divergence problems. In fact, activations represent the input features of the next layer. The internal computation of this layer adjusts the parameters in order to infer patterns from features (i.e., activations).The internal computation of this layer adjusts the parameters in order to infer patterns from features (i.e., activations). By analogy to the input normalization, the distribution of the intermediary detected features (inside the DNN) is important to ensure an effective optimization using backpropagation of loss gradient through layers' parameters. More formally, the index-free partial derivative formula~\ref{eq_loss_W_idx_free} shows well how the magnitude of activations affects directly the magnitude of weight updates. Researchers~\cite{ba2016layer}~\cite{batchnorm} have proposed different techniques to normalize the outputs of hidden layers and obtain activations with zero mean and unit standard deviation. Concretely, these additional internal scaling transformations are important to control the magnitude of the gradients and improve, formally, the $\beta$-smoothness and the Lipschitzness of the estimated loss.
\subsection{Optimization-related issues}
The optimization of DNN's learnable parameters consists in minimizing, iteratively, the loss, \ie{} empirical error of DNN's predictions regarding supervised training data. Actually, gradient-based algorithms such as SGD, Momentum, and Adam, are the preferred way to optimize the DNN's internal parameters. Next, we discuss several issues that impede the optimization process, while describing our proposed verification routines to catch them earlier. 
\subsubsection{Wrong or Inappropriate Performance Measurements}
The iterative optimization of parameters often converges to an equilibrium behavior of the DNN. At this point of equilibrium, the optimal or near-to-optimal DNN status is reached. To find this best-fitted DNN (i.e., highest accuracy or lowest absolute error), the DNN training algorithm acts indirectly by minimizing a loss function estimated on the training data with hope of improving the performance of the on-training DNN. Hence, the loss is primarily designed to measure the distance between predictions and real outputs, while it should respect fundamental properties of an objective function for first-order gradient optimization. Empirical loss minimization for DNN training works well when the minimized loss represents the fitness of the DNN relative to the data. Thus, a wrong loss function with regards to true model risk, misleads the training algorithm that, despite its success to reduce the loss, could not improve the target performance measure (e.g., classification accuracy). For instance, inadequate choice of loss function like choosing mean squared error (MSE), which is a standard loss for regression problems, to compute the deviation between predicted probabilities and target class in a classification problem (e.g., SO posts \#38319898 and \#50641866). Another common fault in relation with the loss is the use if ineffective loss reduction strategy like in these SO posts where there are no reduction at all (\#36127436) or a sum instead of mean reduction (\#43611745 and \#41954308). Indeed, the reduction strategy allows to aggregate the losses computed for all of the data instances into a scalar loss value. The aggregation could be the average or the sum, however, mini-batch gradient descent variants are commonly used for minimizing the DNN's non-convex loss function. Hence, mean reduction is better than sum reduction, because averaging losses over the mini-batch would keep the magnitude of loss independent of the batch size and of other hyperparameters that are also sensitive to the magnitude of loss gradients like the learning rate. 
Besides, a mistaken performance metric also regresses the expected covariance between both training quality measures (e.g., loss and accuracy). For instance, bad choice of accuracy metric with respect to the problem would yield illogic performance measurements and it can be the use of classification accuracy rate for a regression problem, the use of multiclass accuracy metric for a binary classifier, or inversely, the use of binary classification accuracy metric for a multinomial classifier, as evidenced by the SO posts respectively, \#62566558, \#62354952 and \#42821125.
\subsubsection{Inadequate Learning rate}
As shown in the update equations~\ref{Eq5} for weights and biases, the predefined learning rate controls the magnitude of update at each step; so setting the learning rate too high or too low can cause drastic changes to the optimization process and cause several erroneous behaviors. A high learning rate would push the layer's parameters changing rapidly in an unstable way; preventing the model from learning relevant features. The intuition is that the parameters are a part of the estimated mapping function, so we risk overfitting the current processed batch of data when we try to strongly adapt the parameters in order to fit this batch. An excessively-high learning rate may lead to convergent loss minimization and numerical instability by having NaN loss or output values. Inversely, a low learning rate can slow down the parameters changing; making it difficult to learn useful features from data, and consequently, the minimization process may not converge to a steady state and may even experience a non-decreasing loss value during training. Starting by ineffective learning rates is very common when the DL developer is dealing at first time with a learning problem as evidenced by these SO posts \#42264716, \#62381380, \#55718408, \#34743847, \#47245866, \#40156629 and \#59106542.
\subsubsection{Unstable Gradient Problem}
The loss gradient equations~\ref{Eq3} and~\ref{Eq4} show that the gradient of a layer is simply the product of errors back-propagated from all its next layers (i.e., following the forward direction). Intrinsically, the layers tend to learn at different speeds and deeper neural networks can be subject to unstable situations if no advanced mechanism is applied to balance out the magnitude of gradients. However, the unstable gradient problem could be more severe and could manifest in the form of vanishing or exploding gradients, as described below, due to poor design choices of initializations and hyperparameters.
\paragraph{Vanishing Gradient}
In this case, the gradient tends to have smaller values when it is back-propagated through the hidden layers of the DNN. This causes the gradient to vanish in the earlier layers, and consequently, it would be nullified or transformed to undefined values such as Not-a-Number (NaN) caused by underflow rounding precision during discrete executions on hardware. The problem of vanishing gradient can lead to the stagnation of the training process and eventually causing a numerical instability. As an illustration, we take the example of a DNN configured to have sigmoid $\sigma$ as activation function and a randomly initialized weight using a Gaussian distribution with a zero mean and a unit standard deviation. The sigmoid function returns a maximum derivative value of $\sigma'(0)=0.25$ and the absolute value of the weights product is less than $0.25$ since they belong to a limited range between $[-1, 1]$. Hence, it is apparent that earlier hidden layers (i.e., closer to the input layer) would have very less gradient resulting from the product of several terms that are less or equal to $0.25$. Therefore, earlier layers receiving vanishing gradients would be stagnant with low magnitude of weights' changes.
\paragraph{Exploding Gradient}
The exploding gradient phenomenon can be encountered when, inversely, the gradient with respect to the earlier layers diverges and its values become huge. As a consequence, this could result in the appearance of $-/+\infty$ values. Returning to the previous DNN example, the same DNN can suffer from exploding gradients in case the parameters are large in a way that their products with the derivative of the sigmoid keep them on the higher side until the gradient value explodes and eventually becomes numerically unstable.

Therefore, advanced mechanisms like batch~\cite{batchnorm}, layer~\cite{ba2016layer}, weight~\cite{weight_norm} normalizations and tuning of optimization hyperparameters such as learning rate or momentum coefficients, are needed to provide adaptive gradient steps and to establish relatively similar learning speed for all the neural network's layers.
\subsection{Regularization-related issues}
The regularization strategy prevents the model from overfitting the data, while allowing the DNN to acquire enough learning capacity to learn useful patterns and fit the data properly. In the following, we introduce potential issues related to incorrect and ineffective regularization techniques.
\subsubsection{Lack of-or-incorrect Regularization}
Regularization techniques~\cite{DL_ebook_2016}, including penalty cost on the weights magnitude and specialized DL regularization like dropout, discourages the optimization from exploring complex models and exploiting spurious correlations in favor of reducing further the loss. Thus, lack of regularization leads to a noiseless training process with high capacity modeling that risks capturing even residual variations in the given sample of data and tends to overfit it quickly. Concretely, as shown in Equation~\ref{reg_loss}, a zero or very low $\lambda$ makes the penalty cost useless with no effect on the objective loss function, which enables the free-growth of weights and intensifies the threat of overfitting even coincidental noises in the sampled batches used for training the parameters. Regarding dropout, a high retainment probability of neurons ($pkeep$) for wide dense layers or large activation maps for convolutional layers decreases the size of the dropped subset of neurons, which eliminates the randomness effect introduced between the input features at each inference calculation. Concerning batchnorm, a common mistake (e.g., SO posts \#43234667 and \#52279892) consists in forgetting or misusing the routine that ensures the continuous estimation of the moving average $E[x]$ and variance $Var[x]$ during the training. Moreover, batchnorm makes the loss function dependent on the batch size because an instance of the batch can affect the batch mean and variance estimated, and consequently, affect both activations and loss values for other instances in the batch. For example, the use of unit batches, as reported in the SO post \#59648509, should not be applied because the batch variance would be zero and relatively small batches would increase randomness and make the statistics estimation noisy. Thus, the DNN fails to perform any normalization on the intermediary calculation results at the inference time, which means that the regularized version of DNN becomes incorrect and its inner calculations are non-representative. In case of mixing dropout and batchnorm, Li et al.~\cite{disharmony_batch_drop} reported a disharmony issue, named variance shift, between dropout and batchnorm when applying dropout first. Indeed, the population statistical variance estimated by batchnorm on the entire DNN training becomes inconsistent and non-representative because of the shift variance of weights done by the dropout when the DNN is transferred from the training to the testing mode. In other words, dropout proceeds by randomly removing the information coming from a subset of neurons to prevent possible neurons' co-adaptation. Thus, we have to pass the cleaned information (i.e., after dropping out some neurons) through batchnorm statistics estimator, otherwise, the statistics would be biased by considering all the neurons, i.e., dropped ones included. A common symptom for all the above issues is that validation/testing error rates are higher than training error rates. However, this symptom is quite connected to overfitting situations; so more fine-grained symptoms are necessary to guide the users towards the occured issue. 
\subsubsection{Over-Regularization}
Inversely, a too strong regularization (with high $\lambda$) can significantly reduce the magnitude of weights, which may result in underfitting; leading to useless, dead weights, as discussed in the SO posts \#51028324 and \#51028324. Regarding dropout, a low $pkeep$ will be considered as a strong regularization because it introduces too much randomness and may prevent the DNN from convergence to a stable state; so $pkeep$ should be tuned carefully considering the underlying DL problem and the depth of the designed neural network. For instance, many SO posts \#60591577, \#44832497, \#46515248, \#44695141, and \#64289118 reported poor training convergence and model performance resulting from inappropriate configuration of dropout layers. Indeed, it has been shown that $pkeep$ cannot go below the minimum of $0.5$, which represents the maximum additive regularization~\cite{dropout}. In practice, the fixed $pkeep$ for hidden layers, and especially, layers close to the output layer, is recommended to be within $[0.5,0.8]$. However, $pkeep$ for the input layer should be kept to about $0.8$ or higher (i.e., closer to $1.0$). In fact, during the test time, to compensate for disabling the dropout, the learned parameters are scaled by the $pkeep$ factor. However, Gal et al.~\cite{dropout_prob} has shown that this approximation at test time becomes noisier and less accurate as the underlying layer is far from the output; which can explain the high risk of instability induced by low neurons' retainment probability for the earlier layers, especially, convolutional layers in CNN and the input layer. Therefore, the common effect of strong regularization is having low error rates for training/validation/testing data, however, this indicates an underfitting situation in general. In our debugging methodology, we propose verification routines to target accurately these over-regularization issues. 
\section{DNN Training Program : Property-Based Debugging Approach}
\label{approach}
In the previous section, we have presented DNN training pitfalls related to misconfigurations and coding bugs, organized by problematic components, to comprehend their symptoms and their negative effects on the dynamics of training and the quality of optimization. In this section, we introduce an adaptation of property-based software testing (PBT) that assembles the verification routines that we propose for debugging DNN training programs. 
\subsection{Property-Based Model Testing}
It has become mainstream research to focus on DL testing approaches that explicitly check for model properties. Regarding security~\cite{SecAndPrivDL} and robustness~\cite{robustDL}, the learned function should be Lipschitz, such that small perturbations to the input are guaranteed to spawn bounded changes to the output. In a similar way, other important properties have been proposed to satisfy the desired requirements in terms of privacy~\cite{SecAndPrivDL} and fairness~\cite{fairnessTesting}. Indeed, the validation of a trained DL model's properties is quite similar to traditional property-based testing (PBT) because it consists of verifying that the implemented/trained deterministic function, mapping the inputs to the outputs, satisfies the desired properties for all the valid inputs. However, the straightforward application of PBT to the DNN training program, which is a stochastic data-driven optimization algorithm, appears to be quite challenging. In fact, the DNN training consists of a data-driven, iterative process guided by the backpropagation equations that starts from an initial DNN state and evolves the state following a trajectory dictated by the dynamics of the update step equations until converging hopefully to an equilibrium state at which the DNN state becomes stable. Thus, PBT can be applied on In this section, we introduce an adaptation of property-based software testing (PBT) that assembles several properties of training programs that should be satisfied by the initial state (i.e., pre-training conditions), maintained for the intermediate states of the on-training DNN (i.e., proper fitting conditions), and validated for converged DNNs (i.e., post-training conditions). Besides, an adaptation of input shrinking, as detailed in~\ref{shrinking-prog-states}, can be applied to the DNN state to determine at which level and in which component the property is violated. 
\subsection{Origins and Types of DNN Training Properties}
Over the last decade, we have witnessed a wide adoption of DL technology in various industrial domains that represent the outgrowth of huge efforts and dedications from the DL community into the knowledge transfer. Indeed, applied DL researchers and experts vulgarize DL fundamentals and research advances in sort of principles, techniques and tricks in order to get more practitioners involved into applying DL technology for solving learning task problems. This gave birth to popular applied DL textbook~\cite{DL_ebook_2016}, academic lectures~\cite{mitDLCourses, stanfordCourse}, articles~\cite{DLFixes, CNN_design_patterns, CNN_principles, practical_CNNs}, and industry courses~\cite{GoogleDLCourse, CourseraDLAI}, as well as experts' blogs~\cite{slavReasons, ceceliaCheckList, dishankPitfalls, troubleshootingD4J, karpathyRecipe} and DL practitioners' forum discussions~\cite{reddit,twitter} on DNN troubleshooting techniques and strategies.
\begin{figure}[ht]
\centering
\includegraphics[scale=0.45]{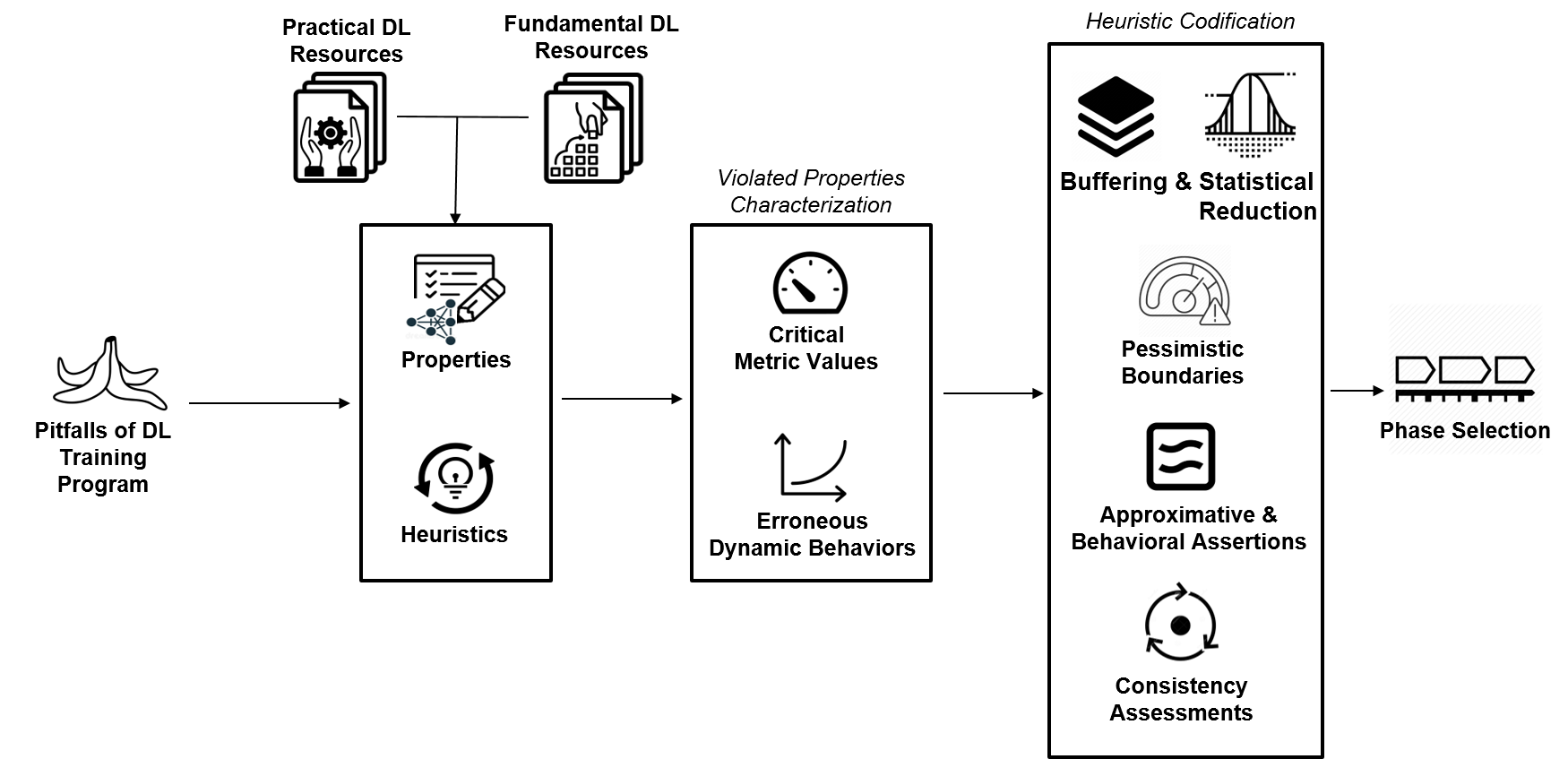}
\caption{Main Steps of Development Process for Verification Routines}
\label{fig:Verif_Routine_Steps}
\end{figure}
Figure~\ref{fig:Verif_Routine_Steps} summarizes the steps that we follow to construct automated verification routines for the identified DL training pitfalls. From the fundamental DL resources, we distill the main properties and design principles for each neural network component for which issues have been detected. For instance, we find the properties in relation to the initial random parameters, the hidden and output activation functions, or the gradient-based optimization routines. Basically, DL training programs are sensitive to the DL bugs that violate some of the involved components' properties. Nevertheless, the detection of properties' violations in a suspicious DL training program is quite challenging. That's why, we explore the heuristics and troubleshooting strategies elaborated by DL experts, which rely mainly on plotting histograms of model internals and curves of performance metrics, in order to spot unexpected distributions and irregular curve shapes. For instance, Glorot and Bengio~\cite{glorot2010} watched the activation distribution to detect any possible layer saturation when they studied different random weight initializations. Then, we analyze how these heuristics are able to characterize the violations of properties. Generally, the heuristics specify critical values for metrics that shed light on a faulty DL program's state, such as high ratio of null activations or high magnitude updates of parameters. Besides, the heuristics can also describe erroneous training behaviors that would manifest in the dynamics of the metrics, such as unanticipated fluctuating or diverging loss. Even if these abnormal behaviors could be captured and illustrated by experts through visualizations (e.g., histograms of activations, curve of losses over epochs, etc.), the codification of automated verification routines, that detect the violation of statistical learning properties and not-recommended instability for an on-execution DL training program, is challenging. Indeed, the inherent iterative nature and stochasticity of DNN training algorithms makes the regular deterministic test assertions impractical because a single property-violating state is not sufficient for asserting the occurrence of an issue. For instance, the current state may trigger a dead layer (i.e., more than $50\%$ of ReLUs in a layer are null), but the next state following the updates can avoid the problematic situation by reducing the inactive ReLUs. Hence, the persistence of the property-violating state, catched by the heuristics, should be taken into consideration to avoid overwhelming warnings and misleading false alarms during the debugging sessions. Below, we explain the developed guidelines to codify the DL experts' heuristics into robust and dynamic verification routines that are designed to assess persistent behavioral issues.
\subsubsection{Buffering and Statistical reductions}
The parameters and internal computations in a deep neural network are volatile multi-dimensional arrays. Thus, we define buffers to store the last intermediary states, including the hidden activations, the predictions, the losses, etc., in order to validate the heuristics on a set of recent states instead of a single one. Then, we calculate different statistics on their distributions over different axis of interest to reduce the dimensions and create the most appropriate data views/metrics to handle in the codification of the rule. For instance, the checks on the activation distribution would run periodically on all the activations stored on the buffer, which are obtained from the last training iterations (i.e., anticipated to be using different parameters and batches of data). By default, we set up a buffer size that equals $10$. Therefore, given the same example of dead layers, we reduce the buffered activations, from $10$ values per neuron to a single one, using $95^{th}$ percentile, which is a robust maximum estimation (highest value under the top 5\%). Then, our proposed check would flag the layers with more than a half of dead neurons, which have returned $95\%$ of outputs below the minimum threshold of $1e-5$ for the last $10$ training iterations.
\subsubsection{Pessimistic boundaries}
As \name{} consists of a debugging method, we set up pessimistic thresholds to spot critical values and erroneous behaviors that are probably caused by a DL bug. This conservative strategy can effectively reduce many possible false alarms in relation to ineffective training traits that usually manifest in earlier iterations or on complex learning problems. Nevertheless, we keep the thresholds as user-configurable settings in order to adjust the sensitivity of the verification routines on specific DL architectures according to user interests.
\subsubsection{Approximative and Behavioral assertions}
\label{dynamic_heuristics}
The heuristic-guided DL program diagnostic requires the implementation of approximative assertions including almost numerical equal assertions for floating numbers and statistical significance tests to identify if the obtained program state is outside the anticipated set of possible states by the experts. For instance, there are no best initial parameters, however, DL experts have shown the importance of sampling the random parameters at the first iteration from a carefully-designed distribution, depending on the neural network characteristics. Besides, crafting verification rules for diverse abnormal DNN states such as stagnated loss, huge weights or vanished gradients, etc., can be difficult because the involved metrics’ thresholds would vary between models and problems. Thus, it is important to focus on characterizing the erroneous behaviors: stagnation, diverging, or vanishing instead of the resulting faulty state to which the buggy DNN training program would converge. Indeed, we enable the detection of abnormal trends through the assessment of their evolution over consecutive iterations, i.e., by considering a window of steps, generally, window size would be in-between $3$ and $5$. In the following, we describe our proposed behavioral assertions for the common erroneous behaviors resulting from the identified DL training pitfalls:
\paragraph{Stagnation} It is the opposite of changing and moving quantity. Hence, we define a minimum percentage difference by which the quantity of interest should change at each step within the window; otherwise, we flag it as stagnated. More specifically, the change direction is already known to compute a relative percentage of increase or decrease, e.g., we expect that the loss keeps decreasing until convergence to a minimum.
\paragraph{Diverging or vanishing} They represent unexpected huge increases and decreases in a quantity over time. They can be simulated as exponential growth and decay, $q_t = q_0 \times r^t$ where, respectively, $r > 1$ and $r < 1$. Thus, we can approximate the rate of change $r_t = \frac{q_t}{q_t-1}$ for a window of recent steps, then, we consider that a quantity is diverging if the calculated $r_t$ are higher than a low bound, or it is vanishing if all the $r_t$ are lower than a high bound. In our constructed verifications, we used $2$ as low\_bound, meaning that the quantity should constantly double its value at minimum to be considered a diverging quantity and inversely, we used $\frac{1}{2}$ as high\_bound. The component is flagged when it maintains this unanticipated non-linear evolution for a predefined window of steps.
\subsubsection{Consistency assessments} 
Several experts' heuristics, that are incorporated into \name{}'s verification routines, help recognize the unstable distributions of parameters, activations and gradients, as well as unexpected optimization updates and loss curves. However, the riskiness of these unstable learning situations increases when the identified issue persists or becomes more severe over the iterations. Hence, we smooth the verification logic by \textit{(1)} computing the rule’s metrics on the aggregation of previously-obtained states from the buffer; \textit{(2)} considering a forbearance period, which is a prefixed number of steps to wait, despite the persistent failure of the verification rule, before flagging the occurrence of the issue.

\begin{figure}[ht]
\centering
\includegraphics[scale=0.85]{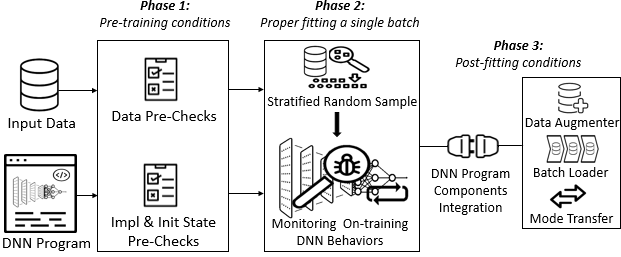}
\caption{Overview of \name{} Debugging Phases}
\label{fig:Phases}
\end{figure}

Finally, we select the debugging phase during which the codified verification routine would run depending on their input DNN states. Indeed, we mainly separate between the initial state verification, the validation of the under-test DL program running on a batch of data, and the need for longer training over larger datasets or comparison of multiple trained models. In the following, we describe the chronological sequence of the debugging phases, as illustrated in Figure~\ref{fig:Phases} and we detail the heuristic and logic of all the verification routines included in each phase of \name{}’s debugging session.

\subsection{Phase 1: Pre-training conditions}
The first phase of our debugging process occurs before starting a training session. It enables running static pre-checks of the input data and the starting initial state of the on-training DNN. The benefit of these preliminary verifications is to validate, from the start with a null training cost, the quality of feeding data (i.e., input features, labels), the correctness of essential implemented components (i.e.,  gradient, custom operations), and the adequacy of the starting state (i.e., initial parameters, first loss). In the following, we describe the pre-checks and their related training pitfalls.
\subsubsection{Data Distribution}
DL practitioners often perform linear re-scaling of the input and output features, in order to adjust their distribution into a common scale without distorting differences between the ranges of original values. In fact, the two most common data scaling techniques are: \textit{(1) standardization} consists in transforming the inputs into z-scores, which means that the transformed data should have a zero mean and a unit standard deviation; and \textit{(2) normalization} consists in re-scaling each input feature using its maximum and minimum elements, to have values within a predefined small range such as $[0, 1]$ or $[-1, 1]$.
\paragraph{Scaled Data Verification}
We extract the data, inputs and outputs, that are fed to the training program; i.e., the final data that have been going through the preprocessing pipeline, and then, verify that the data is scaled properly. In our verification routine, we start by detecting the constant features that have a zero variance. Then, we support checking whether the features are zero-centered with unit standard deviation, or they belong to one of the two well-known normalized range of values: $[0, 1]$ or $[-1, 1]$~\cite{lecun2012efficient}. Moreover, this helps detect if the data accidentally includes undefined or non-finite quantities (i.e., NaNs and Infs).
\paragraph{Unbalanced Labels Verification}
We compute the Shannon equitability index~\cite{shannon1948mathematical}, which summarizes the diversity of a population in which each member belongs to a unique group, to estimate the balance between the frequencies of labels. On a data set of $N$ instances, if we have $K$ labels of size $N_k$, we can Shannon equitability index as follows, $\frac{-\sum_{i=1}^{k}{\frac{N_k}{n}\log(\frac{N_k}{n})}}{\log(K)}$, and it will be zero when there is one single label. Thus, it tends to zero when the dataset is very unbalanced. In our verification routine, we use by default the minimum threshold of $0.5$ to flag the labels data as unbalanced. This will be reported to the user as warning and will affect the verification of the initial bias.
\subsubsection{Starting DNN state} 
Starting from different initial states, the optimization algorithm follows different trajectories and can terminate at different equilibrium states. Thus, a poor initial state adversely affects the optimization routines, and consequently, the optimality of the equilibrium state. 
\paragraph{Initial Weights Verification}
First, we verify that there are substantial differences between the parameter's values by computing the variance of each parameter's values and checking if it is not equal to $0$. Next, one can make sure that, given the chosen activation function, the distribution of initial random values are sampled from a uniform or normal distribution with a careful tweak, i.e., by calibrating attentively the variance because the distribution of the outputs from a randomly initialized neuron has a variance that grows with the number of inputs. The equality between the actual variance of each weight and its recommended variance given the input size is verified using f-test~\cite{f_test}. In the following, we describe the recommended variances depending on the activation function of the corresponding layer.
\begin{itemize}
  \item[-] LeCun~\cite{lecun2012efficient} proposes a heuristic that initializes each neuron's weight as either $\mathcal{N}(0, \sqrt{1/fan_{in}})$, i.e., normal distribution with zero mean and $1/fan_{in}$ of variance or $\mathcal{U}(-\sqrt{3/fan_{in}},+\sqrt{3/fan_{in}})$, i.e., uniform distribution within $[-limit,limit]$ and $limit= \sqrt{3/fan_{in}}$, where $fan_{in}$ is the number of inputs. This guarantees that all the initial neurons' weights have approximately the same output distribution, and its empirical evaluation on sigmoid layer activation shows a significant improvement on the rate of convergence.
  \item[-] Glorot et al.~\cite{glorot2010} recommend the following initialization (especially when tanh is used as activation function) which consists of neuron's initialization following \\ either $\mathcal{N}(0, \sqrt{2/(fan_{in}+fan_{out})})$ or $\mathcal{U}(-\sqrt{6/(fan_{in}+fan_{out})},+\sqrt{6/(fan_{in}+fan_{out})})$, where $fan_{in}$, $fan_{out}$ are the number of inputs and outputs.
  \item[-] He et al.~\cite{HE_init} also proposed an initialization specifically for ReLU neurons. They suggest that the variance of neurons in the network should be $2/fan_{in}$, which gives an initialization of either $\mathcal{N}(0, \sqrt{2/fan_{in}})$ or $\mathcal{U}(-\sqrt{6/fan_{in}},+\sqrt{6/fan_{in}})$. All of the above initializations have been discovered empirically and proven to be optimal in classic well-known CNNs like LeNet~\cite{LeNet} and modern architectures such as VGG~\cite{VGG} and GoogleNet~\cite{GoogleNet}.
\end{itemize}
\paragraph{Initial Biases Verification}
As a baseline, we verify that the bias exists, and initially is set to $0$ in case of well distributed labels. Nevertheless, we proceed by more advanced check on the last bias initialization in case the pre-check on unbalanced labels for classification was fired. Indeed, we make sure that the bias set for the output layer reflects the bias already found in the distribution of outcomes in the given ground truth data. In our implemented verification routine, we consider the unbalanced classification problem, where it is usually effective to set each bias unit $b_i$ to $\log(p_i/1-p_i)$, where $p_i$ is the proportion of training instances of the label corresponding to the bias $b_i$ of unit $i$~\cite{hinton2012practical}. Concerning the regression problem, if the coefficient of variation w.r.t each output $j$ (i.e., the ratio of the standard deviation to the mean) is low (e.g., our default threshold is $0.1\%$), we verify that its corresponding $b_j$ is set to $m_j$, the mean value of the supervised target $j$. This eases the optimization by transforming the regression problem into predicting the deviation against the baseline (i.e., the mean value).
\paragraph{Cold Start Loss Verification}
An unexpected loss at the iteration $0$ with an untrained model (cold start), can indicate numerous issues including faulty loss function, ineffective loss reduction strategy, as well as buggy parameters' initializers~\cite{CS231n}. Indeed, we compute the loss at cold start with an increasing size of batches in order to verify that the obtained losses are not proportionally increasing. This validates that the loss estimation is an average-based expectation and it is not a sum-based reduction. In our verification routine, we duplicate a random batch of data by doubling its size ($\times 2$), then, we check if their corresponding losses at cold start are doubling ($\times 2$) as well. Next, we verify that the optimizer starts well at the expected initial loss (i.e., the one estimated at the first run with random internal parameters). It is always possible to derive approximately the correct initial loss for a given DNN program configuration. For instance, the cross-entropy loss for a balanced classification problem should start with uniformly distributed probabilities of $p(label)=1/L$ and initial value of $loss = -\log(1/L)$, where $L$ is the number of target labels. 
\paragraph{Tensor-Based Operation Verification}
A careless developer can introduce mistakes when transforming math formulas or pseudo-code from white papers to tensor-based operations using basic DL libraries calculation APIs. As the DNN tensor-based computations include intensive broadcasting and reduction operations in order to perform individually calculation over all the instances/neurons at each level, then, reduce the calculations to define aggregative scores towards summarize all into a scalar cost that represent how well the DNN performs on the data. We found that common implementation mistakes miss or add unnecessary dependencies to network components (instances, neurons, scores,..etc). Thus, it is important to test that a written operation depends only on its related components. For example, an activation function should be applied separately on all the neurons, which means the activation output $i$ should depend only on the neuron input $i$, or a distance calculation between prediction and actual values should contain independent components, where the distance value $i$ depends on only the prediction $i$ and the ground truth label $i$ as well. To validate the correctness of dependencies of all the math operations for a computed quantity within the neural network, we use the gradients flowing in the network to debug the dependencies between each operation's components given the fact that the gradient of a function w.r.t an independent component is always zero~\cite{karpathyRecipe}. For instance, let's consider a newly-implemented loss function, we can extract only the loss obtained for the outputs of a data point $i$, and then perform a full backward pass to the input data in order to make sure that only the gradient w.r.t the $i$-th input data is not null. A violation of this condition signals that overlapping dependencies exist, which means that the on-watch average loss for DNN performance is wrong and misleading. 
\paragraph{Computed Gradient Verification}
The backbone of backpropagation implementation lies in the computation of gradients with respect to different DNN operations, including linear weighted sum and non-linear activations. Whenever DL developers add hand-crafted math operations and gradient estimators, we perform a numerical gradient checking that consists of comparing between the analytic and numerical calculated gradients, respectively, the gradient produced by the analytic formula and the centered finite difference approximation, $\frac{f\left(x+h\right)-f\left(x-h\right)}{2h}$. Both gradients should be equivalent, approximately equal, for the same data points. The following steps are recommended by~\cite{CS231n, DL_ebook_2016} to improve the effectiveness of this gradient checking process (and the detection of faulty gradients).

\textbf{\textit{Relative error comparison:}} The difference between numerical gradient $f_n^\prime$ and analytic gradient $f_a^\prime$ represents the absolute error that should not be above a predefined threshold. However, it is hard to fix a common threshold of absolute error for DNN because its internal computations are usually composed of multiple functions; so the errors build up through backpropagation. Thus, it is preferred to use a relative error, $\frac{|f_a^\prime-f_n^\prime|}{max(|f_a^\prime|,|f_n^\prime|)}$, that might be acceptable below $1e^{-2}$.

\textbf{\textit{Sampled data instances:}} Sampling a few data instances for numerical gradient checking reduces the risk of crossing kinks, which are non-differentiable areas of the loss landscape. For instance, ReLU has zero gradient at the origin; but the numerical gradient can cross over the kink and produce a value different from zero. Besides, DNN's parameters are large with thousands of dimensions; so the computation error could be on a random subset of each gradient dimension. Therefore, a random sampling among both data points and dimensions makes the finite-difference approximations less error-prone and faster in practice.

\textbf{\textit{No regularization:}} The standard regularization can render large errors, misleading the numerical gradient checking when the penalty term added overwhelms the original loss (i.e., the gradient is mostly related to penalty cost). Moreover, advanced regularization techniques such as dropout induce non-determinism in the DNN internal computations, which enhances the error-proneness of numerical gradient checks.

\textbf{\textit{Prior burn-in training:}} A short burn-in training during which the parameters take better and more representative values than randomly initialized ones is recommended. It is also recommended to avoid the gradient checking at cold start since it could introduce pathological edge cases, masking a buggy implemented gradient. 
\subsubsection{Fitness of a single batch of data}
Given a tiny sample of data, the target problem becomes easy to solve and the training algorithm should be able to converge to a DNN that fits the data without any issue, as every well-designed DNN should be able to fit a small dataset. This is a main pre-check for DNN training routines because it is a necessary condition, where its non-satisfaction indicates a misconfiguration or a software bug.
\paragraph{Input Dependency Verification: }
We confirm that training programs on zeroed batches of data perform worse than those on real samples of data. This check was initially proposed by Karpathy~\cite{karpathyRecipe}, as a verification that the model outperforms an input-independent baseline. For debugging, this improvement over the input-independent baseline shows that the training program is successfully leveraging the input to optimize the DNN parameters. 
\paragraph{Overfit Verification: }
We verify that the optimization mechanism is working well on a controlled sample of data with reduced size (i.e., a few data points for each class)~\cite{DL_ebook_2016, CS231n}. The acceptance criteria is that the DNN achieves $100\%$ accuracy or near-to-zero absolute errors (AEs) on continuous outputs (by default, we consider AE in the order of $10^{-3}$). A failure to achieve this performance would signal the presence of issues regarding the DNN optimization routines. 
\paragraph{Regularization Verification: }
As above-mentioned, the controlled experiment of fitting a single batch would lead to overfit the provided few data points in normal situations, however, the loss should be greater than zero~\cite{DL_ebook_2016} when there is quite regularization. This check can be improved by watching furthermore the smoothness of the loss curve to spot a lack of noise in the optimization, and consequently, it reinforces doubts about the absence of active regulation. In our verification routine, we set up $1e-5$ and $0.95$ as default thresholds for loss and smoothness ratio~\ref{smoothness} to recognize suspicious loss curves that are smoothly decreasing towards a very low loss. This shows the model's propensity to overfit quickly the training data caused by a lack of regularization, which often implies high risks of capturing useless residual variations in the given features. 

Regarding pre-check of a single batch of data fitness, we concentrate on the functioning of the training algorithm, and precisely, its ability to converge (i.e., reaching an optimal equilibrium state) given a reduced size problem. However, the DNN training may still contain inefficiencies that did not prevent it from solving a small problem size but would affect its performance when it trains on larger problem size. Moreover, the failure at these batch fitness prechecks do not provide indications about the possible root causes behind this incapacity to successfully pass them. Therefore, the next phase in our debugging process aims to guarantee the ``proper functioning" of the training algorithm given a reduced size problem. What we mean by ``proper functioning" is the ability to converge with a valid accomplished trajectory (i.e., passing through valid intermediary DNN states). 

\subsection{Phase 2: Proper fitting a single batch of data}
At this phase, a monitored training is launched on a representative sample of data (i.e., a single batch of data resulting from a stratified sampling). The monitoring routines serve to detect early and precisely the potential issues with diverse automated verifications that watch periodically for the DNN components' misbehaviors to spot and report properties' violations. In case of a DNN with major issues, i.e., already failed the precheck on single batch fitness, this phase allows to steer the user towards problematic components and provide meaningful messages on the violated properties that restrict further the potential root causes. Otherwise, DL practitioners can still leverage this debugging phase for pre-examined DNNs to spot inefficiencies of their design choices that cause training instability in regards to learning updates or activation distributions, which might prevent the DNN from reaching optimal steady states and lead to a wastage of time and resources on unnecessary long training sessions with the full datasets. 
\begin{figure}[ht]
\centering
\captionsetup{justification=centering}
\includegraphics[scale=0.5]{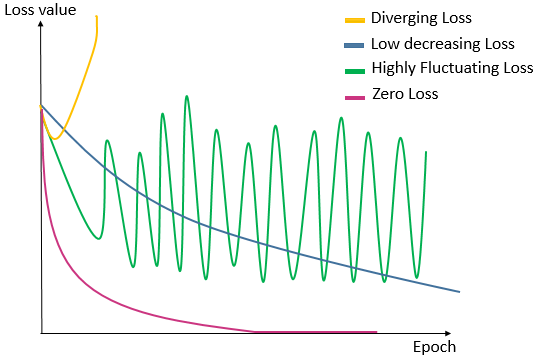}
\caption{Illustration of loss minimization issues}
\label{fig:loss_min_issues}
\vspace{-15pt}
\end{figure}
\subsubsection{Abnormal Loss Curvature}
A loss curve is a plot of model loss value over time in terms of epochs or iterations. The shape and dynamics of a loss curve are useful to diagnose the behavior of the optimizer against the target learning problem. Figure~\ref{fig:loss_min_issues} shows the abnormal loss curves, detailed-below, which indicate different pathologies of statistical learning from data. The anomalous loss evolution~\cite{CS231n} can be detected using continuously updated metrics that are cheap to compute and which can reveal anomalies effectively. 
\paragraph{Non- or Slow-Decreasing loss}
A flat or low slope down curve shows that the loss is either non-decreasing or decreasing very slowly which means that the model is not able to learn at all or has a low learning capacity. This could be due to an inadequate loss function or a low learning rate. As introduced in~\ref{dynamic_heuristics} for the stagnation test, we verify that the loss is decreasing at an acceptable decay rate for a window of steps, e.g., we set up $5$ steps by default. For the loss decreasing verification, we proceed by watching that the percentage difference between consecutive losses' values, $loss\_decay\_ratio = \frac{previous\_loss-current\_loss}{previous\_loss}$, is greater or equal to a predefined minimum percentage difference (by default, we set up $5\%$) by which the loss should decrease, $loss\_decay\_ratio > prefixed\_min\_loss$.
\paragraph{Diverging loss}
A curve with a high slope indicates that the loss is diverging with wildly increasing values which means that the optimization problem is turned into a loss maximization instead of minimization. This could be the result of a high learning rate or a buggy gradient. Therefore, we verify that the loss has no increasing tendency instead of decreasing one. This verification can be done automatically by updating a reference minimum loss ($lowest\_loss$), and then, watching that the absolute ratio of loss, $abs\_loss\_ratio = \frac{current\_loss}{lowest\_loss}$ (i.e., we call it absolute ratio because it is computed w.r.t the lowest obtained loss) is not dramatically diverging, as described in~\ref{dynamic_heuristics}, during the training.
\paragraph{Highly-Fluctuating loss}
A noisy curve with random fluctuations demonstrates that the loss is not converging to a line of stability, hence, the optimization is facing difficulties preventing it from converging normally. Potential reasons behind this issue could be: strong regularization that gives rise to noisy loss estimation, or high learning rate that produces large updates keeping the optimizer jumping over the local minimum without converging. Hence, we compute the smoothness ratio of the loss curve as follows:
\begin{equation*}
\label{smoothness}
smoothness\_ratio = \frac{N\_samples - N\_direction\_changes}{N\_samples}
\end{equation*}
Where $N\_samples$, $N\_direction\_changes$ denote, respectively, the sampled iterations count and the number of alternations in the loss evolving direction. Then, we check periodically that the smoothness ratio is not lower than a predefined threshold, otherwise, it indicates a high amount of fluctuations (e.g., we fix $0.5$ as a default threshold, which means that more than $50\%$ of consecutive sampled losses are having altered directions).
\subsubsection{Performance Metrics Correlation}
DL practitioners should select or implement the right loss measure (i.e., mean squared error, cross-entropy, etc.) that will be set as objective function for parameters optimization, while they keep watching a target performance metric (i.e., R-squared, Accuracy, etc.). For this verification routine, we compute, continuously, the absolute value of correlation coefficient between the optimized loss and the target performance measurements between training steps. The latter describes the magnitude of the relationship between two variables within the interval of $[0,1]$ and should not become lower than a predefined threshold (e.g., our verification routine reports an absolute correlation coefficient of less than $0.5$). This provides an indicative metric of how representative the optimized loss is with respect to the target objective and vice-versa. Thus, a poor choice of loss function would have high chances to be uncorrelated with the performance metric, similarly, a buggy performance calculation would yield incorrect values with low correlation with the loss evolution. 
\subsubsection{Unstable Gradient}
We propose to detect unstable gradient issues by examining, continuously, the evolution of estimated gradient's values with respect to each DNN's layer. More specifically, we check the growth and decay rate of the absolute average of the layer's gradients, estimated for the last iterations, to detect if the latter is following an unstable evolution trend, i.e., it is exploding or diminishing, as described in~\ref{dynamic_heuristics}. 
\subsubsection{Magnitude of Regularization Penalty Cost}
In addition, there is an issue with the addition of a high amount of regularization to the computed loss, whether by using a high $lambda$ value or a poorly-designed penalty regularization cost. Indeed, Park et al.~\cite{park2018gradient} highlight that the learning suddenly fails when the magnitude of gradients from $loss(W,b,D)$ decreases faster than that from $\Omega(W)$; so the penalty term gradient overwhelm the loss data gradient. In such problematic situations, the weights' updates become mainly related to the regularization term, which causes the failure of the model to learn.\\
One should ensure that the regularization is not too strong. The regularization term gradient should not dominate and suppress the loss gradients w.r.t the weights.To ensure this, we recommend watching continuously, the proportion of the magnitude of the penalty terms' gradients w.r.t the magnitude of loss data in order to validate that it is not diverging, as described in~\ref{dynamic_heuristics}.
\subsubsection{Parameters States and Dynamics}
The main DNN's on-training parameters, weights and biases, should be optimized towards better solving the learning problem over the training sessions. In the following, we propose different verification routines on the parameters current states and update dynamics that could indicate convergence issues and non-optimality of the on-going estimation.
\paragraph{On-Training Parameters Verification} 
A preliminary sanity check for neural network parameters would be verifying their changing estimations over the training iterations. We make sure that all defined trainable layers have their parameters updated. Indeed, a non-zero difference between the values of trainable parameters before and after the execution of a few training passes (i.e., optimization updates) confirms that the dependencies between the layers are correctly set up and that all the trainable parameters are getting optimized.
\paragraph{Dead and Over-Negative Weights Verification} 
A layer's weights are considered dead or over-negative, when respectively, the ratio of zeros or negative values in the tensor elements is very high~\cite{dead_W_U}. These two states of weights are likely to be problematic for the learning dynamics. Indeed, given a common DNN setup (e.g., normalized inputs within $[0,1]$ and a variant of ReLU as hidden layers activations), null or negative learned weights within hidden layers represent connections to intermediate features that do not contain any relevant information for the target task. Thus, if a layer's weights are mostly full of null or negative weights, their corresponding activation layers are likely to stagnate on a non-optimal flat region and consequently, the DNN would start facing dead layers (i.e., ReLUs mostly outputting the value zero) and frozen layers (i.e., no updates of the weights). In our verification routine, we flag any tensor of layer's weight ($W$), having either the ratio of very low values (i.e., lower than $1e-5$) or negative values is higher than a predefined threshold (i.e., $95\%$ is used by default), as respectively, dead or over-negative weights.
\paragraph{Stable Parameters Update Verification}
Deep neural networks introduce challenges regarding learning stability compared to shallow networks. In the training pitfalls section~\ref{catalog}, we discussed some practices such as tuning the learning rate and adding activation or weight normalization to balance out the learning speeds for the hidden layers. Thus, it is important to make sure that the parameters' updates are stable. Hinton~\cite{hinton2012practical} and Bottou~\cite{leon2015} proposed the following heuristic, the magnitude of parameter updates over batches should represent, respectively, $0.1\%$ or $1\%$, of the magnitude of the parameter itself, not $50\%$ or $0.001\%$. Therefore, we propose a verification routine to detect unstable learning parameters by comparing the magnitude of parameters' gradients to the magnitude of the parameters themselves. More specifically, following the recommendation to keep the parameter update ratio around $0.01$ or $0.001$ (\ie{} $-2$ or $-3$ on base 10 logarithm), we compute the ratio of absolute average magnitudes of these values and verify that this ratio doesn't diverge significantly from the following predefined thresholds:
\begin{equation*}
-4 < \log_{10}\left(\overline{|W^{(i+1)}-W^{(i)}|}/\overline{|W^{(i)}|}\right) < -1
\end{equation*}
The proposed verification mechanism reports irrelevant layers (i.e., where updates are unstable) and frozen layers (i.e., where updates are stalled) to the user. 
\paragraph{Parameters Diverging Verification} 
Worse than unstable learning, weights and biases risk divergence, and may go towards $+/-\infty$. For instance, high values of initial weights or learning rate with a lack of--or--insufficient regularization provokes highly-increasing weights' updates, leading to bigger and bigger values, until reaching $\infty$ (this is caused by overflow rounding precision). In addition to that, biases can also become huge in certain situations where features could not explain enough the predicted outcome or might not be useful in differentiating between the classes. Therefore, we automate a verification routine that watches continuously the absolute averages of parameters are not diverging, as described in~\ref{dynamic_heuristics}. 
\subsubsection{Activations Distribution}
\paragraph{Out-of-Range Activation Verification}
Given a newly-implemented activation function, we include a baseline verification routine to watch that the produced activations are within the expected range of values. This would be useful to find computation mistakes or misconceptions causing out-of-range outputs. 
\paragraph{Validation of Output Activation Domain}
The last layer's activation represents the outputs of the neural network, which should produce valid outcomes while covering the whole distribution of the possible ground truth labels. We implement a verification routine to check that the outputs of classifiers are probabilities, i.e., positive values within $[0,1]$ and sum-to-one in case of multidimensional output. For regression, we empirically verify that the predicted outputs over the iterations were able to satisfy necessary conditions derived from ground truth boundaries, i.e., non-zero variance, can be negative, can exceed $1.0$. Under these conditions, the common identified faults of using the wrong activation function can be detected.
\paragraph{Saturated Bounded Activation Detection}
To detect saturation issues in DNNs, we compute single-valued saturation measure $\rho_B$ proposed by Rakitianskaia and Engelbrecht~\cite{saturationNN} if the hidden activations are bounded functions such as sigmoid or tanh. This measure is computed using the outputs of an activation function and is applicable to all bounded activation functions. It is independent of the activation function output range and allows a direct statistical measuring of the degree of saturation between NNs. $\rho_B$ is bounded and easy to interpret: it tends to $1$ as the degree of saturation increases and tends to zero otherwise. It contains a single tunable parameter, the number of bins $B$ that converges for $B \geq 10$, \ie{} it means splitting the interval of activation outputs into $B$ equal sub-intervals. Thus, $B = 10$ can be used without any further tuning. Given a bounded activation function $g$, $\rho_B$ is computed as the weighted mean presented in Equation~\ref{rhoB}.
\begin{equation}
\label{rhoB}
\rho_B = \frac{\sum^B_{b=1}|\bar{g}'_b|N_B}{\sum^B_{b=1}N_B}
\end{equation}
Where, $B$ is the total number of bins, $\bar{g}'_b$ is the scaled average of output values in the bin $b$ within the range $[-1,1]$, $N_b$ is the number of outputs in bin $b$. Indeed, this weighted mean formula turns into a simple arithmetic mean when all weights are equal. Thus, if $\bar{g}'_b$ is uniformly distributed in $[-1, 1]$, the value of $\rho_B$ will be close to $0.5$, since absolute activation values are considered, thus all $\bar{g}'_b$ values are squashed to the $[0, 1]$ interval. For a normal distribution of $\bar{g}'_b$, the value of $\rho_B$ will be smaller than $0.5$. The higher the asymptotic frequencies of $\bar{g}'_b$, the closer $\rho_B$ will be to $1$.\\
This verification routine can be automated by storing for each neuron its last $O$ outputs' values in a buffer of a limited size. Then, it proceeds by computing its $\rho_B$ metric based on those recent outputs. If the neuron corresponding value tends to be $1$ (e.g., a threshold of $0.95$ is used in practice to spot this tendency), the neuron can be considered as saturated. After checking all neurons for saturation, we compute the ratio of saturated neurons per layer to alert the DL developers about layers with saturation ratios that surpass a predefined threshold (e.g., we fix $50\%$ by default). 
\paragraph{Dead ReLU Activation Detection}
By definition, a given neuron is considered to be dead if it always returns zero~\cite{lu2019dying}. Hence, we detect practically dead ReLUs by reducing the last outputs for each neuron stored in the limited size buffer into a single $95^{th}$ percentile, which is more robust than the maximum reduction against outliers (e.g., a non-zero stored values from earlier training iterations). Thus, we mark all the neurons with a $95^{th}$ percentile less than a predefined threshold (i.e., by default, we set up $1e-5$). Next, we proceed by calculating the ratio of the marked dead neurons per layer and we flag the layers with a number of dead neurons higher than a predefined threshold (e.g., we fix a default threshold of $50\%$).
\paragraph{Unstable Activation Detection}
Although this unstable activation issue is more generic than dead or saturation phenomena, DL experts usually watch the histograms of sampled activations from each layer while expecting to have normally-distributed values with unit standard deviation, e.g., a value within $[0.5,2]$ has been shown to be an acceptable variance of activations~\cite{actUnstable}. Thus, we base on this expert's heuristic to statistically validate that the sampled activations of each layer over the last iterations is having a well-calibrated variance scale. Concretely, the test would pass the actual standard deviation ($\sigma_{act}$) belongs to the range of $[0.5,2]$; otherwise, we perform an f-test~\cite{f_test} to compare $\sigma_{act}$ with either the low-bound $0.5$ if $\sigma_{act} < 0.5$ or $2.0$ if $\sigma_{act} > 2.0$.  

DL practitioners can perform a closed feedback loop using this inexpensive and rapid debugging on a single batch of data until fixing all the covered coding bugs and improving the settings in a way that enhances the chances of converging to a more optimal model. Thus, a successful pass at this phase increases the confidence that the training program is devoid of common DNN pathologies such as: vanishing gradients, saturation of activation functions, and inappropriate learning speed. Nevertheless, other components of the training program have not been tested yet. For instance, a training program may pass all the single batch debugging checks, while the data loader can inject too much noise or mismatch features and labels, which yield corrupted batches of data, and consequently, the resulting DNN does not solve the target problem. 
   
\subsection{Phase 3: Post-fitting conditions}
Once the fitting of a single batch step is successfully passed, several post-fitting conditions should be satisfied to guarantee the correctness of the data loader, the data augmentation module, and the advanced regularization techniques that require additional computations during the inference. The following debugging phase validates the behavior of the DNN training program during regular training sessions, i.e., we use the available training and validation data. 
\subsubsection{Distribution-Shifting Augmentation Verification}
We propose a post-check that verifies the validity of the augmentation methods, applying data transformations on the generated batches to enhance diversity and improve the generalizability by smoothing the loss landscape and forcing the model to capture the invariants useful for the target task. A valid augmenter should not introduce a shift in the data distribution that makes the model perform worse on the original dataset. For instance, overwhelming noise injection leads to produce meaningless inputs; so both of ratio and scale of the injected noise should be carefully picked to hold the augmented data distribution close to the original one.\\
To detect this poor design of data transformations, we debug the data augmenter module by comparing the performance and the activation patterns of the DNN trained with-and-without augmentation on a sample of data from the validation set. Concerning the measurement of dissimilarity between the activation patterns, we use Centered Kernel Alignment (CKA)~\cite{act_sim}, which is an optimized and powerful representational similarity measure, allowing the assessment of the differences and the correspondences between patterns learned by different DNNs or same DNN with different data. Indeed, given two matrices $X$,$Y$, where $X\in\mathbb{R}^{m\times n_1}$ is a matrix of activations of $n_1$ neurons for $m$ examples and $Y\in\mathbb{R}^{m\times n_2}$ is a matrix of activations of $n_2$ neurons for the same $m$ examples, a DNN representational similarity index $s(X,Y)$ estimates the similarity between the representations learned in both matrices of activations. Thus, the CKA's empirical validation~\cite{act_sim} shows its ability to compare representations within and across DNNs, in order to assess the effect of changing variation factors on the DNN training. This similarity metric is robust and it does not affect by the stochasticity of the optimizer or the random initializations, which makes it suitable for our verification on the resulting activation patterns instability when the augmented data is noisy and shifted w.r.t the training data distribution.\\
Therefore, the DNN training program would fail the test if there is a degradation of the performance and a substantial difference in the activation patterns between the two trained DNNs. In our implementation, we check for an increase in the loss ($loss_{augm\_data} > loss_{orig_data}$) and a decrease of activation pattern similarity (e.g., we set up a threshold of a minimum CKA index equals to $0.8$, which means a decrease of $20\%$). Indeed, any difference in the activation pattern could be acceptable and might be considered as an improvement in the detected patterns in case the performance is enhanced, however, having both of the activations pattern divergence and the performance degradation indicate strongly that the augmented data induce a distribution shift. 
\subsubsection{Corrupted Labels}
Given a corrupted data shuffler, the paired data (features $X$, labels $y$) are mismatched, where the row label $y_i$ does not correspond to the row features $X_i$. Since the shuffling is executed after each epoch (i.e., full pass over the data), a corrupted data loader will generate a new distribution of supervised training data every time it is called. Thus, the training loss curve would be subject to intermittent spikes because the neural network starts learning on a new distribution at the $1$th-iteration of every epoch. Based on this observation, we propose to collect all the $1$th-iteration losses into a set and perform a statistical test to detect if there is significative difference between them, which means the loss estimated on the first sampled batch of data is improved over the epochs; otherwise, we flag the data loader as corrupted because the DNN successfully passes all the previous verifications, but cannot improve its performance over the epochs; so it is high probable that the data loader is falsifying the batches of data in-between the elements' shuffles.
\subsubsection{Unstable Mode Transfer: From Train to Inference Mode}
Advanced regularization techniques like dropout or batchnorm introduce, respectively, noise-injection and normalization mechanisms in order to grant the DNN training access to sub-model ensembling and well-conditioned loss minimization. They incorporate two functional modes: the training mode and the inference mode; so many bugs can remain silent and hidden during the training mode, but the transfer to the inference mode can reveal them through DNN misbehaviors and divergences induced by the mode transfer~\cite{dishankPitfalls,ceceliaCheckList}. Thus, we construct a verification routine that detects the behavioral shift occurring when transferring the DNN from the training mode to the inference mode. Our implemented behavioral difference assessment is based on the CKA metric for representational dissimilarity between different modes' activations (e.g., we fix a default threshold of $0.75$ as minimum similarity of activation patterns on the same data), and the relative change between the modes' losses (e.g., the threshold is by default set to $50\%$). Then, one can check the different regularizers' internals, including $pkeep$, $E[x]$ and $Var[x]$ to diagnose the root cause.
\subsection{TensorFlow-based Implementation}
To assess the effectiveness of our proposed debugging approach on real faults in DL-based software systems, we implemented a TF-based library that performs the debugging phases on a training program written using TF features. Indeed, we choose to focus on TF-based training programs because of the popularity of TF in the ML community~\cite{braiek2018open}. Nevertheless, the property-based approach proposed in this paper can be adapted for other DL frameworks. In the following, we describe the components of this TF library.
\subsubsection{Setting Up the Testing Session}
The testing of a DNN training program can be more complicated than for a traditional program, because of its non-deterministic aspect. It is difficult to investigate the training issues and identify the root causes when the program exhibits a substantially new behavior for each execution. To reduce the stochasticity of a DNN training program, we fix all the random seeds of all the computational libraries beforehand, which guarantees the reproduction of the same random variables. Furthermore, we offer the option of deactivating the parallelism, if a tester wants to obtain a perfectly reproducible result. As default settings, we allow leveraging multi-cores CPUs and GPUs through multithreading execution. We allow this because a single-thread execution slows down dramatically the training time. Also, \name{} targets major training issues related to erroneous training behaviors caused by the introduction of coding bugs or misconfigurations, which differ from minor training issues caused by non-optimal choices of hyperparameters, leading to near-to-optimal DNNs instead of the best-fitted one. Therefore, the resulting pathologies in the training dynamics would likely be persistent to the possible low magnitude differences between multiple executions of parallel floating-point computations.
\subsubsection{Fetching and Monitoring the Training Program Internals}
TFDBG~\cite{tensorflowDebugger} is the official debugging tool specialized for TF programs that offers features such as inspection of the computational graph, addition of conditional breakpoints and real-time view on internal tensor values of running TF computational graphs. However, it is not practical for our approach since it adds a huge overhead on computation time, as it handles each execution step of the graph, to allow debugging the issues and pinpoint the exact graph nodes where a problem first surfaced. In fact, the implemented verifications fetch the values of tensors, representing parameters, activations, and gradients, by requesting through the provided API the on-running DAG that encodes both the DNN design and the used training method. The routines that continuously monitor the state of the neural network do not need to break neither the feed-forward nor the backward passes since they can access the internals of the intermediate neural networks to detect pathologies in the learning dynamics. This allows the monitored training iteration to be executed in an atomic way and avoid the overhead of using TFDBG. To manage our set of verification routines running simultaneously, we use the monitored session and hooks mechanism; to handle the additional processing injected between training iterations. To do that, we need to perform the following two steps: 
\begin{enumerate}
\item Create one or more Hooks objects that implement methods such \emph{before\_run} and \emph{after\_run} to access the intermediate tensors' values of activations, parameters, and gradients, then, apply the verification logic.
\item Create a \emph{Monitored Session} that handles the execution of hooks' additional treatments before and after running each training pass.
\end{enumerate}
\subsubsection{Buffering the DNN's status data}
The activations and gradients represent intermediary computations over, respectively, forward and backward passes. The weights and biases represent learnable parameters that are updated, continuously. As a result, the internal tensors do not survive between two consecutive training iterations. Thus, we implement buffer data storage to save incrementally the values of watched tensors. By default, we set the size of the buffer to $10$ elements but we keep it a configurable option in the debugging tool. 
\subsubsection{Running the checks}
Conceptually, each check performs the following two steps. First, the instantiation of the property requires the computation of necessary metrics in relation to the targeted violations. Regarding the on-training verification routines, a data preparation using a reduction strategy (i.e., average, norm, and quartiles, etc.) is necessary to aggregate the accumulated tensors in the buffer data storage. Second, the issue detection consists of applying a heuristic-based verification rule, involving the inferred metrics and predefined thresholds, that captures the negative effects and anomalous training behaviors induced by the targeted issue. Indeed, the choice of thresholds would affect the sensitivity and specificity of the issue detection. Additionally, the training pathologies are correlated to each other and a particular bug can be the root cause of multiple of them. Thus, our dynamic debugging strategy alleviates these challenges by leveraging limited-size buffers, continuous verification routines, and informative raised errors. Therefore, \name{} implements a debugging feedback loop that does not stop after finding violations, but it keeps watching the training execution while dynamically producing error messages that steer the DL developer to further narrow down the possible root causes and avoid errors conflicts relying on the persistence of the errors, the chronological order of the raised issues, and the reported information.
\subsubsection{Shrinking the suspicious state and raising errors}
\label{shrinking-prog-states}
Once an issue is detected, \name{} shrinks the state of the on-training DNN to communicate the component where the property violation is found and its corresponding indicators including the reference of the layer, the computed metric, and the predefined threshold. In fact, the reported information should provide an explanation of why the underlying property is considered to be in violation. First, reporting the shrinked neural networks' states helps the user avoid false positives. For instance, the learning speed can start by relatively high updates that can be close or-even slightly larger than the prefixed update magnitude threshold. Thus, a single warning message including the current update magnitude and the surpassed threshold could help the user decide whether or not the unstable learning issue occurs in the current situation. Second, the shrinking of the buggy DNN state is useful to pinpoint the suspected computational units that developers should investigate, and therefore, it helps them identify the occurred issue's root cause. For example, the computational layers send information throughout the DNN by using edges that connect layers to one another during the forward pass and they receive updates from the gradients of the loss flowing reversely over the same edges during the backward. Thus, the level of the dead layer and its amount of dead neurons, as well as, the level of vanishing gradient and its magnitude scale can be used to identify the actual unstable layer, where the information is no longer flowing during either forward or backward pass. Developers would first investigate the underlying layer and then fix the bug within the program using this information.
\begin{figure}[t]
\centering
\includegraphics[scale=0.75]{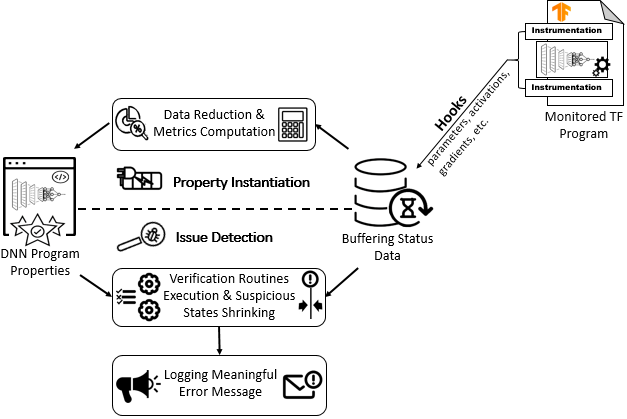}
\caption{Illustration of our Property-based Debugging Approach for TF Programs}
\label{fig:Implementation}
\end{figure}

Figure~\ref{fig:Implementation} summarizes the above-mentioned implemented steps of our property-based debugging approach for TF Programs. 
\section{Evaluation}
\label{evaluation}
The objective of this evaluation is to assess the effectiveness of our proposed property-based debugging method in allowing for the early detection of real bugs that occur in DNN-based software systems. We also conducted a usability study with two professional DL engineers to assess the relevance of \name{}'s error messages at guiding developers in identifying the root cause of bugs and fixing them. 
\subsection{Design of Case Studies}
In this section, we describe the design of our case study that aimed to assess the performance of \name{}. We explain how we selected and reproduced relevant bugs for our evaluation.
\subsubsection{Real Faults in DNN-based Training Programs}
The reproduction of buggy DNN training programs is quite difficult because of the rapid evolution of TF API functions and even infeasible when major code blocks, datasets, or environment settings are missing. In previous research works, program mutation~\cite{mutation} was leveraged to evaluate the quality of DL program debugging approaches. The mutation relies on well-defined rules to change slightly the syntactic structure of the code, or mimic systematically application-specific errors. Then, the debugger should detect and reject mutants, which is called killing the mutant in such analysis. Thus, its effectiveness is measured by the ratio of killed mutants w.r.t the total of generated mutants. Xie et al.~\cite{xieMT} leverage MuJava~\cite{mujava}, which is an automatic java code mutator, to produce defective mutants of Weka’s implementations~\cite{weka} of $k$-Nearest Neighbors (kNN) and Naive Bayesian (NB). Dwarakanath et al.~\cite{DLMTDebug} use MutPy~\cite{MutPy}, an open-source tool for python code mutation, to inject typical programming errors in a clean implementation of deep residual neural networks (ResNET). Both mutation analyses rely on language-specific code mutators that alter randomly the arithmetic operators, logical operators, variable's scope and casting types, etc. However, even if these random code alterations mimic a large variety of coding mistakes, they cannot be representative for the real-world buggy training programs, where the code is relatively short with heavy dependency on tensor-based computational libraries. Besides, Ma et al.~\cite{deepmutation} proposed DeepMutation that defines DL-specific mutation rules for DL programs, including a layer addition, a layer removal, and an activation function removal. Based on our investigation on DL faults, we found that these operators can actually mimic real faults in relation to missing DNN components, such as missing batchnorms (i.e., removing the normalization layer) or redundant softmax (i.e., adding another softmax on the output layer). However, many of the generated mutants using these operators can be equivalent to the original network, or even better for solving the underlying learning problem. This makes the evaluation based on the ratio of killed mutants misleading. Therefore, it is necessary to assess the effectiveness of our debugging approach on detecting the real DL bugs that have been experienced and reported by the DL practitioners. As shown in Figure~\ref{fig:Pitfalls_Process}, we have collected concrete instances of real DL bugs~\cite{DLfaults} that cause training issues without crashing the DL program. Thus, we rely on their identified root causes and symptoms to inject each DL fault into clean DL programs to force the creation of valuable synthetic buggy versions. In the following, we detail the two high-level categories of common root causes for the non-crashing DL bugs.
\subsubsection{Coding Bugs in DNN-based Training Programs}
Like any software system, DNN training programs may contain the missing and wrong code statements that cause a deviation between the designed DNN and its corresponding written code (see Table ~\ref{bugs_root_causes}). In fact, the DNN training program is implemented using conventional programming languages, which may include coding faults. The lack of oracle for internal variables and the stochastic nature of the DNN optimization process, make most of these coding mistakes hidden without disrupting the flow of the program's execution. For instance, the majority of these hidden bugs are incorrect math operations such as flipped sign result, inverted order of calculations, wrong data transformations. We found another type of coding bugs in TF programs that consist of a TF API misuse committed by developers who misunderstood the implicit assumptions regarding how to use these configurable routines. In fact, modern DL libraries provide rich APIs covering more and more state-of-the-art techniques. Consequently, the API routines integrate more and more configurable options and set up default values to make them ready-to-use for quick prototyping. However, they assume that their users are capable of configuring them properly, which is unfortunately not always the case. Some of the built-in data loaders, for example, automatically perform standard pre-processing of numerical data, such as normalization. This misleads some rookies to blindly perform a double linear scaling afterward. Another common API misuse is related to the recent versions of loss functions. Indeed, numerically stable implementations regarding some of the loss functions require merging the loss and output activation formula together to re-write them carefully without any potential $\log(0)$ or $\exp(\infty)$. However, users may ignore this gap between theoretical loss function and built-in numerical stable ones; which may result in redundant activations.   
\subsubsection{Misconfigurations of DNN-based Training Programs}
Modern DNN training programs are highly-configurable software built using routines from DL libraries. Their correct settings, given the context, becomes a challenging task and if an incident occurs due to misconfiguration, the on-training DNN may produce misleading performance faults. These misconfigurations assemble all the wrong and poor choices for the configuration of DNN-based software systems, including the DNN design and the training method (see Table~\ref{bugs_root_causes}). The lack of understanding of DL fundamentals is the main reason behind the occurrence of these configuration issues, especially, when dealing with a novel DL technique or facing an unfamiliar target problem. For instance, numerous misconfigurations in relation with random initializers, loss functions, normalization methods and optimization hyperparameters, lead to training pathologies. Others are related to the DNN design and structure that lead to performance degradation, whether the DNN is underfitting the data (i.e., low training and validation errors) or overfitting it (i.e., low training error and high validation error). The misconfiguration of a DNN training program impacts the effectiveness of the training process, and consequently, the quality of the trained DNN. However, they manifest themselves during the model learning process; so the debugging of the training program should help catch these undercover failures that are hard to detect at inference executions during the model testing. 
\begin{table}
\hspace*{-1cm}
\caption{Real Bugs in DNN-based Software System}
\label{bugs_root_causes}
\begin{tabular}{|c|c|c|} 
\hline
\textbf{Category} & \textbf{Bug} & \textbf{Common Root Cause(s)}\\
\hline
\multirow{10}{*}{\textbf{\shortstack{Coding \\ Bug}}}&\multirow{2}{*}{missing preprocessing}&missing input normalization\\
\cline{3-3}
&&missing output normalization\\
\cline{2-3}
&wrong preprocessing &redundant data normalization\\
\cline{2-3}
&wrong optimisation function &gradients with flipped sign\\
\cline{2-3}
&missing softmax &missing softmax activation\\
\cline{2-3}
&redundant softmax &softmax out-and in-the loss\\
\cline{2-3}
&\multirow{3}{*}{wrong type of activation}
&softmax for hidden activations\\
\cline{3-3}
&&softmax for $1$-dim output\\
\cline{3-3}
&&over-restricted output domain\\
\cline{2-3}
&wrong softmax implementation &softmax over wrong axis\\
\cline{2-3}
&\multirow{3}{*}{wrong loss function }&CE over wrong axis\\
\cline{3-3}
&&inverted CE's mean and sum\\
\cline{3-3}
&&MSE with wrong broadcasting \\
\cline{2-3}
&wrong data batching &shuffle only the features\\
\cline{2-3}
&wrong data augmentation &invalid data transformation\\
\hline
\multirow{7}{*}{\textbf{\shortstack{System \\ Misconfi- \\ guration}}}&\multirow{2}{*}{wrong initialization }& constant weights\\
\cline{3-3}
&& dummy random weights\\
\cline{2-3}
&wrong loss selection & use of MSE instead of CE\\
\cline{2-3}
&\multirow{2}{*}{suboptimal learning rate }& a low learning rate\\
\cline{3-3}
&&a high learning rate\\
\cline{2-3}
&epsilon for optimiser too low & an Adam epsilon $\epsilon<10^{-8}$\\
\cline{2-3}
&\multirow{2}{*}{missing normalization layer }& missing batch-norms \\
\cline{3-3}
&&no-update of batch-norm globals \\
\cline{2-3}
&\multirow{4}{*}{missing regularisation }
&low $\lambda$ for norm penalties\\
\cline{3-3}
&& high $\lambda$ for norm penalties\\
\cline{3-3}
&& high $keep\_p$ for dropouts\\
\cline{3-3}
&& low $keep\_p$ for dropouts\\
\cline{2-3}
&\multirow{1}{*}{unbalanced dataset }
&Labels are not equally distributed\\
\hline
\end{tabular}
\end{table}

\subsubsection{Synthetic Tensorflow Buggy Programs}
\label{synthetic_examples}
\begin{figure}[ht]
\centering
\includegraphics[scale=0.45]{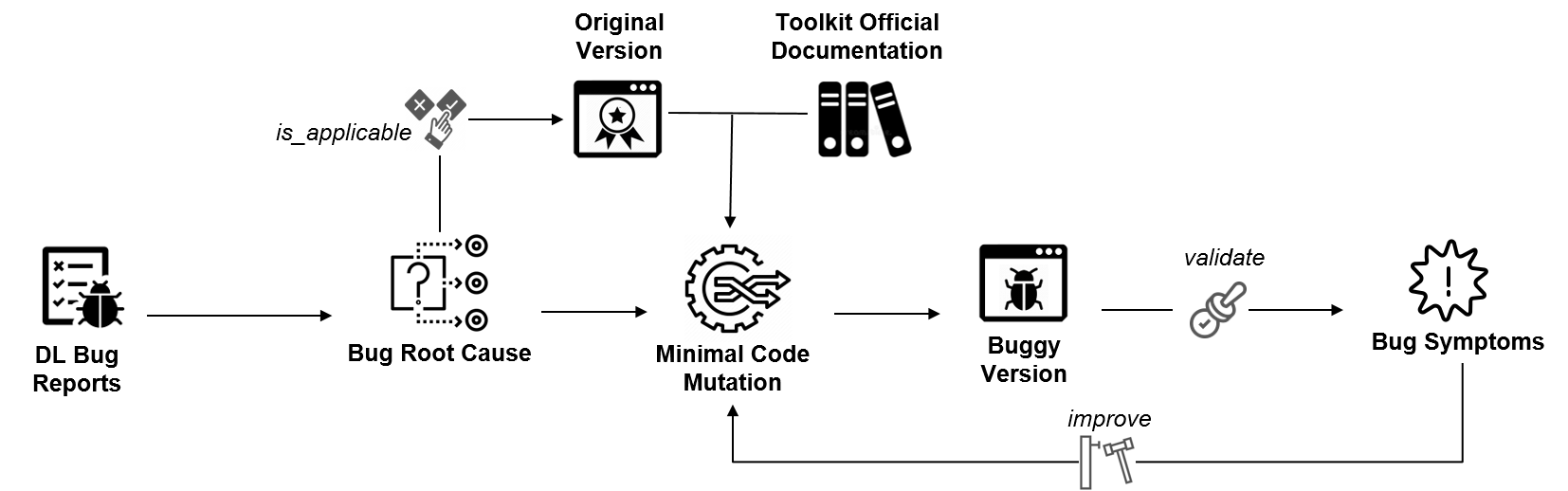}
\caption{Overview of Synthetic DL Programs' Creation Steps}
\label{fig:Syn_Prog_Process}
\end{figure}

Liu et al.~\cite{liu2016towards} designed a base CNN that represents a typical CNN for image classification. Then, they derived diverse ineffective variants by poorly re-designing some parts of the CNN to evaluate the capacity of their visual diagnosis tool, CNNVis, in detecting the effects of the added poor design choices. Indeed, \name{} gauges different statistics and metrics on the DNN internals to detect fine-grained symptoms on the dysfunctioning or instability of DNN's components. The heuristic-based verification rules identify the occurred bug based on its effects on the training routines and flag the defective component relying on its current metrics' status. Therefore, our first evaluation of \name{} consists of an assessment on synthetic buggy DNN programs. Figure~\ref{fig:Syn_Prog_Process} shows our systematic approach, following the same methodology of Liu et al.~\cite{liu2016towards}, to create synthetic mutant DL programs, containing the above-mentioned DL faults. First, we select the Base DNN training programs for which the identified DL fault is applicable. Indeed, we prepare a mixed set of Base NNs in order to cover different architecture and activation functions that are related to a particular learning problem (i.e., either classification or regression), as well as, advanced techniques to regularize and stabilize the training of complex DL models (i.e., increasing the depth of the neural network enhances its complexity). However, adding arbitrarily hidden layers to have higher learning capacity would be unnecessary and may induce issues if it is applied on simple learning problems. Thus, we set up two base CNNs, ShallowCNN and DeepCNN, that have been proposed to solve two well-known classification problems with increasing complexity. Next, the injection of the DL bug in the Base program consists in mutating minimally the original source code based on its main root cause and the toolkit official documentation (i.e., Tensorflow). Then, we validate the presence of the DL bug-related symptoms; otherwise, further refinements should be performed. Generally, the symptoms observed for non-crashing bugs are any unexpected low model performance (i.e., accuracy or average error) or slow learning process. The bug reporters were able to perceive the symptoms based on their past DNNs training experiences or their comparisons with the original source (e.g., a research paper, or a tutorial). In a similar fashion, the reference performances of model and learning speed for our Base DL programs is already known and can be leveraged to check the success of the bug injection. Following the steps described in Figure~\ref{fig:Syn_Prog_Process}, we are able to create a buggy synthetic DNN program for each matched pair of a base neural network and a single DL fault. This helps isolate the DL faults and assess the sensitivity and specificity of \name{} regarding each injected fault separately. At this level of controlled experiments, we could separate the valid fired checks that highlight the foreknown negative effects of the injected fault, and the false positives that point out to other irrelevant side effects (which could mislead the users during the debugging).

In the following, we describe the three Base NNs that we identified based on empirically-evaluated architecture, officially-debugged TF implementations, and publicly-available datasets.

\begin{figure}[ht]
\centering
\includegraphics[scale=0.55]{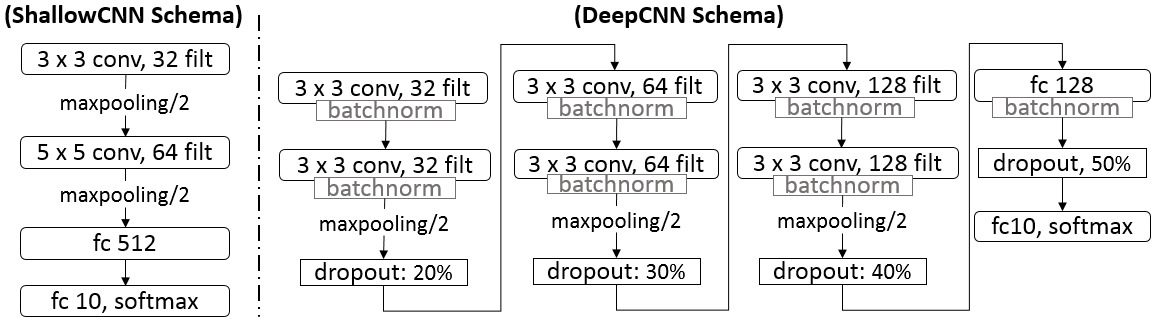}
\caption{Schema of the two BaseCNNs Architecture}
\label{fig:BaseCNNs}
\end{figure}

\textbf{RegrFNN.} The RegrFNN represents a basic feedforward neural network inspired from Google official examples~\cite{regrNN}, which contains two hidden fully-connected layers of $64$-neurons with ReLU activation. It uses mean squared error (MSE) to compute the loss for regression problems. For regularization, we add both ridge regression (L2-norm of weights). Regarding the optimization, we use ADAM as a variant of gradient descent algorithm. It was trained on the classic Auto MPG Dataset~\cite{Duad} with the aim of predicting the fuel efficiency of late-1970s and early 1980s automobiles. The RegrNN reached less than $1.85$ of mean absolute error on unscaled outputs within the range of $[10, 47]$.

\textbf{ShallowCNN.} The ShallowCNN is a LeNet~\cite{LeNet} variant, containing two convolutional layers and two fully connected layers (more details in Figure \ref{fig:BaseCNNs}). A max-pooling layer is set next to each convolutional layer. In addition, we select the widely used activation function, rectified linear unit (ReLU). As we target to solve a multi-class classification problem, we employed softmax and cross-entropy to be respectively, the last output activation and the loss function. For regularization, we add both ridge regression (L2-norm of weights) and lasso regression (L1-norm of weights) as penalty terms to the loss function. Regarding the optimization, we use stochastic gradient descent (SGD), which remains the standard optimizer for training neural networks. It was trained on MNIST handwritten digits benchmark dataset~\cite{mnist}, which includes $60,000$ labeled gray-scale images of size $28\times28$ in $10$ labels. The dataset was split into a training set containing $50,000$ images and a test set containing $10,000$ images. The ShallowCNN achieved $99.35\%$ accuracy rate on the test set.

\textbf{DeepCNN.} The DeepCNN is a VGG~\cite{VGG} variant, containing stacking three blocks of two convolutional layers followed by a max pooling layer and two fully-connected layers (more details in Figure \ref{fig:BaseCNNs}). We also use ReLUs, softmax, and cross-entropy for activation, output and loss functions. However, we employ advanced regularization techniques to enable the training of this deep NN. We put a batchnorm layer next to each layer that stabilizes and accelerates the optimization process. In addition, each block or dense layer ends with a dropout layer using an increasing dropout rate (going from $20\%$ to $50\%$) in order to offset the learning acceleration. Our optimization method is based on Adam, an advanced variant of gradient-based that can automatically adapt its learning rate within each optimized parameter. The DeepCNN was trained on CIFAR10~\cite{cifar10}, which consists of $60,000$ labeled color images of size $32\times32$ in 10 different classes (e.g., airplane, bird, and truck). The dataset was split into a training set containing $50,000$ images and a test set containing $10,000$ images. The DeepCNN achieved $89.15\%$ accuracy rate on the test set.

\subsubsection{Real TensorFlow Buggy Programs}
Zhang et al.~\cite{DL_bugs_1} reproduce several defective TF programs extracted from SO posts and GH projects. Among the categories of studied DL bugs, we found the Incorrect Model Parameter or Structure (IPS), which includes inappropriate modeling choices and algorithm configurations, degrading the quality of the training. Indeed, the major symptom of IPS bugs is low effectiveness, i.e., the performance of the on-training model does not improve as expected. After filtering out redundant TF programs and other incomplete models from their SO dataset, we ended up with $10$ unique and full IPS buggy TF programs including data, model and training algorithm implementation. Table~\ref{SO_Programs_Fixes} presents the selected programs alongside the recommended fixes extracted from their related SO post discussions. Moreover, we clone $10$ buggy versions of GH projects that correspond to bug-fixing commits in relation with IPS bugs. Table~\ref{GH_Programs_Fixes} shows the versions of GH projects with the implemented fixes that have been identified from the bug-fixing commit. Indeed, we emphasize that the fixes with the buggy examples are added to mention the fault identified by the SO users or the GH project contributors, but there is no guarantee that it is the only bug or inefficiency in the project at that version. Technically, we follow the ‘how to’, libray’s version, and official datasets, as described in the replication package of the empirical study~\cite{DL_bugs_1} on Tensorflow Bugs, to prepare our set of buggy DL training programs.
\begin{table}
\hspace*{-1cm}
\caption{SO buggy TF-based training programs}
\label{SO_Programs_Fixes}
\begin{tabular}{|c|c|c}
\hline
\textbf{Program}&\textbf{Recommended Fixes}\\
\hline
IPS-1&switch to a numerical stable loss\\
\hline
IPS-4&add mean to the loss\\
\hline
IPS-5&change gradient descent by Adam\\
\hline
IPS-7&add an output activation and a numerical stable loss\\
\hline
IPS-11&remove the useless ReLU on the logits\\
\hline
IPS-12&fix a typo in the accuracy function\\
\hline
IPS-13&set lower learning rate($\eta$) and norm penalty($\lambda$)\\
\hline
IPS-14&set a low learning rate($\eta$)\\
\hline
IPS-15&set a low learning rate($\eta$)\\
\hline
IPS-17&set a low learning rate($\eta$)\\
\hline
\end{tabular}
\end{table}
\begin{table}
\hspace*{-1cm}
\caption{Github buggy TF-based training programs}
\label{GH_Programs_Fixes}
\begin{tabular}{|c|c|c|}
\hline
\textbf{Program}&\textbf{Recommended Fixes}\\
\hline
DLT\_0edb182&remove redundant softmax layers\\
\hline
DLT\_20d1b59&add a softmax layer\\
\hline
DLT\_437c9c2&improve the loss reduction\\
\hline
DLT\_726b371&add $\epsilon$ for a numerical stable loss\\
\hline
DLT\_ded6612&improve the parameters intialization\\
\hline
FCN-CTSCAN\_b170a9b&fix a mistake in the loss function\\
\hline
TFE\_333&set an adequate weight initialization\\
\hline
TFE\_368&set a low learning rate($\eta$)\\
\hline
TFE\_742675d&improve the loss reduction\\
\hline
TFE\_bc09f95&improve the loss reduction\\
\hline
\end{tabular}
\end{table}

\subsubsection{Rule-based Debugger Baseline}
As a baseline for automated training issue detection, we use SageMaker Debugger(\textit{SMD})~\cite{smd}, which is a fully managed debugging and monitoring service within the Amazon SageMaker platform for scalable machine learning in the cloud. \textit{SMD} represents a framework-agnostic system to collect, query, and analyze data from ML model training and to automatically capture issues using a set of built-in rules. Indeed, SageMaker automatically creates the training instance, pulls the training image from the Container Registry, and downloads data and training scripts into the container. Once the training is launched, \textit{SMD} retrieves asynchronously the model data at specific intervals and uploads them to S3 bucket. Then, \textit{SMD} runs the activated built-in or user custom rules in independant processing jobs on separate containers in a way that they do not interfere with the training job. Finally, users can set up alarms within Amazon Cloudwatch, which is a real time monitoring and observability service, to indicate when a rule is triggered. Table~\ref{SMD_Rules} summarizes the \textit{SMD}’s built-in rules that are applicable for feedforward neural networks.
\begin{table}
\hspace*{-1cm}
\caption{SMD Built-in Rules Applicable to FNNs}
\label{SMD_Rules}
\begin{tabular}{|c|c|c|c|}
\hline
\textbf{Component}&\textbf{Rules}&\textbf{\#id}\\
\hline
\multirow{2}{*}{Data}&Class Imbalance&$R_0$\\
\cline{2-3}
&Non-normalized image input&$R_1$\\
\hline
\multirow{5}{*}{Loss}&Not Decreasing&$R_2$\\
\cline{2-3}
&Unchanged&$R_3$\\
\cline{2-3}
&Overfitting&$R_4$\\
\cline{2-3}
&Underfitting&$R_5$\\
\cline{2-3}
&Overtraining &$R_6$\\
\hline
\multirow{4}{*}{Weights}&Poor Initialization &$R_7$\\
\cline{2-3}
&Abnormal Update Ratio &$R_8$\\
\cline{2-3}
&All Zeros&$R_{9}$\\
\cline{2-3}
&Abnormal Variance&$R_{10}$\\
\cline{2-3}
&Exploding&$R_{11}$\\
\hline
\multirow{2}{*}{Activations}&Neurons Saturation&$R_{12}$\\
\cline{2-3}
&Dying Neurons&$R_{13}$\\
\hline
\multirow{2}{*}{Gradients}&Vanishing&$R_{14}$\\
\cline{2-3}
&Exploding&$R_{15}$\\
\hline
\end{tabular}
\end{table}
Although \textit{SMD} and \name{} target a common subset of training issues, the logic of debuggers’ rules are quite different, as \textit{SMD} relies exclusively on the model data collected offline during the user-configured training session. However, \name{} is an online debugger framework that takes control of the ML training program and the datasets to orchestrate the running of a sequential multi-phase debugging workflow, including preliminary independent checks, a monitored single-batch training, and comparisons of multiple trained models. Indeed, the \name{}’s debugging workflow involves verification rules that validate necessary properties of the training program and assess the status of heuristic-based detectors for common DL model training issues.
\subsection{Case Study Results}
In the following, we present the results of debugging sessions using \name{} on both synthetic and real-world buggy programs. First, we evaluate the capacity of \name{} in detecting faults injected in the base NNs, whether the fault is a coding bug or a misconfiguration of the system. Then, we assess the effectiveness of \name{} in debugging real-world buggy TF programs that may contain multiple hidden bugs.
\subsubsection{\textbf{Debugging DNN Software System that contains Coding Bugs.}}
\label{coding_bugs_section}
Table \ref{buggy_code_checks} shows the results obtained on the synthetic buggy training programs. For each injected bug, we report all the checks fired by \name{}, as well as the performance of the trained model on test data and the \textit{SMD}'s rules that detected issues.

\textbf{Accuracy of \name{} on coding bug detection. }Given the fault injected in the synthetic example, we put in bold the fired check(s) that are considered to be conceptually connected to the underlying issue. Indeed, these bolded checks could guide the user towards recognizing the occurred issue and fixing the buggy training program. Moreover, we study the other fired checks for a full assessment as we keep running the \name{}, but we do not need to wait for all the fired checks to spot and fix alerted DL bugs. Thereby, we also provide a user-defined boolean setting (\textit{failed\_on=True/False}) that would enable an exception to be raised whenever \name{} encounters any of these verifications. In practice, this helps save useless running time and computational resources. Thus, we calculate the number of True Positive (TP) checks, False Positive (FP) checks and False Negative (FN) checks. True positives are represented as \textit{a}+\textit{b}, where \textit{a} and \textit{b} are the number of checks, respectively, defined to catch precisely the injected bug and to identify generic training difficulties that can be correlated to various DNN issues. Although the proposed generic checks, counted in $b$, spot fine-grained inefficient training traits, we do not consider the bug was detected in case of $a=0$. Therefore, \name{} has a precision of $90\%$ and a recall of $96.4\%$ in detecting the synthetized DL coding bugs.\\
\textbf{Comparison with Baseline.} As can be seen, missing code statements can result in the dysfunction of a component in the DNN training program. In such cases, our verification routines find some violations of properties associated with the expected output and normal behavior of the buggy component. First, missing or redundant input normalization and over-scaled outputs are detected by \name{} using the following prechecks on the loaded data: unscaled inputs and unscaled outputs. Next, unapplying softmax over outputs and unshuffling the labels would trigger the following verification routines, respectively, invalid predictions (i.e., do not respect the probability laws) and corrupted labels (i.e., labels do not match with the features). Another studied data preprocessing bug is the inappropriate data transformers that induce a shift between original and augmented datasets. \name{} was able to detect such data shifts from the perspective of the trained model behavior by comparing the activation patterns of the neural networks trained on augmented/original datasets. In relation to the data, \textit{SMD} supports only a rule ($R_1$) for unnormalized images detection. For the missing activation, \textit{SMD} triggers vanishing gradients ($R_{14}$), which is a quite generic issue for DNN training.\\

Concerning wrong coding statements, \name{} is still accurate for \textit{mistaken axis in tensor-based operations}, thanks to the gradient-based dependency verification that point out any overlapping between instances (i.e., rows) caused by an incorrect calculation. However, the dependency verification fails on the reproduced wrong broadcasting of MSE because the error does not induce an overlap between instances. In most of these wrong coding statements, \textit{SMD} triggers rules indicating the difficulty of training such as vanishing gradients ($R_{14}$) and non-decreasing loss ($R_{14}$). Both \name{} and \textit{SMD} succeed in revealing coding faults that change the training algorithm's behavior, like flipped signs in the gradient calculation that turns the training process into a loss maximization. Nevertheless, \name{} is capable of detecting a wrong cross-entropy function with switched mean and sum operations based on the initial loss value which is larger than expected, and consequently, indicate the presence of a wrong reduction function for the losses over data inputs. \textit{SMD} reports vanishing gradients ($R_{14}$) for only the ShallowNN that could not train the model with a wrong loss reduction (i.e., the accuracy of the trained model is equivalent to random guess).

Regarding the misuse of API functions, we reproduced the situation of redundant softmax in-and out-the cross-entropy loss for both studied Base CNNs. Despite the fact that \name{} does not contain a specialized check to pinpoint accurately the issue, it could successfully detect its presence in the training program for both buggy examples, contrary to \textit{SMD} that remains silent. In fact, \name{} generated an error message reporting learning issues in the last dense layer of both ShallowCNN and DeepCNN containing the softmax redundancy within the loss; more specifically, a slowness learning issue due to either low updates' ratio. These slowness learning issues, when spotted at the last layer (i.e., softmax activation), strongly indicate a waste of information and an obstruction of the back-propagation of the error through the two consecutive softmax activations. Hence, the relation between the fired checks and the occurred bug is not straightforward (i.e., we do not count it among the main fired checks (i.e., the number \textit{a} in True Positives). Nevertheless, it is worth mentioning that \name{} generated a warning for the user about the slow weight update encountered exclusively in the last dense layer, which can lead him to review this ending part of the DNN design which contains the issue (i.e., redundant/useless last activation function). Besides, \name{} reported symptoms in relation to difficulties faced in the loss minimization: slow decreasing loss for both subjects and additionally highly-fluctuating loss for DeepCNN. 

Overall, \name{} reported misleading checks which are considered as false alarms. First, it mistakenly reports non-representative loss (i.e., loss measure is not correlated with performance metric) in some DeepCNN cases. Indeed, the obtained highly-fluctuating loss caused by the activation redundancy significantly reduced the magnitude of the correlation between the resulting noisy loss and the classification accuracy, and triggered the verification routine related to non-representative loss. This highlights the difference between shallow and deep CNNs and how the loss landscape is more complex for deeper NNs and can be substantially affected by these relatively minor changes in the math calculations involved in the DNN mapping function. Second, the traditional regularization check consistently triggers an alert of overwhelming regularization gradient cost over the data gradient loss, but our inspection leads us to the fact that actual issue was the vanishing gradient problem reaching exactly $0$; we consider as misleading because the check was designed to capture over-regularization cost and it can mislead the user towards inspecting unnecessarily the regularization. On the other hand, we unbolded the \textit{SMD}'s rules that have been fired even for the clean training program. Indeed, these rules, $R_8$ and $R_{10}$, were fired during the early training iterations on the DeepNN that includes advanced regularizers to smooth the loss landscape and be able to train several hidden layers. This can be explained by the high sensitivity of these rules that alert quickly about abnormal weights' update ratio and variance starting from the warm-up period required by these regularizers to stabilize the training. Besides, \textit{SMD} reports a false alarm of poor weight initialization ($R_7$) on a wrong cross-entropy calculation. In fact, \textit{SMD} relies on the application of rules' functions on the periodically-saved summary statistics. Hence, $R_7$ checks the variance of activation inputs across layers, the distribution of gradients, and the loss convergence for the initial steps to determine if a neural network has been poorly initialized. These training inefficiencies can be the result of many issues in the training program, which increases the risk of misleading alerts.

\begin{table}
\hspace*{-1cm}
\caption{Results of debugging coding bugs in DNN-based software systems}
\label{buggy_code_checks} 
\resizebox{.95\textwidth}{!}{
\begin{tabular}{|c|c|c|c|c|c|c|c|c|}
\hline
\textbf{Faults} & \textbf{Base NN} & \textbf{Perf.} & \textbf{SMD Rule(s)} & \textbf{Fired Check(s)} & \textbf{TP} & \textbf{FP}& \textbf{FN}\\
\hline 
\multirow{3}{*}{\textbf{missing input normalization}}
&Regr&$24.20$&-&\shortstack{\textbf{Uns-Inps}\footnotemark, PI-Loss\footnotemark \\ Un-Fit-Batch\footnotemark, Uns-Act-HS\footnotemark}&1+3&0&0\\
\cline{2-8}
&Shallow&$11.35\%$&\bm{$R_1$},\bm{$R_8$},\bm{$R_{14}$}&\shortstack{\textbf{Uns-Inps}, PI-Loss, Un-Fit-Batch \\ Div-Loss\footnotemark, Div-W\footnotemark, Div-B\footnotemark, Div-Grad\footnotemark}&1+6&0&0\\
\cline{2-8}  
&Deep&$85\%$&\bm{$R_1$},$R_8$,$R_{10}$&\shortstack{\textbf{Uns-Inps}, PI-Loss, Uns-Act-HS, \cancel{NR-Loss\footnotemark}}&1+2&1&0\\
\hline
\textbf{over-scaled outputs}&Regr&$20.14$&\bm{$R_2$}, \bm{$R_{12}$}&\shortstack{\textbf{Uns-Outs}\footnotemark, SD-Loss\footnotemark, Dead-ReLU\footnotemark, Uns-Act-HS}&1+3&0&0\\
\hline 
\multirow{3}{*}{\textbf{redundant input normalization}}
&Regr&$2.86$&-&\shortstack{\textbf{Uns-Inps}, SD-Loss, Uns-Act-LS\footnotemark, Un-Fit-Batch}&1+3&0&0\\
\cline{2-8}
&Shallow&$33.75\%$&\bm{$R_8$}, \bm{$R_{14}$}&\shortstack{\textbf{Uns-Inps}, SD-Loss, W-Up-Slow\footnotemark, Uns-Act-LS}&1+3&0&0\\
\cline{2-8}
&Deep&$77.5\%$&-,$R_8$,$R_{10}$&\textbf{Uns-Inps}, Uns-Act-LS&1+1&0&0\\
\hline
\multirow{3}{*}{\textbf{gradients with flipped sign}}
&Regr&$1.72e7$&-&Un-Fit-Batch, \textbf{Div-Loss}, Uns-Act-HS&1+2&0&0\\
\cline{2-8}
&Shallow&$9.8\%$&\bm{$R_{11}$},\bm{$R_{14}$}&\shortstack{Un-Fit-Batch, \textbf{Div-Loss}, Div-W,\\ Div-B, Uns-Act-HS, Van-Grad\footnotemark}&1+5&0&0\\
\cline{2-8}    
&Deep&$10\%$&\bm{$R_{11}$},\bm{$R_{14}$}&Un-Fit-Batch, \textbf{Div-Loss}, Uns-Act-HS, NR-Loss\footnotemark&1+2&0&0\\
\hline
\multirow{2}{*}{\textbf{missing softmax activation}}&Shallow&$9.8\%$&\bm{$R_{14}$}&\shortstack{PI-Loss, \textbf{Inv-Outs}\footnotemark, SD-Loss W-Up-Slow, \\ Van-Grad, Un-Fit-Batch, \cancel{Over-Reg-Loss}\footnotemark}&1+5&1&0\\
\cline{2-8}
&Deep&$11.48\%$&\bm{$R_{14}$},$R_8$,$R_{10}$&\shortstack{PI-Loss, \textbf{Inv-Outs}, Van-Grad}&1+2&0&0\\ 
\hline
\multirow{2}{*}{\textbf{softmax out-and in-the loss}}&Shallow&$99.29\%$&-&SD-Loss, W-Up-Slow(Dense)&0+2&0&1\\
\cline{2-8}
&Deep&$83.24\%$&-,$R_8$,$R_{10}$&SD-Loss, HF-Loss\footnotemark, W-Up-Slow(Dense), \cancel{NR-Loss}\footnotemark&0+2&1&1\\
\hline 
\multirow{2}{*}{\textbf{softmax over wrong axis}}&Shallow&$99.45\%$&\bm{$R_{14}$}&\shortstack{PI-Loss, \textbf{Inv-Outs}, \textbf{Inv-Out-Dep}\footnotemark, Inv-Loss-Dep\footnotemark}&2+2&0&0\\
\cline{2-8}
&Deep&$85.86\%$&\bm{$R_{14}$},$R_8$,$R_{10}$&\shortstack{PI-Loss, \textbf{Inv-Outs}, \textbf{Inv-Out-Dep}, Inv-Loss-Dep}&2+2&0&0\\
\hline 
\multirow{2}{*}{\textbf{CE over wrong axis}}&
Shallow&$8.92\%$&\bm{$R_{2}$},\bm{$R_{7}$}&\textbf{PI-Loss}, \textbf{Inv-Loss-Dep}&2+0&0&0\\
\cline{2-8}
&Deep&$86.79\%$&-,$R_8$,$R_{10}$&\textbf{PI-Loss}, \textbf{Inv-Loss-Dep}&2+0&0&0\\
\hline
\textbf{MSE with wrong broadcasting}&Regr&$7.02$&\bm{$R_{2}$}&Un-Fit-Batch, SD-Loss, Van-Grad&0+3&0&1\\ 
\hline
\multirow{2}{*}{\textbf{inverted CE's mean and sum}}&Shallow&$11.34\%$&\bm{$R_{14}$}&\textbf{PI-Loss}&1+0&0&0\\
\cline{2-8}
&Deep&$87.08\%$&-,$R_8$,$R_{10}$&\textbf{PI-Loss}&1+0&0&0\\
\hline
\multirow{2}{*}{\textbf{shuffle only the features}}
&Regr&$7.27$&-&\textbf{Corrupted Labels}&1+0&0&0\\
\cline{2-8}
&Shallow&$11.35\%$&-&\textbf{Corrupted Labels}&1+0&0&0\\
\cline{2-8}
&Deep&$10.09\%$&-,$R_8$,$R_{10}$&\textbf{Corrupted Labels}&1+0&0&0\\
\hline
\multirow{2}{*}{\textbf{invalid data transformation}}
&Shallow&$99.24\%$&-&\textbf{Shifted-Augmented-Data}&1+0&0&0\\
\cline{2-8}
&Deep&$86.28\%$&-,$R_8$,$R_{10}$&\textbf{Shifted-Augmented-Data}&1+0&0&0\\
\hline
\end{tabular}
}
%\vspace{-15pt}
\end{table}
\footnotetext[1]{Unscaled Inputs}
\footnotetext[2]{Poor Initial Loss}
\footnotetext[3]{Unable to fit Single Batch}
\footnotetext[4]{Unstable Activation with High STD}
\footnotetext[5]{Diverging Loss}
\footnotetext[6]{Diverging Weights}
\footnotetext[7]{Diverging Biases}
\footnotetext[8]{Diverging Gradient}
\footnotetext[9]{Non-Representative Loss}
\footnotetext[10]{Unscaled Outputs}
\footnotetext[11]{Slow-Decreasing Loss}
\footnotetext[12]{Dead ReLU}
\footnotetext[13]{Unstable Activation with Low STD}
\footnotetext[14]{Weight Update Slowly}
\footnotetext[15]{Vanishing Gradients}
\footnotetext[16]{Invalid Outputs}
\footnotetext[17]{Overwhelming Regularization Loss}
\footnotetext[18]{High-Fluctuating Loss}
\footnotetext[19]{Invalid Output Dependency}
\footnotetext[20]{Invalid Loss Dependency}
\subsubsection{\textbf{Debugging Misconfigured DNN Software System.}}
Table \ref{misconfigured_DNN_checks} presents the debugging results of \name{} on misconfigured synthetic training programs, following a similar structure as Table \ref{buggy_code_checks}. 

\textbf{Accuracy of \name{} on misconfigurations detection. }We calculate true positive, fault positive, and fault negative rates alongside the list of fired checks. Overall, \name{} achieved $77\%$ precision and $83.3\%$ recall, when detecting misconfigurations in the studied DNN training programs.

\textbf{Comparison with Baseline. }As can be seen, inappropriate initial weights, i.e., constant and inept randomness, correctly trigger the following \name{}'s checks, \emph{unbreaking symmetry} and \emph{poor weights initialization}. \textit{SMD} uses multiple indirect criteria to recognize a bad initialization through activations' variances, gradients' distributions, and the loss curve. It spots the unbreaking symmetry issue, but it was less effective in detecting the dummy random weights with inappropriate variance.

Similarly, \name{} was able to detect missing stabilization components, including missing the whole batchnorm layers and missing the update of their global statistics, through respectively, high unstable activations and unstable transfer from train to inference mode. Also, weak regularization triggers the checking rule of zero loss, which implies that the optimizer likely overfits the given training batches and strong regularization can lead to overwhelming weight norm penalty (in case of l2-norm) or unstable transfer from train to test mode (dropout). Nonetheless, \textit{SMD} was only capable of detecting the negative effects of strong l2-norm regularization on the weights' variances and dying ReLUs.

Moreover, architecture- or problem-dependant issues like using ineffective loss function are challenging to detect for developers (e.g., we refer to SO posts \#36515202, \#49322197, \#56013688, and \#62592858); their identification depends heavily on the knowledge of the developer on Deep Learning and his experience in implementing DNN programs. Indeed, these issues can have severe effects on the convergence and stability of the DNN training process, especially on relatively high capacity DNNs and complex statistical learning problems. When it comes to regression problems, the use of cross-entropy(CE) instead of mean squared error(MSE) hinders the learning; so both approaches were able to detect that. In case of classification problems, the use of MSE instead of CE had less effects on the training performance, and consequently, harder to automatically detect. Thus, \name{} alerted only inefficient training traits including vanishing gradients and slow updating parameters for the deepCNN. Inversely, \textit{SMD} alerted abnormal training curve, including non decreasing and unchanged loss for the ShallowCNN, but the resulting test performance of the model is high enough to just conclude that this steadiness was due to the optimization convergence.

With less implicit issue-verification connections, poor choices regarding the optimization routines, including inadequate magnitude of the learning rate or the epsilon of null divider prevention, influence negatively the speed of parameters learning. Only \name{} successfully detected the substantial difference in the magnitude of parameters' updates. It detected slow updating parameters in case of low learning rate and fast updating parameters in case of high learning rate and low Adam epsilon. Indeed, \name{}'s unstable weight update detection strategy reposes on DL researchers' recommendations and it is sensitive enough to report precisely the low or high magnitudes of weight updates that can guide DL users to adjust the optimizer's configuration touching the extent of updates. Nevertheless, both compared debugging methods were able to the divergence of ShallowCNN caused by the high learning rate that manifests through a training stagnation. Indeed, the high learning rate quickly induces a large update step towards the gradient direction, which takes the weights to nonoptimal regions, and consequently, provokes dead ReLUs and a highly-fluctuating loss optimization without convergence (unable to overfit the batch). Since the inadequate optimizer's jump likely happens at the very first iterations, there would not be fired checks in relation with the speed of learning; so the users cannot identify the root cause easily based on all this checking report.

As above-discussed for DL coding bugs~\ref{coding_bugs_section}, \textit{SMD} keeps consistently reporting both of $R_8$ and $R_{10}$ rules, as well as, it reports a false alarm of poor weight initialization ($R_7$) on a inappropriate loss selection (CE instead of MSE). On the other hand, \name{} reports false positives of non-representative loss, overwhelming regularization loss paired with, respectively, high fluctuating loss and vanishing gradients. This reinforces the fact that these false positives are caused by the non-consideration of potential dependencies between DL training issues.

In the future, we plan to analyze in-depth the sequence of fired checks for the different DL bug types and extend \name{} to consider recurrent patterns of negative verifications in order to not only reduce the number of false positives that we discover during the assessment, but also avoid overwhelming the users by several correlated training issues. Indeed, we observe in both results (i.e., Tables \ref{misconfigured_DNN_checks} and \ref{buggy_code_checks}) repetitive co-occurrences of multiple checks. For instance, we see that various errors make the weight updates unstable, so the weights can potentially have more and more negative values (i.e., over-negative tendency). In this case, the layers' weighted sum would produce mainly negative quantities, which will turn into zero activations by the ReLUs (i.e., dead). Then, the back-propagated gradients, as described in Equation~\ref{eq_loss_W_idx_free}, starts vanishing quickly and, as a result, the weights' updates will tend to have too low magnitude and could cause the DNN to freeze (i.e., triggering likely a slow- or non-decreasing loss).

\begin{table}
\hspace*{-1cm}
\caption{Results of debugging misconfigurations in DNN-based software systems}
\label{misconfigured_DNN_checks}
\resizebox{.95\textwidth}{!}{
\begin{tabular}{|c|c|c|c|c|c|c|c|}
\hline
\textbf{Faults} & \textbf{Base NN}& \textbf{Perf.} & \textbf{SMD Rule(s)} & \textbf{Fired Check(s)}&\textbf{TP}&\textbf{FP}&\textbf{FN}\\
\hline
\multirow{2}{*}{\textbf{constant weights}}
&Regr&$2.53$&\bm{$R_7$},\bm{$R_{10}$}&\shortstack{\textbf{Un-Sym-W}\footnotemark, SD-Loss, Uns-Act-LS\\ W-Up-Slow, Un-Fit-Batch}&1+4&0&0\\
\cline{2-8}
&Shallow&$11.35\%$&\bm{$R_7$},\bm{$R_{10}$},\bm{$R_{13}$},\bm{$R_{14}$}&\shortstack{\textbf{Un-Sym-W}, SD-Loss, Neg-W\footnotemark, \\ Over-Reg-Loss, Uns-Act-LS, Dead-ReLU,\\ W-Up-Slow, Van-Grad, Un-Fit-Batch}&1+8&0&0\\
\cline{2-8}
&Deep&$75.92\%$&\bm{$R_7$},\bm{$R_{14}$},$R_8$,$R_{10}$&\shortstack{\textbf{Un-Sym-W}, SD-Loss, Uns-Act-HS, W-Up-Slow}&1+3&0&0\\
\hline
\multirow{2}{*}{\textbf{dummy random weights}}
&Regr&$2.17$&-&\textbf{PI-W}\footnotemark, Uns-Act-LS&1+1&0&0\\
\cline{2-8}
&Shallow&$99.18\%$&\bm{$R_2$},\bm{$R_{7}$}&\shortstack{\textbf{PI-W}, PI-Loss, SD-Loss \\ Dead-ReLU, Uns-Act-LS}&1+4&0&0\\
\cline{2-8}
&Deep&$71.89\%$&-,$R_8$,$R_{10}$&\shortstack{\textbf{PI-W}, Uns-Act-HS, \cancel{NR-Loss}}&1+1&1&0\\
\hline
\multirow{2}{*}{\textbf{use of MSE instead of CE}}
&Shallow&$99.17\%$&\bm{$R_2$},\bm{$R_3$}&-&0&0&1\\
\cline{2-8}
&Deep&$69.52\%$&-,$R_8$,$R_{10}$&SD-Loss, HF-Loss, W-Up-Slow&0+3&0&1\\
\hline
\textbf{use of CE instead of MSE}&Regr&$49.46$&\bm{$R_3$},\bm{$R_7$},\bm{$R_14$}&Un-Fit-Batch, Van-Grad(dense), \cancel{Over-Reg-Loss}&0+3&1&1\\
\hline
\multirow{2}{*}{\textbf{low learning rate}}
&Regr&$5.48$&-&Un-Fit-Batch, SD-Loss, \textbf{W-Up-Slow}&1+2&0&0\\
\cline{2-8}
&Shallow&$98.96\%$&-&SD-Loss, \textbf{W-Up-Slow}&1+1&0&0\\
\cline{2-8}
&Deep&$53.73\%$&-,$R_8$,$R_{10}$&SD-Loss, \textbf{W-Up-Slow}, \cancel{NR-Loss}&1+1&1&0\\
\hline
\multirow{2}{*}{\textbf{high learning rate}}
&Regr&$2.55$&-&\textbf{W-Up-Fast}\footnotemark, SD-Loss&1+0&0&0\\
\cline{2-8}
&Shallow&$11.34\%$&\bm{$R_13$},\bm{$R_14$}&\shortstack{Un-Fit-Batch, HF-Loss, Dead-ReLU, \\ Uns-Act-LS, \cancel{NR-Loss}}&0+4&1&1\\
\cline{2-8}
&Deep&$86.29\%$&-,$R_8$,$R_{10}$&Uns-Act-HS, \textbf{W-Up-Fast}&1+1&0&0\\
\hline
\textbf{Adam epsilon $\epsilon<10^{-8}$
}&Deep&$86.75\%$&-,$R_8$,$R_{10}$&Uns-Act-HS, \textbf{W-Up-Fast}&1+1&0&0\\
\hline
\textbf{missing batch-norms}&Deep&$80.79\%$&-&\shortstack{SD-Loss, \textbf{Uns-Act-LS}, W-Up-Slow, \cancel{NR-Loss}}&1+2&1&0\\
\hline
\textbf{no-update of batch-norm globals}&Deep&$84.35\%$&-,$R_8$,$R_{10}$&\shortstack{\textbf{Uns-Mode-Tr}}&1+0&0&0\\
\hline
\textbf{low $\lambda$ for norm penalties}&Regr&$2.39$&-&\textbf{Zero-Loss}&1+0&0&0\\
\hline
\textbf{low $\lambda$ for norm penalties}&Shallow&$99.27\%$&-&\textbf{Zero-Loss}&1+0&0&0\\
\hline
\textbf{high $\lambda$ for norm penalties}&Regr&$7.05$&\bm{$R_10$},\bm{$R_13$}&\textbf{Over-Reg-Loss}, Uns-Act-LS, Un-Fit-Batch&1+2&0&0\\
\hline
\textbf{high $\lambda$ for norm penalties}&Shallow&$64.88\%$&\bm{$R_10$},\bm{$R_13$}&\textbf{Over-Reg-Loss}&1+0&0&0\\
\hline
\textbf{high $keep\_p$ for dropouts}&Deep&$73.15\%$&-&\textbf{Zero-Loss}&1+0&0&0\\
\hline
\textbf{low $keep\_p$ for dropouts}&Deep&$78.74\%$&-,$R_8$,$R_{10}$&HF-Loss, \textbf{Uns-Mode-Tr},\cancel{NR-Loss}&1+1&1&0\\
\hline
\multirow{2}{*}{\textbf{unbalanced dataset}}
&Shallow&$99.24\%$&bm{$R_0$}&\textbf{Unbalanced Labels}&1+0&0&0\\
\cline{2-8}
&Deep&$86.28\%$&bm{$R_0$},$R_8$,$R_{10}$&\textbf{Unbalanced Labels}&1+0&0&0\\
\hline 
\end{tabular}
}
\end{table}
\footnotetext[1]{Unbreaking Symmetry Weight}
\footnotetext[2]{Poor Initial Weight}
\footnotetext[3]{Weight Update Quickly}
\footnotetext[4]{Unstable Transfer Mode}
\subsubsection{\textbf{Debugging Runtime Evaluation}}
\name{} enables the debugging of DNN training programs through performing a stack of checks prior, during, and after the program execution. Executing these checks can be expensive.\\
In this section, we assess the execution cost of the checks on a DNN training program. The execution was done using a single machine having a CPU Intel i7-8750H 6-cores and a GPU NVIDIA GeForce RTX 2080. The evaluation was done using the above-described base neural networks. To provide insights on the execution cost of \name{} during a debugging session, Table~\ref{overall_time_table} reports the average time spent using \name{} on each phase. As can be seen, the pre- and post-training phases, (1) pre-training conditions check and (3) post-fitting conditions check, varies according to the dimensionality of the input data and the complexity of NN as they include normalization tests, tensor-based operations dependency, and even regular training of DNN for several epochs (we set up $50$ by default) to compare metrics over epochs and make some activation patterns comparison. Next, the phase of proper fitting on a single batch requires only few iterations (i.e., it remains a setting option for running the test and by default, a maximum iterations equals to $200$), but with short periodicity of verification routines (i.e., it is also a setting option and by default, we fixed a period equals to $10$ iterations). These default setting options are derived from our experimentation, however, the configuration choices should take into consideration the complexity of the DNN under test and the default settings enable a sufficient amount of monitored iterations to detect the issues given the complexity of the studied neural networks. For higher complexity NNs, an increase of these parameters' values may enhance the issue detection capability of \name{} as it will make more intensive verification during the testing iterations. Given the average execution time of a regular training iteration and a monitored training iteration, we find that the verification routines running in-between the training iterations increased the training iteration runtime by averagely $10\times$. In contrast, \textit{SMD} hooks multiple internal data recorders to the full training session in order to fetch and save periodically tensors including activations, weights, and gradients. \textit{SMD} incorporates several optimizations to improve I/O performance and sets up relatively long, by default, save intervals (i.e., around $500$ steps). In fact, our experimentation on AWS instance, \textbf{ml.p2.xlarge}, which contains 1-GPU and 4 virtual processors, yields, averagely, an overhead of no more than $13\%$ for a training iteration monitored by $4$ rules. Then, \textit{SMD} verifies the rules by offloading data inspection, shared into separate containers in a way that users can run an arbitrary number of rules without impacting the training process itself. Nevertheless, this multi-job processing and I/O data offloading adds an overhead, even for simple DNN programs. During our debugging sessions on \textbf{ml.p2.xlarge} with $4$ activated rules, we wait for $3-5$ minutes to have the final rules check reports.

\begin{table}[H]
\caption{Execution cost of \name{} during different debugging phases (average time in seconds)}
\label{overall_time_table}
\begin{tabular}{|c|c|c|c|c|} 
\hline
 &  \textbf{Pre-train Check} & \textbf{Fitting Check} & \textbf{Post-Fit Check}\\
\hline
\textbf{ShallowCNN} & 16.63 & 20.61 & 751.70\\
\hline
\textbf{DeepCNN} &  20.81 & 82.74 & 1641.59\\
\hline
\textbf{RegrCNN} &  6.75 & 0.996 & 5.70\\
\hline
\end{tabular}
\end{table}
\subsubsection{\textbf{Assessment of \name{} on Real Buggy DNN Software}}
Table~\ref{SO_Programs_checks} and Table~\ref{GH_Programs_checks} show the debugging results of both \name{} and \textit{SMD} on real-world buggy TF programs extracted, respectively, from Stackoverflow and GitHub. Indeed, we add all the turned-on verification checks/rules on each tested DNN program, but we highlight the ones that are considered to be related to the actual fixed bug. For \textit{SMD}, we rely on its official built-in rules documentation~\cite{smdRulesDocs} about their logic and targeted issues in order to decide if the fired rules have  This allows us to compute the success rate of \name{} in detecting the bugs that have been fixed by the SO users or the GH project maintainers. Indeed, \name{} succeeds in $70\%$ of the SO buggy code and $80\%$ of the buggy versions of GH projects. However, \textit{SMD} succeeds in $60\%$ of both SO and GH buggy examples. \textit{SMD} generally alerts the user with high-level indicators of abnormal/suspicious on-training neural network state, but \name{} often reports broken properties that are connected to a narrower scope of DL faults and help users identify the main root cause. For instance, lines of code~\ref{IPS7} shows that the NN's output layer, \textit{y\_}, has no activation function, which causes an incorrect calculation of cross entropy using logits instead of probabilities. Moreover, the cross entropy formula involves a risky use of log on possibly zero values.
\begin{lstlisting}[language=Python, caption=Lines 18;21 of IPS-7,label=IPS7]
y_ = tf.matmul(h1,W_out)
cross_entropy = tf.reduce_sum(-y*tf.log(y_)-(1-y)*tf.log(1-y_),1)
\end{lstlisting}
\name{} reports invalid output layer because the verification routine on the last layer requires yielding probabilities when NN solves a classification problem. It also alerts about the diverging loss caused by the NaNs of \textit{tf.log(0)}. Although abnormal variance of weights reported by \textit{SMD} reflects an unstable learning process, it cannot be connected to the missing activation or the unstable loss issues.
\begin{lstlisting}[language=Python, caption=Lines 37-38;45-46 of IPS-12,label=IPS12]
Yhat = tf.matmul(l3, W5) + b5
Ypred = tf.nn.sigmoid(Yhat)
# ...
correct_prediction = tf.equal(tf.greater(Y, 0.5), tf.greater(Yhat, 0.5))
accuracy = tf.reduce_mean(tf.cast(correct_prediction, tf.float32))
\end{lstlisting}
Code~\ref{IPS12} shows another buggy code snippet, where \textit{SMD} reports only abnormal variance of weights, contrary to \name{} which spots the non-correlation anomaly between the loss and the accuracy. Indeed, the user mistakenly used the logits \textit{Yhat} instead of the sigmoid outputs \textit{Ypred} in the inference of predictions with a threshold of $0.5$, which leads to a wrong calculation of accuracy. Only \name{} cover the coding mistakes in the performance functions through the validation of their correlation coefficient over the training iterations. Thus, the use of the logits instead of probabilities would always yield the class $1$ for predictions; which would break the property of performance metrics correlation as the progress of accuracy metric would be uncorrelated with the loss value.
\begin{lstlisting}[language=Python, caption=Lines 20-28;46-51 of IPS-11,label=IPS11]
def fc_layer(input, size_in, size_out, name="fc"):
    with tf.name_scope(name):
        w = tf.Variable(tf.truncated_normal([size_in, size_out], stddev=0.1))
        b = tf.Variable(tf.constant(0.1, shape=[size_out]))
        activation = tf.nn.relu(tf.matmul(input, w) + b)
        # ...
        return activation

def mnist_model(learning_rate, path):
    # ...
    logits = fc_layer(fc1,1024,10,name="fc2")
    probabilities = tf.nn.softmax(logits)

    with tf.name_scope("xent"):
        xent = tf.reduce_mean(
            tf.nn.softmax_cross_entropy_with_logits(logits=logits,labels=y))
\end{lstlisting}
Even if \name{} has no specific property that would be broken by the fault, \name{} can trigger training difficulty symptoms equivalent to \textit{SMD}'s rules, as shown in the lines of Code~\ref{IPS11}. Indeed, the DL developer unified all the fully-connected layers by a custom function, but he mistakenly applied it over the last fully-connected layer. This induces a useless ReLU activation on the logits before computing softmax-based scores or losses. As a result, the double activation obstructs the information from flowing smoothly and causes a slowness of weight update and vanishing of gradients. Although the same fault-related symptoms were reported by both debugging tools, \name{} reports the \textit{unable to overfit the batch} that strongly indicate a serious model fitting problem.

Nonetheless, \name{} was more verbose triggering several positive checks for each buggy DL training program, and relying on the accepted answer on SO post or the changes in the bug-fixing commit does not allow us to compute the false positives, which represent fired checks without corresponding issue, and false negatives, which represent the existed faults without corresponding fired checks, on these real-world DNN programs. Moreover, finding fired verification routines that are, by definition, related to the actual fixed bug, does not imply that users can fix the bug correctly using the \name{} diagnostic reports. In the next evaluation of usability, we will ask two experienced DL engineers to perform a closed-feedback debugging and fixing loop using \name{} on each of the buggy snippets of code found on StackOverflow. 
\begin{table}
\hspace*{-1cm}
\caption{Debugging Results of StacOverflow TF-based training programs}
\label{SO_Programs_checks}
\begin{tabular}{|c|c|c|}
\hline
\textbf{Program} & \textbf{\name{}'s Fired Check(s)} & \textbf{SMD Rule(s)}\\
\hline
IPS-1&PI-W, PI-b, PI-Loss, Uns-Act-LS
&$R_7$, $R_{10}$\\
\hline
IPS-4&\shortstack{PI-W(1), PI-b(1), \textbf{PI-Loss}(1),\\ Uns-Act-HS(1), Zero-Loss(2)}
&\bm{$R_2$}, \bm{$R_{14}$}\\
\hline
IPS-5&\shortstack{Uns-Inps, \textbf{Un-Fit-Batch}, \textbf{Div-Loss},\\ \textbf{W-Up-Fast}, W-neg}
&$R_{9}$, \bm{$R_{10}$}\\
\hline
IPS-7&\shortstack{PI-W(1), Miss-b(1), PI-Loss(1), \textbf{Inv-Outs}(1),\\ \textbf{Div-Loss}(1), Un-Fit-Batch(1), Sat-Sigmoid(1-2)}
& $R_{10}$ \\
\hline
IPS-11&\shortstack{PI-W(1), PI-b(1), PI-Loss(1), \textbf{Van-Grad}(1),\\ \textbf{W-Up-Slow}(1), \textbf{Un-Fit-Batch}(1), Zero-Loss(2)}
& \shortstack{\bm{$R_8$}, \bm{$R_{10}$}, \bm{$R_{14}$}}\\
\hline
IPS-12&\shortstack{Uns-Inps(1), PI-W(1), PI-Loss(1), \textbf{NR-Loss}(1),\\ HF-Loss(2), Dead-ReLU(2), Uns-Mode-Tr(2)}
&$R_9$, $R_{10}$\\
\hline  
IPS-13&\shortstack{Uns-Inps(1), PI-W(1), PI-Loss(1), \textbf{W-Up-Fast}(2),\\  \textbf{Over-Reg-Loss}(2), \textbf{Un-Fit-Batch}(1-2)}
& - \\
\hline
IPS-14&PI-W, PI-Loss, Un-Fit-Batch
&\shortstack{\bm{$R_8$}, \bm{$R_9$}, \bm{$R_{10}$}}\\
\hline
IPS-15&\shortstack{Uns-Inps, Miss-b, \textbf{Div-Loss},\\ \textbf{Div-W}, \textbf{Div-Grad}, Un-Fit-Batch}
&\bm{$R_9$}, \bm{$R_{10}$}\\
\hline
IPS-17&\shortstack{Uns-Inps(1), Div-Loss(1),\\ Un-Fit-Batch(1), W-Up-Slow(2)}
&\bm{$R_9$}, \bm{$R_{10}$}\\
\hline
\end{tabular}
\end{table}
\footnotetext[1]{Poor initial bias}
\footnotetext[2]{Missing bias}
\footnotetext[3]{Saturated Sigmoid} 
\begin{table}
\hspace*{-1cm}
\caption{Debugging Results of Github TF-based training programs}
\label{GH_Programs_checks}
\begin{tabular}{|c|c|c|}
\hline
\textbf{Program} & \textbf{\name{}'s Fired Check(s)} & \textbf{SM Debugger's Fired Rule(s)}\\
\hline
DLT\_0edb182&PI-W, PI-B, Uns-Act-LS, SD-Loss
& \bm{$R_{14}$} \\
\hline
DLT\_20d1b59&\shortstack{PI-W, PI-B, PI-Loss, Uns-Act-LS,\\ SD-Loss, NR-Loss, Un-Fit-Batch}
& \bm{$R_{14}$} \\
\hline
DLT\_437c9c2&Un-Sym-W, \textbf{PI-Loss(Huge Err)}
&\shortstack{$R_8$, $R_9$, $R_{10}$}\\
\hline
DLT\_726b371&\shortstack{ PI-W, PI-B, PI-Loss,\\ \textbf{Div-Loss}, Uns-Act-LS}
& $R_7$, $R_{10}$ \\
\hline
DLT\_ded6612&\shortstack{\textbf{PI-W}, \textbf{PI-B}, PI-Loss,\\ Uns-Act-LS, SD-Loss}
& $R_2$, \bm{$R_{10}$}, $R_{14}$\\
\hline
FCN\_b170a9b&\shortstack{\textbf{PI-Loss}, Uns-Act-LS, Dead-ReLU,\\ Van-Grad, Un-Fit-Batch, \textbf{SD-Loss}}
&$R_{8}$, \bm{$R_{14}$}\\
\hline
TFE\_333&\shortstack{\textbf{PI-W}, PI-B, PI-Loss,\\ W-Up-Slow, SD-Loss, NR-Loss}
& - \\
\hline
TFE\_368&\shortstack{PI-W, PI-B, PI-Loss,\\ \textbf{W-Up-Fast}, Zero-Loss}
& \bm{$R_{10}$}\\
\hline
TFE\_742675d&Un-Sym-W, \textbf{PI-Loss(Huge Err)}
& \shortstack{$R_8$, $R_9$, $R_{10}$}\\
\hline
TFE\_bc09f95&\shortstack{PI-W, PI-B, \textbf{PI-Loss(Huge Err)}, LR-Loss\\ Un-Fit-Batch, W-Up-Slow, Van-Grad}
&\shortstack{$R_8$, $R_{10}$, \bm{$R_{14}$}}\\
\hline
\end{tabular}
\end{table}
\subsection{Usability of \name{}}
In this section, we report about a usability study performed with two professional DL engineers, namely $E_1$ and $E_2$, with the aim to assess the relevance of \name{}'s error messages at guiding developers in identifying the root cause of bugs and fixing them. The two DL engineers involved in the study have 3 years of experience working with TensorFlow. They are currently employed in AI software development teams in technology companies, building DL-based software systems. To assess the relevance of  \name{}'s error messages, we provided the two engineers with $10$ buggy DL training programs (i.e., the ones reproduced from SO posts in Table~\ref{SO_Programs_checks}) and asked them to use \name{} for debugging them and fixing the identified bugs. We made it clear to them that we are not evaluating their ability to detect the bugs based on their knowledge and asked them to only follow the clues contained in the debugging logs generated by \name{}. We also asked them to explain how they inferred the root cause of faults based on the information provided by \name{} and to suggest fixes. We made the decision to ask participants to fix detected bugs because to allow them to explore as many bugs as possible. The DL programs used in our study contain more than one bug, and to progress in the debugging process participants have to fix detected bugs. For instance, unnormalized inputs may cause the divergence of the training and turn the loss quickly to NaN. Hence, it obstructs the training dynamics, and consequently, the \name{}'s debugging session. Thus, the DL engineer should normalize the inputs to fix the issue and restart the debugging session. In Table~\ref{SO_Programs_checks}, we added ($n$) next to verification routines to identify the debugging session during which the routine was triggered. For example, Zero-Loss(2) means that during the second debugging session ($2$) after fixing some of the issues, \name{} newly reports the Zero-Loss warning that indicates strongly a lack of regularization in the DNN.\\
Although the bug fixes suggested by the two engineers are quite different, they have performed similar sequence of debugging sessions where they focus on fixing the same training issues at each session and have received mostly the same amount of notification messages (from the fired verification routines) at the different debugging steps for each given buggy program. 

Table~\ref{SO_Programs_Engineers_Fixes} shows the fixes suggested by the engineers based on the error messages generated by \name{}.
As discussed in the paragraph~\ref{shrinking-prog-states}, many training issues are correlated and induced by the same bug. In Table~\ref{SO_Programs_Engineers_Fixes}, based on the explanations provided by the engineers, we present the main issues reported by \name{} that lead them to identify and localize the root cause of the faults, whether it being caused by a coding bug or a misconfiguration.\\
As can be seen, most of the found faults are common (almost $96.5\%$ of cases), which reinforces the argument that our verification routines are quite precise, regarding the problematic component and its occurring symptoms. However, given the recommended fixes from SO post's answers (see Table~\ref{SO_Programs_Engineers_Fixes}), we can see that the majority of fixes provided by the engineers are different. In the following, we discuss these differences in detail. 

First, we observe that there are emergent fixing patterns followed by the community, which are not always efficient. Indeed, we can consider them as technical debts because they enable the convergence of training and fitting the DNN, but the main root cause of the issue is not solved, which provokes the same issue following any further changes on the DNN program or the inputs data. For instance, the buggy TF programs, \textit{IPS-5}, \textit{IPS-14}, \textit{IPS-15} and \textit{IPS-17} share the main issue of diverging loss problem, which turns its value to NaN and obstructs the training process. The initial fixes recommended by the community consists in improving the optimization routines, including the decrease of learning rate or the substitution of regular gradient-descent by advanced variants with internal adaptive learning rate like Momentum or Adam. When using \name{}, our two engineers were able to find the main root cause of diverging loss in buggy TF programs, \textit{IPS-4}, \textit{IPS-15} and \textit{IPS-17}, which is the unnormalized inputs. In the case of \textit{IPS-14} the problem was both the inefficient initial random weights and the poorly designed loss (i.e., using sum over the instances' errors instead of average). Without a fine-grained analysis tool like \name{} it was difficult for Stack Overflow users (who suggested solutions) to uncover this. The main lesson that can be derived from this example is that multiple poor design choices and coding mistakes can induce well-conditioning to the loss minimization problem, and as result, may be the origin of its divergence. Therefore, tuning the learning rate blindly will only make the training program run at its minimum capacity, and hence, the real bugs will remain hidden. By decreasing the learning rate in \textit{IPS-15} TF program from $0.5$ to $0.0005$ to enable learning under the condition of unnormalized inputs, SO users only introduced a technical debt in their program. However, \name{} steers the engineers towards fixing permanently the root cause problem, which is the inappropriate scale of features' values. Concerning the same bug of unnormalized inputs in the TF program \textit{IPS-17}, \name{} reports in the second debugging session following the normalization of the inputs that the weights is slowly updating, which lead both our two engineers to fix it by increasing further the learning rate (i.e., both engineers ended up with taking actions that are totally the inverse of the initial recommended fix).

Second, we found that \name{} spots the major bug preventing the training program from fitting the model in regard to the buggy programs \textit{IPS-4}, \textit{IPS-11}, \textit{IPS-12}, and \textit{IPS-13}. Indeed, our engineers, $E_1$ and $E_2$, confirmed that the poor initial loss check alerted them to the fact that the loss is not a scalar, which led them to add the average as loss reduction strategy in \textit{IPS-4} program. In \textit{IPS-11}, they mentioned that the vanishing gradient problem starting from the first training iterations at the last dense layer, guides them to inspect the last layer (logits). They found that the program uses the same implemented function \textit{fc\_layer} that performs ReLU as non-linear activation for all fully connected layers. However, this useless non-linear activation erases relevant learned information and obstructs the training, because the nullified negative values make all their corresponding labels share the same probability after applying the softmax. In the case of \textit{IPS-12}, the non-representative loss check that remains active over the iterations displaying increasingly smaller correlation, persuaded both DL engineers that either the accuracy or the loss function is mistaken, so they check them out carefully and found a typo in the accuracy function (i.e., passing logits instead of probabilities as predictions). Regarding the \textit{IPS-13} program, the engineers adjust the learning rate and the norm penalty values to make the program satisfy the verification routines related to standard regularization risks and unstable learning of parameters that triggered, respectively, fast updated weights and overwhelming regularization loss verification routines. However, the two DL engineers failed to correctly fix the major issues contained in \textit{IPS-1} and \textit{IPS-7}. They proposed to change the sum reduction strategy by the average and to pass the probabilities instead of logits to the cross-entropy loss, to correct, respectively, the poor initial loss in \textit{IPS-1} and diverging loss in \textit{IPS-7}. However, the real bug reported by the user was the loss turning into NaN values. Because this exception is raised infrequently, the problem cannot be always detected easily. Also, \name{} considers the NaN loss as diverging loss and cannot provide further indications about any potential root cause. As a result of this limitation of \name{}, both engineers could not identify the root cause of the issue by relying on the message generated by \name{}, which claims that the cross-entropy loss function contains the expression $t\times\log(y)$, which renders NaN ($0\times\log(0)$) when $t=0$ and $y$ approaches to $0$. In fact, the recommended fix was adding an epsilon ($\epsilon>0$) to avoid the undefined expression, but a more appropriate repair for the loss numerical instability is to use, instead of hand-crafted loss, the recent TF built-in logit-based loss function, including both softmax and cross-entropy, which is numerically stable. By definition, our property-based debugging process relies on the available data and tries to catch properties' violations through watching the execution of the training program. Therefore, it cannot detect numerical instabilities that occur in particular ranges of values. Odena and Goodfellow~\cite{tensorfuzz} proposed a coverage-guided fuzzing testing tool that is able to find mutated inputs triggering erroneous TF program's behaviors including NaNs raised by unstable math computation. Their evaluation shows that original and even randomly synthetic data have low chances to trigger such corner-case behaviors and expose these numerically unstable math functions. Therefore, to the best of our knowledge, fuzzy testing approaches are more suitable for detecting numerical instabilities in DNN training programs.

Besides, the results show that \name{} guided the DL engineers towards detecting other issues that do not prevent the program from training, but which should be addressed to improve performance and avoid all the non-optimal local minima in the loss curve. As can be seen in Table~\ref{SO_Programs_Engineers_Fixes}, most of these additional detected issues are related to poor initial parameters, lack of regularization, and unstable learning velocity between the layers. Nevertheless, we observed that the DL engineers proposed some different repairs to fix the same issue identified through the debugging sessions using \name{}. This means that there are multiple possible fixes for the same issue identified by \name{} and that the choice of a specific fix depends on the knowledge and experience of the engineer. Indeed, we found that for some issues, our engineers, $E_1$ and $E_2$ were able to turn off the alert and improve the performance of the training using different techniques. For instance, \name{} spots unstable activations with low variance regarding the first convolutional layer in the trained DNN of \textit{IPS-1} program. Engineer $E_1$ understood that the first convolutional layer was not optimally learning the features, which led him to carefully increase the learning rate and solve the problem. In another example, Engineer $E_2$ understood that the difference in magnitude of updates between intermediate layers causes a problem of internal covariate shift, which can be solved by adding batch normalization following each intermediate layer. He went on and implemented this fix. In the future, we will examine further the fixes proposed by DL practitioners to overcome the studied training issues and analyze their impact on the quality of the code and the performance of the DNN training program.
 \begin{table}
\hspace*{-1cm}
\caption{The repairs suggested by DL Engineers ($E_1$ and $E_2$) for real-world buggy TF programs}
\label{SO_Programs_Engineers_Fixes}
\resizebox{.95\textwidth}{!}{
\begin{tabular}{|c|c|c|}
\hline
\textbf{Program} & \textbf{Fired Checks} & \textbf{Suggested Fixes}\\
\hline
\multirow{3}{*}{IPS-1}&PI-W, PI-b&change $W$ and $b$ initializers\\
\cline{2-3}
&PI-Loss&set average instead of sum for loss reduction\\
\cline{2-3}
&Uns-Act-LS&$E_1$: increase $\eta$ | $E_2$: add batch-norms\\
\hline
\multirow{4}{*}{IPS-4}&PI-W, PI-b&change $W$ and $b$ initializers\\
\cline{2-3}
&PI-Loss&set average as loss reduction strategy\\
\cline{2-3}
&Zero-Loss&add dropout layers\\
\cline{2-3}
&Redundant-Layers&remove a dense layer\\
\hline
IPS-5&Uns-Inps, Div-Loss&normalize the data\\
\hline
\multirow{4}{*}{IPS-7}&PI-W, Miss-b&change $W$ initializers and add null $b$\\
\cline{2-3}
&Inv-Outs&add output activation layer\\
\cline{2-3}
&Div-Loss&passing the probas instead of logits to the loss\\
\cline{2-3}
&Sat-Sigmoid&change hidden activations (Sigmoid to ReLU)\\
\hline
\multirow{3}{*}{IPS-11}&PI-W, PI-b&change $W$ and $b$ initializers\\
\cline{2-3}
&Van-Grad (last layer)&remove ReLU on the logits\\
\cline{2-3}
&Zero-Loss&add dropout for the dense layer\\
\hline
\multirow{4}{*}{IPS-12}&Uns-Inps&normalize the data\\
\cline{2-3}
&PI-W&change $W$ initializers\\
\cline{2-3}
&NR-Loss&fix typo in the accuracy\\
\cline{2-3}
&HF-Loss, Uns-Inference&increase the $keep\_p$ for dropout layers\\
\hline
\multirow{4}{*}{IPS-13}&Uns-Inps&normalize the data\\
\cline{2-3}
&PI-W&change $W$ initializers\\
\cline{2-3}
&W-Up-Fast&decrease the learning rate $\eta$\\
\cline{2-3}
&Over-Reg-Loss&decrease the norm penalty $\lambda$\\
\hline
\multirow{2}{*}{IPS-14}&PI-W&change $W$ initializers\\
\cline{2-3}
&PI-Loss&set average instead of sum for loss reduction\\
\hline
\multirow{2}{*}{IPS-15}&Uns-Inps&normalize the data\\
\cline{2-3}
&Miss-B&add null $b$\\
\hline
\multirow{2}{*}{IPS-17}&Uns-Inps&normalize the data\\
\cline{2-3}
&W-Up-Slow&increase the learning rate $\eta$\\
\hline
\end{tabular}
}
\end{table}
\section{Threats to Validity}
\label{threats}

\textbf{Selection Bias. }
The selection of the subject DL training programs could be an internal threat to validity. In this paper, we try to counter this issue by using two complementary empirical evaluations on synthetic buggy programs and real buggy programs. We used diverse base DL training programs that solve regression and classification problems. They encode DL models with different architectures, complexities, and techniques. Moreover, the base synthetic DL programs have official implementation references and run on widely-studied datasets (i.e., Auto-MPG, MNIST, CIFAR10). Then, the buggy versions of these synthetic DL training programs were created to mimic the DL faults reported and studied in empirical studies on DL programs’ defects. The discussion on the results of the evaluation on the synthetic buggy programs provides insights on how the properties and heuristics deployed in the verification routines were able to detect the behavioral training issues caused by the injected bug. Indeed, we leveraged totally-disconnected workflows for the construction of the synthetic buggy programs and the verification routines. Figure shows how the abstraction of DL faults to synthetize buggy subjects was done essentially on former empirical studies' datasets and our manual inspection of TF-related SO posts. On the other hand, Figure shows how the design of verification routines was guided by applied DL research works and technical DL expert reports about troubleshooting to codify fundamental properties of DL programs and practical heuristics on proper DL training dynamics. It was crucial to keep synthetic bugs and verification routines separate in order to avoid hard-coded verifications that target the mainstream DL faults identified by the community. Although the injection of DL faults was done on reference DL models with minimal code changes, it is still a human-crafted process that may contain imperfections. Furthermore, each synthetic buggy program represents a well-designed base program with only a single bug injected. Thus, we complement the evaluation using $20$ real-world buggy programs from the dataset provided by Zhang et al., which are related to the scope of our targeted DL bugs, and represent more realistic conditions of debugging since they may contain different issues simultaneously, and even issues that are not uncovered by maintainers yet.

\textbf{Settings' Generalizability and Transferability. }The setup of heuristic-based thresholds could be an external threat to validity. In our design and implementation of property-based verification routines, we relied on the original documentation sources~\cite{glorot2010,hinton2012practical,ceceliaCheckList,troubleshootingD4J} that described the issues, to set up the thresholds, when they are indicated. Nonetheless, some properties were always presented and studied through visual plots comprehensible by humans. This makes the design of metrics and their thresholds challenging, but we focus on the abstract violation traits of the properties rather than the concrete studied instances, in order to be able to construct a verification routine based on the foreknown training misbehaviors, as discussed in the implementation strategies (section~\ref{dynamic_heuristics}). Therefore, given the dynamic and continuous aspects of \name{}, it was possible to set up intuitively pessimistic thresholds, to be our default configuration, in order to ensure a high coverage with less false alarms when enough monitored training iterations were executed. Indeed, we avoided the empirical tuning of these thresholds because we did not have access to a large benchmark of reproducible buggy programs. Besides, the shrinking of suspicious program state includes the computed metrics and thresholds that were behind the fired checks, which is a mitigation strategy to help the user ignore false positives. Given the variables’ instatiations in the violated rules, users can also assess the sensitivity of the thresholds on their DL application domain, which allows them to set up a more precise custom configuration for \name{}.
\section{Limitations and Future Work}
\label{limits_and_future}
\textbf{DL knowledge remains crucial in the debugging of DNN training programs. }Many of our verification routines are implemented using statistical metrics and heuristics that are very related to inefficient training traits, so they are often connected to multiple possible root causes. In the inverse direction, most of the training pitfalls would trigger multiple fired checks because a faulty component often violates its related properties, and consequently, leads to other training properties’ violations. For example, a bad initialization of weights violates the required asymmetry between neurons, however, the resulting unbreaking symmetry would lead to other issues, like over-negative weights, Dead ReLUs, and vanishing gradients. Indeed, all the neurons will receive identical gradients and evolve throughout training, effectively preventing different neurons from learning different things. Thus, it is likely that a non-optimal gradient update, from the starting iterations, would be applied symmetrically to all the neurons, and consequently, would cause the stagnation of the neural network.\\
\name{} incorporates dynamic verifications with periodic inspection reports, narrowing down the space of suspicious states, which help the user recognize fault patterns and identify the root cause by analyzing the timeline of fired checks (i.e. their chronological order) and the reported information (i.e., positions, metrics, thresholds, etc.). Nevertheless, these mitigation strategies require sufficient DL knowledge and skills. Developers need to be sensitive to such details in order to be able to efficiently interpret the debugging reports.\\
In the future work, some engineering efforts are needed to fill-in the messages with relevant links to resources and documentation in relation to the detected properties’ violations. Besides, larger datasets of buggy DL training programs could serve to better learn the patterns of the common DL faults and be able to propose DL faults’ models that synthesize all these metrics, heuristics and properties, aiming at detecting accurately the main root cause and the observed inefficiency. Furthermore, a static code analyzer can be applied to load the code structure and components into the debugger's state, thereby reducing the number of candidate errors and improving the localization of bugs. This also opens the path to include fixes’ recommendation and study solutions for automatic program repairs.

\textbf{Scoping on Feedforward Neural Network Architecture. }Many of the targeted training pitfalls and the proposed properties are generalizable to other model architectures~\cite{liu2017survey}, but in this paper, we focus on their application for the feedforward architecture, which is, first, widely used in several regression and classification problems, as well as, reinforcement learning tasks. Second, it is the basic neural network model that influences novel architectures, and even represents one of their building blocks. Nonetheless, recurrent neural networks (RNNs) have faced more severe gradient problems~\cite{pascanu2013difficulty}, including vanishing and exploding phenomena. Generative adversarial networks (GANs), which particularly leverage two on-training models in a min-max game, raise novel training issues in relation to the learning stability and convergence, in addition to other GAN-specific problems~\cite{roth2017stabilizing, arjovsky2017towards, karras2017progressive} such as mode collapse, where the generator outperforms quickly the discriminator, without fitting the data distribution, but through simply rotating over few data types. In the future, we plan to examine the training issues experienced by these novel architectures, in order to revisit our verifications in the future versions of \name{} and enlarge our scope of DL model architectures. Indeed, we should adjust their integrated metrics and boundary conditions, as well as, elaborate advanced checks to detect the violations of their specific design properties and statistical learning principles.

\textbf{Usability Study Limited on the Mappings from Checks to Fixes. } Given the real-world DL programs published by SO users, we recruit two DL engineers from two different teams and having different backgrounds for a usability evaluation, focusing on the mappings from the fired checks to the DL program fixes. Nonetheless, other important dimensions such as the time spent on debugging and the relevance of the proposed fixes, could be assessed in comparison with the trial-and-error debugging process without \name{}. Therefore, we plan to set up controlled and monitored experiments of debugging DL-based software programs having different complexity, in both modes, assisted with \name{} and not assisted. This allows to perform a broader usability evaluation on a larger group of participants including students, junior/senior DL engineers, junior/senior data scientists, and junior/senior DL researchers.
\section{Related Work}
\label{relatedWork}
\textbf{Software Testing Approaches for DL Models.}
Leveraging concepts from both software testing and adversarial DL domains, researchers have proposed automated testing methods for DNN-based models~\cite{DLTestReview}. The majority of them focus on testing the model by automatically generating synthetic test inputs with high-fault revealing ability. Pei et al.~\cite{deepxplore} proposed the first white-box test adequacy criterion for DL models, named Neuron Coverage (NC), which is inspired from code coverage in traditional software testing. Next, Ma et al.~\cite{deepgauge} generalized the concept of NC by defining a set of multi-granularity testing criteria, including neuron-level and layer-level coverage. Once the test adequacy criterion is set up, a data generator should systematically produce test cases, enhancing the target test adequacy measure. Thus, the proposed data generators solve this maximization problem relying on gradient-based optimizers~\cite{deepxplore}~\cite{guo2018dlfuzz}, gradient-free optimizers(\ie metaheuristics)~\cite{deepevolution}, and greedy search-based process~\cite{deeptest}~\cite{tensorfuzz}~\cite{deephunter}. As DNNs cannot guarantee $100\%$ of correct answers, we are always capable of finding adversarial inputs for which the DNN's predictions are wrong. Additionally, there is no deterministic relation between mismatched pairs of (input,output) and bugs, i.e., we cannot conclude the existence of a bug in the DNN program by observing unexpected outputs of synthetic test cases. To improve the trustworthiness of a DNN-based software system, we propose to debug carefully the training program from the ingestion of data to the generation of the final model. Our debugging approach, \name{}, aims to assert the correctness of the DNN training's program implementation and configuration, as well as, providing a preliminary validation of its design quality. Once the DNN training program is validated to be bug-free and appropriately configured, testing the trained model against genuine and synthetic test datasets is still required to assert the effectiveness and the robustness of the DNN in performing its target task under different condition.\\
\textbf{Software Testing Approaches for Debugging Training Algorithms.}
In contrast with model testing approaches, few research works have concentrated on the correctness of DNN-based software programs. Dwarakanath et al.~\cite{LastMT} define metamorphic relations (MRs) between data transformations and their resulting effects to validate the correctness of the software system. However, these MRs can be seen as high-level properties of the self-learning software system. For instance, the permutation of the order of data instances or the linear scaling of data features (i.e., train and test datasets), should not affect the performance of the training. Thus, even if these high-level MRs may detect software bugs that substantially affect the behavior of DNN training programs, they do not ensure finding the bugs contained in DNN-based systems, and do not provide guidance towards identifying their root causes. Given the mathematical and statistical nature of DNN-based software systems, Selsam et al.~\cite{bugfreeMLS} propose to use machine-checkable proofs to validate their implementation. However, the proposed approach is difficult to adopt in practice because it is built to test a code written from scratch, and nowadays, developers leverage third-party DL libraries to construct reliable and scalable DNN-based software systems.\\
Contrary to these approaches, \name{} can be applied to code using third party libraries or written from scratch. As shown during the evaluation, the faulty components can be detected by \name{} when they violate fundamental properties of a DNN training algorithm. In addition to its fault detection ability, \name{}'s internal set of verification routines captures the training issue, at its first appearance, and provides rich feedback on the problematic situation, in order to help users identifying and localizing the root cause.\\
\textbf{Interactive VA systems for Diagnosing and Refining DL Algorithms and Models.}
Many VA systems (VAS) provide a model's diagnosis that allow detecting issues on different abstraction levels. Some of them focus on feature importance and model behaviors against real~\cite{zhang2018manifold} or adversarial~\cite{liu2018analyzing} examples; others focus on neuron activations~\cite{kahng2017cti}. Moreover, advanced visualization systems~\cite{liu2016towards}~\cite{deepeyes} go beyond the diagnosis of the DNN and propose refinements required to overcome the detected issues. For instance, Liu et al.~\cite{liu2016towards} constructed an interactive VAS, CNNvis, that extracts successive snapshots of the on-training CNN and analyzes it in-depth using rectangle packing, matrix ordering, and biclustering-based edge bundling in order to cluster the neurons, their interactions, their derived features and roles in relation with the target task. Instead of conducting offline diagnosis, Pezzotti et al.~\cite{deepeyes} proposed an online progressive VAS that provides continuous live feedback on the on-training DNN. Both of these previous works on diagnosis and refinement via visualization demonstrate how rich visual insights can be interpreted by an expert to identify possible modeling issues and make decisions about DNN's refinements.  
Nevertheless, diagnostics via VAS are interactive sessions that require DL engineers to select components to watch beforehand. This makes diagnosis via VAS an expensive process that focuses, particularly, on data and design improvements to enhance the performance results of the trained DNN. In this paper, we propose \name{}, an end-to-end automated debugging approach that is able to detect several implementation bugs and system misconfigurations, as well as poor design choices, with minimum human intervention.
\section{Conclusion}
\label{conclusion}
This paper reports about the design and implementation of \name{}, an end-to-end automated debugging approach for DNN training programs. To develop \name{}, we systematically gather a catalog of pitfalls commonly occurring in the development of DNN training programs. Then, we explore various resources on applied DL researches and technical reports with aim of distilling fundamental properties and practical heuristics that can be codified into verification routines to detect the DL pitfalls' resulting faults and training issues. Next, we develop a property-based debugging approach, named \name{}, that orchestrates the different properties' verification over multiple phases. On the one hand, we evaluate \name{} on synthetic buggy programs that contain each an injected DL fault. The results show its effectiveness at detecting DL coding bugs and misconfigurations with (precision, recall), respectively, equal to ($90\%$, $96.4\%$) and ($77\%$, $83.3\%$). Moreover, we compare \name{} with Amazaon Sagemaker Rule-based Debugger(\textit{SMD}) on real-world buggy programs extracted from SO and GH. The results show that \name{} outperforms \textit{SMD} by detecting $75\%$ rather than $60\%$ of the total of reported bugs in the SO post accepted answer or the bug-fixing commit message. Indeed, \name{} effectively captured the slightest violation of all mandatory training assumptions, even those having only a minor negative effects on the training process, providing sufficient feedback on any problematic issue in DNN program. Using \name{}, two DL engineers were able to successfully locate and fix $93.33\%$ of bugs contained in $10$ buggy TF programs.
In the future works, we plan extent \name{} to include code recipes (i.e., generic snippet of code, one or multiple lines of code, that could characterize a particular fix of a coding bug, an ineffective implementation or a misuse of APIs.)  for common errors in relation to training anomalies with the aim to point DL engineer to the exact lines of code that caused the issues and to include an automatic generation of fixes for these detected issues.
\begin{acks}
This work is partly funded by the Natural Sciences and Engineering Research Council of Canada (NSERC) and the Fonds de Recherche du Quebec (FRQ).
\end{acks}
\bibliographystyle{ACM-Reference-Format}
\bibliography{main}
\end{document}